\documentclass[12pt]{article}

\usepackage{rotating}
\usepackage{geometry}
\usepackage{enumitem}
\usepackage[toc,page]{appendix}
\usepackage{pdflscape}
\usepackage{algorithmic}
\usepackage{enumitem}
\usepackage[nocompress]{cite}
\usepackage{array}
\usepackage[caption=false,font=footnotesize,labelfont=sf,textfont=sf]{subfig}
\usepackage{url}
\usepackage{amsmath}
\usepackage{booktabs}
\usepackage{multirow}
\usepackage{dingbat}
\usepackage{bbding}
\usepackage{pifont}
\usepackage{hyperref}
\usepackage{placeins}
\usepackage{changepage}
\usepackage{longtable}
\usepackage{afterpage}

\usepackage{tikzsymbols}

\graphicspath{figures/}
\DeclareGraphicsExtensions{.pdf}

\usepackage{adjustbox}

\newcolumntype{R}[2]{%
	>{\adjustbox{angle=#1,lap=\width-(#2)}\bgroup}%
	l%
	<{\egroup}%
}
\newcommand*\rot{\multicolumn{1}{R{45}{1em}}}

\renewcommand*\rot[2]{\multicolumn{1}{R{#1}{#2}}}

\begin{document}

\title{A Survey on the Evaluation of Clone Detection Performance and Benchmarking}

\author{Jeffrey Svajlenko and Chanchal K. Roy\\ Department of Computer Science \\ University of Saskatchewan, Canada}

\maketitle
\pagebreak

\tableofcontents
\pagebreak
\listoffigures
\pagebreak
\listoftables
\pagebreak

\section{Introduction}
\label{sec:introduction}

Code clones are pairs of code fragments, within or between software systems, that are similar.  Software developers create clones when they re-use code using copy and paste, although clones can arise for a number of different reasons~\cite{Roy07asurvey}.  Clones can have a negative impact on software development.  They needlessly increase the size of a software system, increasing the costs of software maintenance and re-engineering.  When buggy code is cloned, the bug is duplicated throughout the system, complicating debugging and bug fixing.  Clones can even lead to new bugs when the evolution of a code fragment is not appropriately propagated to its clones.  Cloning can also have benefits, such as accelerated software development and increasing decoupling~\cite{Roy07asurvey}.  However, it is important that developers keep track of their clones in order to manage their negative effects.  Datasets of clones have been shown to have applications in code search, mining for new APIs, bug detection, security vulnerability detection, malware detection, and so on.

The importance and application of code clones has motivated the creation of many clone detection tools and techniques.  We have found at least 184 clone detection tools/techniques in the literature.  However, despite this this interest in clone detection, there has been a lack evaluation of clone detection performance.  In particular, clone detection benchmarks and clone detection tool comparison studies.  Partly, this has been due to difficulty in creating clone benchmarks and standardizing clone detection evaluation.  Specifically, clone detection tools and techniques should be evaluated for their clone detection performance, in terms of recall and precision, and their execution performance, in terms of execution time and scalability.

In this paper, we investigate the state of clone detection tool evaluation.  We begin by surveying the clone detection benchmarks, and performing a multi-faceted evaluation and comparison of their features and capabilities.  We then survey the existing clone detection tool and technique publications, and evaluate how the authors of these works evaluate their own tools/techniques.  We rank the individual works by how well they measure recall, precision, execution time and scalability.  We select the works the best evaluate all four metrics as exemplars that should be considered by future researchers publishing clone detection tools/techniques when designing the empirical evaluation of their tool/technique.  We measure statistics on tool evaluation by the authors, and find that evaluation is poor amongst the authors.  We finish our investigation into clone detection evaluation by surveying the existing tool comparison studies, including both the qualitative and quantitative studies.  

We find that there have historically been very few clone benchmarks, but a few new benchmarks are of high quality.  We find that clone detection tool and technique authors often do not evaluate their tools, and very rarely do they thoroughly evaluate their tools across all four performance metrics.  This may be due to the historical lack of clone benchmarks, meaning that clone detection research has often been accepted on the merits/novelty of its algorithms, and not on empirical evidence of its performance.  We find a few works that have thorough evaluations, that can serve as exemplars for researchers publishing new clone detection papers and for the peer reviewers judging these works.  We hope our evaluation ranking and exemplars will encourage better experimental evaluations by the tool/technique authors.   We find that there is very few tool comparison studies, but a few are high quality works. This study considers publications and benchmarks until 2017. 

This survey paper is organized as follows.  In Section~\ref{sec:cloning} we discuss essential code clone and clone detection theory.  Then in Section~\ref{sec:benchmarkingTheory} we discuss clone benchmarking theory.  Section~\ref{sec:benchmarks} contains our survey on the existing clone detection benchmarks, and our evaluation and comparison of their features.  Section~\ref{sec:toolEvaluations} contains our survey on how authors of clone detection tool and technique authors evaluate their works.  We rank these works, and discuss statistics on the standard of tool evaluation by the authors.  Section~\ref{sec:surveyToolComparisons} contains our survey on the tool comparison studies, both qualitative and quantitative.  We discuss and critique each work.  We conclude our work in Section~\ref{sec:conclusion}.

\section{Cloning Theory}
\label{sec:cloning}

In this section we provide some background on code clones and their detection.  In this survey paper, we focus on the evaluation of clone detection tools.  We provide just a summary of the important clone definitions and background for this purpose.  For general clone knowledge, an excellent survey is provided by Roy et al.~\cite{Roy07asurvey}.  For an extensive survey on clone detection algorithms and techniques, excellent surveys by Roy et al.~\cite{Roy:2009:CEC:1530898.1531101} and Rattan et al.~\cite{rattan} are available.  Zibran et al.~\cite{ZibranSurvey} provide a survey on clone management research.

	\subsection{Code Clones}
	Code clones are instances of similar \textit{code fragments}.  Similarity can take many forms, although typically syntactic and semantic/functional similarity are considered.  Code clones are typically reported as \textit{clone pairs} or \textit{clone classes}.  The following are standard definitions for these terms~\cite{Roy07asurvey}:

	\begin{description}[leftmargin=1cm,labelindent=0.5cm]
	\item[Code Fragment] A contiguous region of source code within a source file. Specified by the triple $(f,s,e)$, including the source file $f$, the line the code fragment starts on $s$, and the line it ends on $e$.
	
	\item[Clone Pair] A pair of code fragments that are similar, for some definition of similarity.  Specified by the tuple $(f_{1}, f_{2}, \tau)$, including the similar code fragments $f_{1}$ and $f_{2}$ and their type of similarity $\tau$.
	
	\item[Clone Class]
	A set of code fragments that are similar.  Specified by the tuple $(f_{1}, f_{2}, ..., f_{n}, \tau)$, including the $n$ similar code fragments and their type of similarity.
	\end{description}

	Clone pairs may be reported with or without the type of similarity, $\tau$, explicitly indicated.  Additionally, some clone detectors mix clones of various types within the same clone class.  The base requirement of a clone class is that every pair of code fragments within the class form a valid clone pair.

	\subsection{Clone Types}
	Researchers agree upon four primary clone types, which are mutually exclusive, and defined with respect to the detection capabilities needed to detect them~\cite{Roy07asurvey}:

	\begin{description}[leftmargin=1cm,labelindent=0.5cm]
		\item[Type-1] Identical code fragments, ignoring differences in white-space, code formatting/style and comments.
		\item[Type-2] Structurally/syntactically identical code fragments, ignoring differences in identifier names and literal values, as well as differences in white-space, code formatting/style and comments.
		\item[Type-3] Syntactically similar code fragments, with differences at the statement level.  The code fragments have statements added, removed or modifiers with respect to each other.
		\item[Type-4] Syntactically dissimilar code fragments that implement the same or similar functionality.
	\end{description}

	The Type-1 and Type-2 clones are well defined, while the Type-3 and Type-4 clones are fuzzy.  While researchers agree upon these definitions, they may disagree upon what is the minimum syntactical similarity of a Type-3 clone, or the minimum functional similarity of a Type-4 clone.  We discuss these primary clone types in more detail in the following.
	

		\subsubsection{Type-1 Clones}
		Type-1 clones are pairs of identical code fragments, when we ignore trivial differences such as extraneous white-space, code formatting/style and commenting.  Type-1 clones can occur when a code fragment is copy and pasted, with only trivial modifications to match the destination's formatting and code style, perhaps with the addition, removal or modification of comments.  
		
		Clone detectors detect Type-1 clones by parsing the source-code in a way that removes or normalizes the allowed differences, and detects those code fragments that are textually/syntactically identical.  For example, a clone detector can tokenize the source code, and search for the identical token sequences to detect Type-1 clones.  Tokenization retains only the language tokens, and drops the commenting, white-space and formatting.  Note in some languages, such as Python, some white-space are tokens as they have syntactical meaning beyond token separation

		\subsubsection{Type-2 Clones}
		Type-2 clones are structural clones: pairs of code fragments that are syntactically identical when we ignore differences in identifier names and literal values, in addition to white-space, layout/style and comments.  Type-2 clones can occur when a developer has reused a code fragment by copy and paste, and has renamed one or more variables to better match the destination.  In addition to renamed variables, Type-2 clones can have renamed/changed constants, classes, method names, and so on, as well as changes in literal values and types.  Type-2 clones can be detected by normalizing identifier names and literal values, in addition to the Type-1 normalizations, and detecting syntactically/structurally identical code fragments.  For example, a clone detector could tokenize the code, and replace each identifier token with a common token (e.g., `identifier') and similar for each literal token (e.g., `literal'), and detect the token sequences that are identical as Type-2 clones.
		
		This is a broad definition of a Type-2 clones, and there are many kinds of Type-2 clones.  For example, a Type-2 clone may be syntactically and semantically identical code fragments except for the systematic renaming of a single variable.  Another valid Type-2 clone is a pair of code fragments that are structurally identical, but have completely different identifier names, including different class types and method names.  In this case, the two code fragments could be semantically different, but structurally the same after normalization.

		\subsubsection{Type-3 Clones}
		Type-3 clones are those that contain statement-level differences, with the code fragments containing statements added/removed or modified with respect to each other.  Type-3 clones can occur when a code fragment has been duplicated and then modified at the statement-level to satisfy new requirements.  The duplicated code fragment could be extended with new statements to add a new feature, statements could be removed to remove an unneeded feature, or statements could be modified to adjust an existing feature for the new use-case.  The Type-3 clone definition also allows for Type-1 and Type-2 differences to occur between the code fragments.  Significant Type-2 differences means a Type-3 clone could be a pair of structurally similar code fragments, that contain significant differences in identifier names and literal values.
		
		Clone detectors can detect Type-3 clones by using clone similarity metrics that measure the syntactical or structural similarity of two code fragments, perhaps after Type-1 and Type-2 normalizations, then reporting those that satisfy a given minimum similarity threshold.  Another method is to detect nearby Type-1/Type-2 clones that are separated by a dissimilar gap, and merge these to form Type-3 clones.  Researchers do not agree on how much modification can be performed on the copied code fragment before it is no longer a clone of its original.  Since clones can arise for reasons other than copy and paste, for example programming language limitations and repeated coding styles~\cite{Roy07asurvey}, researchers are concerned with what is the minimum syntactical similarity of a valid Type-3 clone.
	
		\subsubsection{Type-4 Clones}
		Type-4 clones can occur when the same functionality has been implemented multiple times using different syntactical variants.  Most programming languages allow the same functionality to be specified using different syntax.  For example, a switch statement could be replaced with an if-else chain, or a for-loop could be replaced by a while loop.  Often the statements of a code fragment can be re-ordered without changing functionality but significantly varying the code fragment's syntax and structure.  As a more extreme example, two implementations of merge-sort, one recursive and one iterative, could be considered as a Type-4 clone.
		
		Type-4 clones are a relatively unexplored clone type, with very few tools targeting their detection.  It is also difficult to separate the Type-3 and Type-4 clones as many Type-3 clones have the same or share functionality, while Type-4 clones will often share some degree of syntactical similarity.  There is also the question of how similar must the functionality of two code fragments be for them to be a Type-4 clone, or if implementations of the same functionality (e.g., stable sort) using different algorithms (e.g., bubble sort and merge sort) is a Type-4 clone.

	\subsection{Syntactic and Semantic Clones}
	A code clone is a pair of code fragments that are similar, for some definition of similarity.  Generally, we are interested in code fragments that are similar by their syntax and/or semantics.  The clone types reflect this, with Type-1, Type-2 and Type-3 being concerned with syntactical similarity and Type-4 being concerned with semantical similarity.  While clones of the first three types are defined with respect to syntactical similarity, it is often also the case that syntactic clones share semantics.  In contrast, Type-4 clones are those that are specifically syntactically dissimilar but semantically similar.  A syntactic clone is therefore any clone of the first three types, while a semantic clone is a pair of code fragments that implement similar functionality, but can be any of the four primary clone types.  Both semantic and syntactic clones are of interest to software developers and clone researchers.
	
	Syntactic clones often indicate the incidence of copy and paste source-code reuse.  When source code is re-used by copy and paste, this duplicates any existing bugs, decreasing the quality of the software system and increasing the costs of bug-fixing and other maintenance tasks.  Duplicate code can also lead to new bugs when one code fragment is evolved without appropriately duplicating that evolution to the code fragment's clones.  Many syntactic clones are also semantically similar.  However, there are also syntactic clones that share no semantics.  In particular, the cases of significant or total differences in identifier names and literal values of Type-2 and Type-3 clones.  These \textit{structurally} similar clones can be cases of coincidental similarity, but many may still be interesting when they indicate reuse of program and algorithmic structuring which may still be prone to reuse bugs.
	
	Semantic clones indicate duplicate functionality that should ideally be re-factored into common code.  Otherwise bugs in the semantics need to be addressed in multiple locations, and changes or evolution to semantic requirements need to be implemented in multiple locations.  Both cases leads to redundant maintenance and evolution efforts, and can cause new bugs when not done properly.  As clones are not typically documented~\cite{Roy07asurvey}, this is real concern.
	
	Syntactic clones are often also semantic clones.  Most clone detectors measure syntactical similarity to detect both syntactic and semantic clones.  Very few clone detectors target semantic clones that are not syntactically similar, and those that do are not very proficient.  As such, Type-4 clones are still an open area in clone detection research.	
	
	\subsection{Clone Granularity, Boundaries}
	The general clone definition does not place any constraints on the boundaries of a code fragment except that it is a contiguous sequence of lines.  Therefore clones can exist at various granularities in source code~\cite{Roy07asurvey, ZibranSurvey}.  The most common granularities studied include:
	
	\begin{description}[leftmargin=1cm,labelindent=0.5cm]
		\item[File Clone]  A pair of similar source files.
		\item[Class Clone] A pair of similar class definitions (in object-oriented code).
		\item[Function Clone] A pair of similar functions (or class methods, constructors or destructors).
		\item[Block Clone] A pair of similar code blocks (indicated by a matching pair of opening and closing braces, or sequence  of statements at the same indentation indentation, and so on depending on the programming language).
		\item[Arbitrary Statement Clone] A pair of similar code-statement sequences.
	\end{description}

	File, class, function and block clones have precise boundaries.  The code fragments start and end on the boundaries of the respective granularity.  For example, a code fragment of a function clone starts on the line the function definition begins on and ends on the line the function definition ends on.
	
		\subsubsection{Arbitrary Statement Clone Boundaries}
		The code fragments of an arbitrary statement clone do not have precise boundaries.  They start on the line containing the first statement in the sequence and end on the line containing the last statement in the sequence.  However, a number of rules for high-quality arbitrary statement clone reporting are followed by most tools.  
	
		The code fragments of arbitrary statement clones should not overlap.  The start and end lines of the code fragments should be within the same scope.  For example, the code fragments should not start within an if-statement and end outside of the if-statement.  In other words, the code fragments of arbitrary statement clones should not split higher-order code statements (e.g., if statements, loop statements, switch statements, code blocks, and so on); they must include the entire higher-order code statement or not at all.  The code fragments should not start in one function and end within another, or subsume multiple functions.  Functions are complete and independent logical units, and their position/order within a source file is not generally relevant, so it does not make sense for a code fragment to encompass multiple functions (except for the case of class and file clones).
		
		These are general rules for high-quality arbitrary statement clone reporting for human inspection and software maintenance tasks.  Not all clone detection tools follow these guidelines, often because they can be difficult or computationally expensive to enforce.  Additionally, there may be domain-specific applications of clones that violate these rules.  For example, clone detection for code compaction.

	\subsection{Clone Size}
	\label{sec:cloningTheory_cloneSize}
	The clone definitions do not put any constraints on the size of clones.  Clone size is typically measured as the maximum (or average) length of its code fragments measured in original source lines, pretty-printed source-clones and/or by token.  While there is no minimum size of a clone, very small clones are often spurious.  For example, a pair of identical tokens are not considered a clone, neither is pair of statements that are the same after normalization.  Clone detection tools typically require a minimum clone size configuration in order to filter those smaller identical/similar code fragments that are likely to be spurious or uninteresting.  Typical minimum clone sizes are 6-15 original source lines~\cite{bellon,nicad,moderntools,bigclonebench_evaluation}, and 30-100 tokens~\cite{iclones,ccfinderx}.
	
	\subsection{Clone Detection Tools}
	\label{sec:cloningTheory_cloneDetectionTools}
	Clone detection tools are used to detect clones within a collection of source code.  Clone detection can be defined as shown in Eq.~\ref{eq:cloneDetection}.  The clone detector, $T$, takes a collection of source code $S$, and a configuration of its detection algorithms, $C$, and outputs a set of detected clones, $D$.
		
	\begin{equation} \label{eq:cloneDetection}
	T(S,C) \rightarrow D
	\end{equation}
		
	The collection of source code can be a single software system, a collection of software systems, or just a collection of source files.  The tool configuration can include the language(s) of the source files to process, the granularity(ies) to report clones at (e.g., arbitrary statement, block, function or file), the minimum and maximum sizes of clones to report, the source normalizations to apply during parsing, the similarity threshold or maximum gap size for reporting Type-3 clones, or any other configuration specific to the tool's detection algorithms and implementation.  The tool outputs the detected clones as a collection of clone pairs or clone classes in a clone detection report.  There is no universal standard for clone detection report format or style, so a key aspect of tool evaluation is to convert the clone detection reports into a common format.
	
	All clone detection tools implement the following abstract procedure.  The source code files are parsed, and all of the code fragments are identified, subject to a minimum and maximum code fragment size (by line or token), granularity, boundary constraints, file filters, sliding windows, and so on.  Let $F=\{f_{1}, f_{2}, ..., f_{n}\}$ be the $n$ code fragments found in the source code, after filtering.
	
	This implies a set of \textit{potential} clone pairs, $F \times F$, that the tool should investigate.  The tool may reject some potential clone pairs, such as pairs of the same code fragment, pairs of overlapping code fragments, pairs of code fragments that vary too significantly in size, and so on.  The remaining potential clones pairs are investigated by the tool's detection algorithms and similarity metrics to decide if it is a clone or not.  The tool outputs its \textit{detected clones} (as clone pairs or summarized in clone classes) -- the candidate clone pairs its algorithms have judged to be clones.  This is summarized in Eq.~\ref{eq:abstractDetection}, where judge() either accepts a potential clone pair as a true clone, or reject its as a false clone, as per the tool's judgment.
	
	\begin{equation}\label{eq:abstractDetection}
	D = \{ (f_{i},f_{j}) \in F \times F \ | \ i\neq j \wedge judge(f_{i},f_{j}) \}
	\end{equation}
	
	The tools generally do not implement this abstract procedure explicitly, but their algorithms efficiently implement this procedure implicitly.  Surveys by Roy et al.~\cite{Roy:2009:CEC:1530898.1531101} and Rattan et al.~\cite{rattan} found at least 70 clone detection algorithms and tools in the literature.  These have been classified into various categories based on their detection algorithms~\cite{Roy:2009:CEC:1530898.1531101}, including: text, token, tree, metric, hash, program dependency graph (PDG), hybrid, and so on.  The large number and variety of detection techniques motivates the need for clone benchmarks.
	
	Clone detection tools are not perfect, and their detection reports can contain both true positives and false positives.  Additionally, its reporting of a clone's boundaries may not be precise.  It may include additional code not part of the clone, or miss some code that is part of the clone.

\section{Benchmarking Theory}
\label{sec:benchmarkingTheory}
	Clone detection tools are typically evaluated for their clone detection performance and their execution performance.  Clone detection performance is measured using information retrieval metrics including recall and precision.  Recall measures the proficiency of the clone detector, while precision measures its accuracy.  Execution time and scalability measure the execution performance of a clone detector.

Measuring recall requires an oracle or reference corpora, a set of known true and false clones, which is challenging to build.  Precision can be measured without an oracle, but requires extensive clone validation efforts.  Execution time can be measured by executing the clone detectors for a standard subject system or set of subject systems on standard hardware.

These evaluation metrics are defined as follows:

\begin{description}[leftmargin=1cm,labelindent=0.5cm]
	\item[Recall] The ratio of the true clone pairs within a software system that a clone detector is able to detect.
	\item[Precision] The ratio of the clones detected by a clone detector that are true clones not false positives.
	\item[Execution Time] Given a limited set of computation resources, the length of time a clone detector required to complete its detection of clones within a given subject system or collection of source code.
	\item[Scalability] How large of a subject system or collection of source code a clone detector can be executed for given a limited set of computation resources without crashing due to exceeding the available resources and without requiring unreasonable execution time.
\end{description}

	\subsection{Recall and Precision}
	\label{sec:benchmarkingTheory_recallPrecision}
	
	We now describe how the recall and precision of a clone detection tool can be measured.
	
		\subsubsection{Measuring Recall and Precision with an Oracle}
		Recall and precision can be measured for a specified subject software system as shown in Figure~\ref{fig:recallAndPrecision}.  Here we have the universe, \textbf{U}, of all potential clone pairs (every possible pair of code fragments) within the subject software system.  On the left is the set of all true clone pairs, \textbf{T}, in the subject software system.  This set is determined by a clone oracle, a hypothetical entity that is perfectly able to judge if a pair of code fragments is a clone (in reality, no such entity exists).  The remaining pairs of code fragments, $\textbf{F}=\textbf{U}-\textbf{T}$, is the set of false clone pairs -- the code fragment pairs the oracle decided are not clones.  On the right there is the set of detected clone pairs, $\textbf{D}$, reported by the clone detector when executed for the subject software system.  From the perspective of clone detection, this splits the universe into four regions:
	
		\begin{figure}
		\centering
		\includegraphics[width=0.6\textwidth]{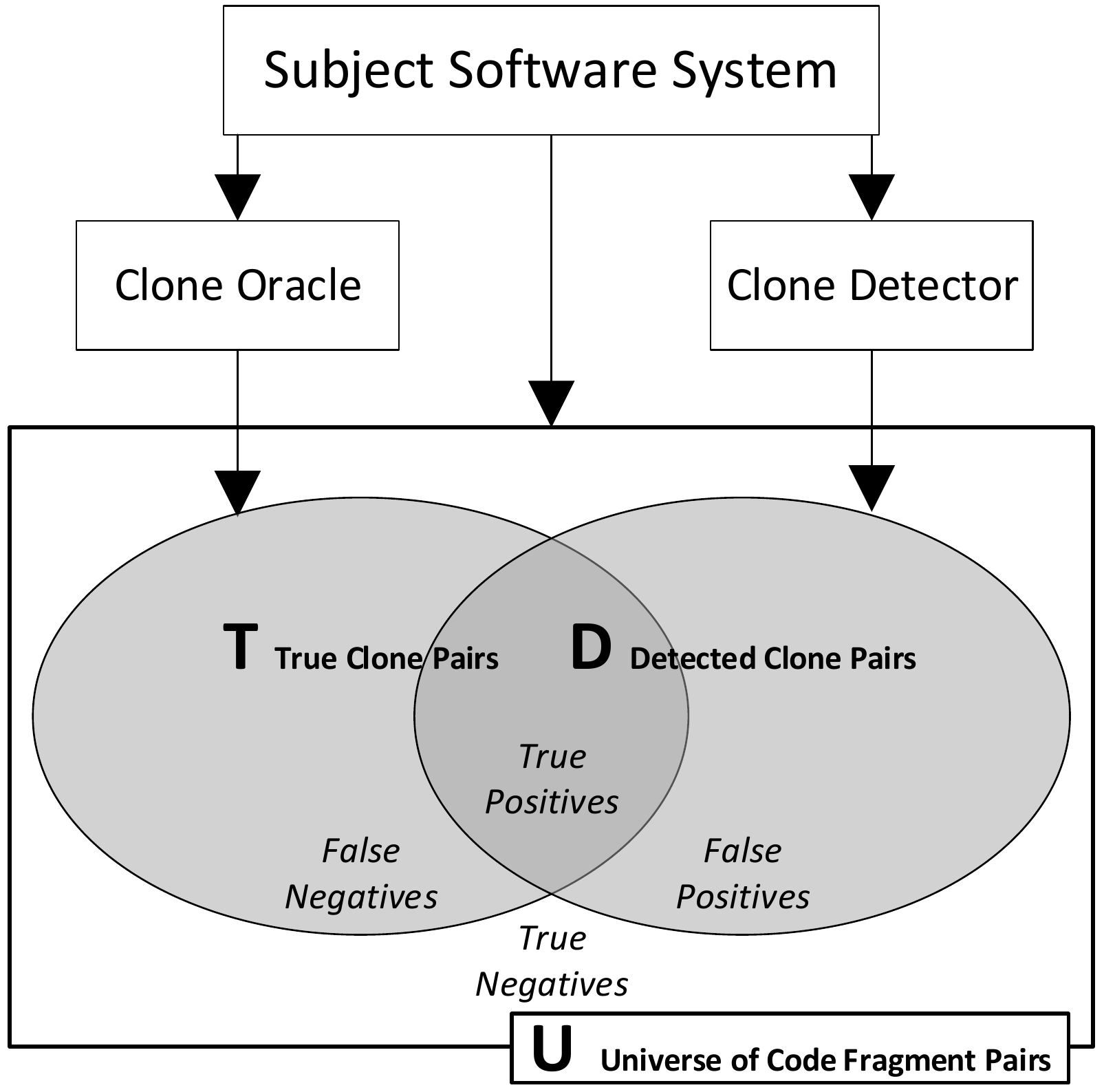}
		\caption{Measuring Recall and Precision with an Oracle}\label{fig:recallAndPrecision}
		\end{figure}
	
		\begin{description}[leftmargin=1cm,labelindent=0.5cm]
		\item[True Positives] The true clones successfully detected by the subject clone detection tool. (Desirable, improves recall.)
		\item[False Positives] The false clone pairs incorrectly identified as true clone pairs by the subject clone detector. (Undesirable, harms precision.)
		\item[True Negatives] The false clone pairs that are (correctly) not reported by the subject clone detection tool. (Desirable, improves precision.)
		\item[False Negatives] The true clone pairs that are not detected (missed) by the subject clone detection tool. (Undesirable, harms recall).
		\end{description}
	
		Recall, as shown in Eq.~\ref{eq:recall}, is the ratio of the true clone pairs that are detected by the subject clone detection tool, i.e., the ratio of \textbf{T} that is intersected by \textit{\textbf{D}}. This is also the ratio of the subject tool's true positives to the union of its true positives and false negatives.  Therefore, to improve recall, a clone detection tool wants to maximize its true positives and minimize its false negatives.
	
		\begin{equation} \label{eq:recall}
		\begin{split}
		recall  = \frac{\left | \textbf{D} \cap \textbf{T} \right |}{\left | \textbf{T} \right |}
		        = \frac{\left | true\ positives \right|}{\left | true\ positives \cup false\ negatives \right |}
		\end{split}
		\end{equation}
	
		Precision, as shown in Eq.~\ref{eq:precision}, is the ratio of the detected clone pairs that are true clone pairs, not false clone pairs, i.e., the ratio of $\textbf{D}$ that is intersected by $\textbf{T}$.  This is also the ratio of the subject tool's true positives to the union of its true positives and false positives.  Therefore, to improve precision, a clone detection tool wants to maximize is true positives and minimize its false positives.
	
		\begin{equation} \label{eq:precision}
		\begin{split}
		precision  = \frac{\left | \textbf{D} \cap \textbf{T} \right | }{\left | \textbf{D} \right | } 
		          = \frac{\left | true\ positives \right |}{\left | true\ positives \cup false\ positives \right |}
		\end{split}
		\end{equation}	
	
		\subsubsection{Challenges in building an Oracle}
		The measurement of recall and precision depends on the identification, or ``oracling'', of all the true clone pairs within a subject system.  This process is extremely effort intensive, as it requires the manual examination of every possible pair of code fragments in a subject system.  Even a small system such as \verb|cook| (51KLOC, 1244 functions) contains on the order of one million code fragment pairs at the function granularity alone~\cite{1287259}.  Not only is this too many potential clones to be examined, the \verb|cook| subject system does not contain a sufficient number and variety of true clone pairs on its own to properly evaluate clone detection recall.  Additional subject systems are required, which adds to the workload issue.  
		
		Additionally, the classification of a potential clone as true or false clone is a subjective process.  Previous studies~\cite{Charpentier:2015} have demonstrated that even amongst clone experts there is disagreement on what constitutes a true or false clone.  A clone expert may even give a different opinion on the same clone when shown it at different times.  There is no universal definition of a true clone, and responses might depend on the current task and goals of the clone-detection user~\cite{1287259}.  As such, it is not only a question of if a potential clone is a true clone, but if it is useful or relevant clone for some clone-related maintenance or development task.
		
		As such, no clone oracle exists, as it is too effort intensive to create.  Likely no true oracle can exist, due to the subjectivity in what constitutes a true clone.  Instead, clone detection researchers must come up with innovative ways to create corpora of validated reference clones that can accurately estimate the recall and precision of clone detection tools without the need to fully oracle multiple subject systems.  To overcome subjectivity, benchmarks must be created with a well-defined perspective and scope such that the results can be properly interpreted.

		\subsubsection{Measuring Recall and Precision with a Reference Corpus}
		\label{sec:becnhmarkingWithRefCorpus}
		Recall and precision can instead be estimated using a \textit{reference corpus}, a set $\textbf{R}$ of the known true and false clone pairs within a subject system, set of subject systems, or some other collection of source code.  The reference corpus can be separated into its set of known true clones, $\textbf{R}_{\textbf{t}}$, and set known false clones, $\textbf{R}_{\textbf{f}}$.  Recall and precision are measured with respect to the know true and false clones.  In most cases, the reference corpus is not complete -- it does not contain every true and false clone within the subject system.  In other words, the reference corpus is a proper subset of an oracle.  Specifically, $\textbf{R}_{\textbf{t}} \subset \textbf{T}$ and $\textbf{R}_{\textbf{f}} \subset \textbf{U}-\textbf{T}$.  The goal in creating a reference corpus (i.e., a clone benchmark) is to efficiently build a large and varied reference corpus for accurate estimation of recall and precision with a minimum of bias in the validation of the clones.
		
		Recall is measured as the ratio of the true clone pairs in the reference corpus that a clone detection tool is able to detect.  This is shown in Eq.~\ref{eq:recallRefCorpus}, where $\textbf{D}$ is the set of detected clone pairs detected by a tool, $\textbf{R}_{\textbf{t}}$ is the known true clone pairs in the reference corpus, and $\textbf{T}$ is the set of all true clones in the subject system(s).  Given a sufficiently large and sufficiently varied reference corpus, we can assume that $\textbf{R}_{\textbf{T}}$ approximates $\textbf{T}$ for the purpose of measuring recall, even if $\left | \textbf{R}_{\textbf{t}} \right | << \left | \textbf{T} \right |$.
		
		\begin{equation}\label{eq:recallRefCorpus}
		recall =   \frac{\left | \textbf{D} \cap \textbf{T}\right |}{\left | \textbf{T} \right |} 
		\approx \frac{\left | \textbf{D} \cap \textbf{R}_{\textbf{t}} \right | }{ \left | \textbf{R}_{\textbf{t}} \right | } \ \ \ \ \ \textbf{R}_{\textbf{t}} \subset \textbf{T}
		\end{equation}
		
		The reference corpus can be used to measure a lower and upper bound on a subject clone detection tool's precision.  The lower bound is measured as in Eq.~\ref{eq:precisionLower}, as the ratio of the tool's detection report that intersect the known true clones in the reference corpus.  The known true clones are used to validate part of the tool's detection report, while the other clones are assumed to be false positives (they either match the known false clones or are unknown to the reference corpus).  The tool's precision cannot be lower than this.
		
		\begin{equation} \label{eq:precisionLower}
		\left \lfloor precision \right \rfloor= \frac{\left | \textbf{D} \cap \textbf{R}_{\textbf{T}} \right |}{\left | \textbf{D} \right |}
		\end{equation}
		
		The upper bound on precision is measured as in Eq.~\ref{eq:PrecisionUpper}, as the ratio of the tool's detection report that are not known false clones.  The known false clones are used to validate part of the tool's detection report as false clones, while the other clones are assumed to be true positives (either match the known true clones or are unknown to the reference corpus).  
		
		\begin{equation} \label{eq:PrecisionUpper}
		\left \lceil precision \right \rceil = \frac{\left | \textbf{D}  \right | - \left | \textbf{D} \cap \textbf{R}_{\textbf{F}} \right |}{\left | \textbf{D} \right |}
		\end{equation}
		
		However, it is challenging to measure precision with a reference corpus without a significant difference between the lower and upper bound.  In most benchmarks, $\left | \textbf{R}_{\textbf{t}} \right | << \left | \textbf{T} \right |$, so the lower bound on precision is likely to be very low.  The number of false clones in a subject system(s) vastly outnumbers the number of clones.  Unless the creator of the reference corpus is very clever and able to mine those false clones that the tool is most likely to mistakenly detect as clones, the detected false positives is unlikely to significantly intersect with the known false clones.  The upper bound is therefore likely to be very high (near-100\%) even for subject tools with poorer precision.  However, it is possible to estimate precision without a reference corpus by manually validating some of the detected clones, which we discuss further in Section~\ref{sec:precisionValidation}.
		
		For this reason, reference corpora typically include only true clones for measuring recall, and leave precision to be measured by manual validation by the tool author.  So we will use reference corpus to mean a collection of known true clones, and treat those also including known false clones as a special case.  We discuss general methods of creating reference corpora in Section~\ref{sec:buildingReferenceCorpus_methods}, and specific methods, including full benchmarks with standard evaluation procedures and utilities, in Section~\ref{sec:benchmarks}.
		
		\subsubsection{Clone Matching Algorithm/Metric}
		The measurement of recall and precision with a reference corpus requires computing the intersection of the detected clones with the true and false clones in the reference corpus.  This requires determining which of the reference clones in the reference corpus are matched by detected clones in the subject tool's clone detection report.
		
		However, clone detection tools do not always report clones perfectly, and may have some some errors in the clone line boundaries.  In particular, off-by-one line errors are common.  So it is not as simple as searching the detected clones for an exact match of a reference clone.  Additional sources of differences in clone boundedness is differing clone reporting styles between the reference corpus and clone detection tool, or even subjectivity in the precise boundaries of the reference clones themselves.
		
		So accommodate these potential discrepancies, a clone matching metric is used to evaluate if a given detected clone is a sufficient match of a specific reference clone for the reference clone to be considered as detected by the subject clone detector.  This matching metric is evaluated for every detected clone until a sufficient match is found, or all detected clones are checked.  This can computed efficiently if the detection report is inserted into an indexed database table and the algorithm is implemented as a database query.
		
		Various clone detection algorithms have been used, that require the detected clone to: exactly match the reference clone (with or without an allowed error), subsume the reference clone, intersect a specified ratio of the reference clone, and so on.  The chosen matching metric usually depends on the way the reference corpus was constructed.

		\subsection{Methods for building a Reference Corpus}
		\label{sec:buildingReferenceCorpus_methods}
		
		Many methods for building a reference corpus have been proposed or attempted, all with different trade-offs and challenges.
		
		\paragraph{Manual Inspection}
		One method is an exhaustive manual search of a subject system.  Every pair of code fragments can be checked to see if they are a true or false clone, and added to the reference corpus.  This is a reasonable approach for very small subject systems.  However, even a small system such as \verb|cook| (51KLOC, 1244 functions) contains nearly one million pairs of code fragments at the function granularity alone~\cite{1287259}, which is far too many candidate clones to check.  A solution is to investigate only a statistically significant random sample of the code-fragment pairs~\cite{KrutzBenchmark}.  However, since the chance that two randomly selected code fragments being a true clone is low, this is an inefficient way to build reference corpus.
		
		\paragraph{Using Clone Detectors}
		A common method is to build the reference corpus using the clone detection tools themselves.
		
		The union method~\cite{bellon} builds a reference corpus as the union of the clones detected by a set of diverse clone detection tools.  The flaw in this approach is it assumes the tools have perfect precision, which is not a reasonable assumption.  The false positives added to the corpus will harm the measure of recall.  Additionally, the true clones not found by any of the tools are missing from the corpus.  The code fragment pairs not found in the union cannot be assumed to be false clones, as it is not reasonable to assume the union of the tools has perfect recall for the subject system.  An additional challenge is the handling of clones that are detected by multiple tools but are reported differently, including different reporting styles and precise line boundaries.
		
		The intersection method~\cite{bellon} builds a reference corpus as the intersection of the clones reported by a set of diverse clone detection tools.  The flaw in this method is it will trivially measure perfect recall for the tools used to build it.  It cannot be assumed that the clones not found by all of the tools are false clones, so it can't be used to measure the precision of the tools either.  It could possibly be used to measure the recall of non-participant tools, but the corpus will likely only contain the clones that are easy to detect (and thus detected by all the tools).
		
		Another method is to combine these approaches~\cite{bellon}, by taking the union, but removing those clones that are not detected by at least $n$ tools, where $n>1$.  The idea is that those clones detected by more than one tools are less likely to be false positives.  However, there is no clear choice of an appropriate value of $n$.  Additionally, it can be the case that $n$ tools report the same false positive, or that $n-1$ tools detect a true clone.  So the resulting corpus may still be of low quality.
		
		A popular approach is to combine the union method with statistical sampling and manual clone validation~\cite{bellon}.  A random sample is selected from the clones reported by each clone detector and manually validated to identify the true and false clones.  The sample size and distribution must be justified to be statistically significant and fair amongst the tools.  The resulting corpora can be used to estimate recall, while the validation efforts can be used to measure the precision of the participating tools.

		\paragraph{Search Heuristics}
		An alternative to using the clone detectors themselves is to use search heuristics that are distinct from the clone detectors.  Ideally, the search heuristic would be designed to have high recall but poorer precision, with the true clones identified by manual inspection before inclusion in the reference corpus.  This is similar to the manual inspection approach, except the search heuristic is used to greatly reduce the manual search space for better efficiency.  However, heuristics could also cause some true clones to be missed and not included in the reference corpus.  The heuristics may be designed to build a corpus with a particular context.
		
		Our implementation of this approach used keywords and source-code patterns to identify code fragments implementing specific functionalities, which revealed large semantic clone classes after manual inspection~\cite{bigclonebench}.  Another implementation used Levenshtein distance to identify true clones as those meeting a specified threshold~\cite{LavoieBenchmark}, without the need for manual inspection.  
		
		\paragraph{Clone Injection}
		A reference corpus can be built by injecting known clones into a subject system, or authoring new clones within that software system.  This is an alternative to mining for clones that already exist within the subject system.  The advantage of this approach is it gives the benchmark creator total control over the clones in their reference corpus.  However, manually creating interesting clones and injecting them into a software system is very effort intensive~\cite{bellon}.  Perhaps only a small reference corpus can be built.  However, the benchmark creator could carefully introduce interesting features into each clone and evaluate how this affects their detection by the subject clone detectors.  Injection could also be automated to study the detection of the clone by the subject tools for different locations of the clone within the source files.

		\paragraph{Artificial Clone Synthesis}
		A reference corpus can be automatically synthesized by programmatically mimicking the creation of a clone by a software developer.  This has been done using source-code mutation operators that mimic the types of edits developers make on copy and pasted code~\cite{Roy:2008:TMA:1370256.1370279,4976382,Svajlenko:2013:MAB:2662708.2662710}.  This is similar to manual clone injection, except the clones are constructed automatically.  The advantage of this technique is a large corpus can be constructed, with custom distribution of clone types.  However, it is difficult to automatically synthesize complex and realistic clones, which is the advantage of the manual clone injection method.

	\subsection{Measuring Precision without a Reference Corpus}
	\label{sec:precisionValidation}
	As mentioned in Section~\ref{sec:becnhmarkingWithRefCorpus}, a reference corpus can only measure a lower and upper bound on precision.  However, since it is very unlikely the reference corpus will contain a significant sample of the possible false positive clones the subject tools will report, it is likely there is a significant spread between the lower and upper bound.  Most reference corpora in fact do not include any known false clones, but rather focus on accurate measurement of recall.  This is not a problem because precision can be estimated without a reference corpus.
		
	The precision of a clone detection tool (for a given subject software system) can be estimated by manually validating a statistically significant sample of its detected clones.  Precision is then the ratio of the validated clones that are judged as true clones not false positives.  The precision of a clone detection tool can vary between software systems, so typically this is repeated for a collection of diverse software systems from a variety of programming domains, and the precision measurement is averaged.
		
	While this procedure is rather simple, there are still some challenges.  Clone validation is a very effort intensive process, and validating detected clones in a variety of software systems can take a significant amount of time.  Svajlenko et al.~\cite{clusterprecision} propose the use of clone clustering to reduce the efforts needed to measure a more generalized precision.  Clone validation is also subjective~\cite{Charpentier:2015}, so precision measured by different individuals could vary significantly.

	\subsection{Execution Time and Scalability}
	\label{sec:benchmarkingTheory_executionTimeScalability}
	
	The execution performance of a clone detector, including execution time and scalability, are also very important.  Execution time is the length of time a clone detector needs to complete its analysis of a subject software system.  Scalability is how large of a subject system the clone detector can be executed for without crashing due to exceeding the available computing resources (e.g., memory and disk-space) and without requiring an unreasonable execution time.  Scalability is also how execution time increases with increases in the size and/or complexity of the subject system.
	
	Execution time is important as users prefer not to wait for analysis results.  When clone detection is integrated into a tool-chain, execution time must be fast enough not to bottleneck the process.  For example, clone detection may be integrated into the analysis step of a build system.  The clone detector must be fast enough not to significantly delay completion of the build, especially in comparison to the development velocity (commit frequency).
	
	As the size of software systems continues to grow, the scalability of clone detectors is becoming a primary concern.  There is also interest in the detection of clones between software systems, between forks, and within an organization's entire portfolio of source code.  Emerging applications of clone detection include the analysis of clones in large inter-project source repositories on the scale of millions or even billions of lines of code or greater.  It is important to see how existing and new clone detectors can handle these scenarios.
	
	Execution time and scalability depend greatly on the target subject system and the hardware the clone detector is being executed on.  Therefore, to compare the execution performance of clone detectors, they must be executed for a standard set of subject software systems on the same hardware.  An experiment will execute a number of clone detectors for a collection of systems on the same hardware configuration.  Of interest is execution performance on standard hardware (e.g., a typical development workstation) as well as on extraordinary hardware (e.g., server computer with many cores and an abundance of memory, or even a computer cluster).

\section{A Survey of Clone Benchmarks}
\label{sec:benchmarks}

In this section, we survey and evaluate the existing clone benchmarks.  A clone benchmark is a reference corpus, or a framework for generating reference corpora, for measuring clone detection recall.  While precision, execution time and scalability are also important metrics, they can be measured without reference data.  Researchers have therefore focused on building benchmarks for measuring recall, although the benchmarks often provide some support in measuring or estimating the other metrics as well.

For this survey, we focus only on the works published as standard benchmarks, or which have been adopted by the community as standard benchmarks.

Tool authors have built impromptu reference corpora for evaluating the recall of their tools (see Section~\ref{sec:toolEvaluations_results_howrecall}).  However, these corpora are often not publicly released, are small, are poorly described, or are built to highlight the capabilities of the featured clone detector.  So we do not consider these to be standard benchmarks.

Similarly, tool evaluation studies (see Section~\ref{sec:surveyToolComparisons}) may build a reference corpus to compare the recall of the participating tools.  These may not be publicly released, or not well described, or are built specifically for the participating tools not as general benchmarks.  So we do not consider these to be standard benchmarks.

We consider only the works whose authors have intended to deliver a general and standard benchmark, irrespective of any particular clone detector(s).  The exception is we consider those benchmarks published in tool comparison studies which have been strongly adopted by the community (specifically, Bellon's Benchmark~\cite{bellon}).

This section is organized as follows.  We begin by describing our survey procedure in Section~\ref{sec:benchmarks_procedure}, and overview the results in Section~\ref{sec:benchmarks_results_overview}.  Then in Section~\ref{sec:benchmark_results} we summarize each benchmark in detail, including our own discussion and critique on these works.  In Section~\ref{sec:benchmarks_evaluation} we formally evaluate and compare the benchmarks.  We propose a set of evaluation criteria, and evaluate each benchmark for them.  We close this section by discussing threats to the survey in Section~\ref{sec:benchmarks_threats}.

	\subsection{Survey Procedure}
	\label{sec:benchmarks_procedure}
	
	To identify the clone benchmarks, we examined the clone research literature.  We began by examining the clone research surveys~\cite{Roy07asurvey,rattan}, which summarize the literature up to at least 2011. As part of our survey of the clone detection tools (Section~\ref{sec:toolEvaluations}), we searched the leading academic databases for clone detection publications for a period of January 2011 to March 2017.  During this search we also identified the clone benchmark publications.  We are confident we have found all or most of the clone benchmark publications.

	\subsection{Overview of Results}
	\label{sec:benchmarks_results_overview}
	
	We found a total of eight clone benchmarks in the literature, which are summarized in Table~\ref{tab:benchmarks_summary}.  We order the benchmarks by their publication year, and provide a short description of each benchmark.  We indicate which of the four primary clone detection performance metrics the benchmark is able to measure, and whether the authors provide an evaluation procedure or utility for the measurement.  We also indicate whether the benchmark is publicly available.
	
	\afterpage{%
		\begin{landscape}
	\footnotesize
		\begin{longtable}{>{\centering\raggedright\arraybackslash}m{4cm}m{2cm}>{\centering\raggedright\arraybackslash}m{10cm}m{0.5cm}cm{0.5cm}m{0.5cm}m{0.5cm}cm{0.5cm}}
			\caption{\normalsize A Survey of Clone Detection Benchmarks} \label{tab:benchmarks_summary} \\
			\toprule
			Benchmark & Year & Description & & \rot{90}{1em}{Recall} & \rot{90}{1em}{Precision} & \rot{90}{1em}{Scalability} & \rot{90}{1em}{Execution Time} & & \rot{90}{1em}{Public Data}  \\
			\midrule 
			\endfirsthead
			
			\toprule

			Benchmark & Year & Description & & \rot{90}{1em}{Recall} & \rot{90}{1em}{Precision} & \rot{90}{1em}{Scalability} & \rot{90}{1em}{Execution Time} & & \rot{90}{1em}{Public Data}  \\
			\midrule 
			
			\endhead
			
			\multicolumn{9}{c}{\checkmark = Can measure, {\ding{60}} = Includes documented evaluation procedure, {\ding{58}} = Includes an utility implementing the evaluation procedure.}
			\endfoot
			
			\multicolumn{9}{c}{\checkmark = Can measure, {\ding{60}} = Includes documented evaluation procedure, {\ding{58}} = Includes an utility implementing the evaluation procedure.}
			\endlastfoot
			
			Bellon's Benchmark~\cite{bellon,bellonbenchmark} & 2007 & Reference corpus of manually validated clones detected by clone detectors. & & \checkmark \scriptsize \ding{58} & \checkmark \scriptsize \ding{58} &  & \checkmark & & \Checkmark  \\ \midrule
			
			Mutation and Injection Framework~\cite{Roy:2008:TMA:1370256.1370279,4976382,Svajlenko:2013:MAB:2662708.2662710} & 2008 & Evaluates using synthetic clones in a mutation-analysis procedure. & & \checkmark \scriptsize \ding{58} & \checkmark \scriptsize \ding{58} & & & & \Checkmark \\\midrule
			
			\textit{Lavoie and Merlo}~\cite{LavoieBenchmark} & 2011 & Automatically validates all pairs of code fragments in a software system using the Levenshtein metric. && \checkmark & \checkmark & & && \\\midrule
			
			
			ForkSim~\cite{6648182} & 2013 & Creates synthetic forks of an existing software system with known similarities and differences.  Can evaluate clone detection performance across software variants. && \Checkmark \scriptsize \ding{60} & & & & & \Checkmark \\\midrule
			
			BigCloneBench~\cite{bigclonebench,bigclonebench_evaluation,bcbeval} & 2014 &  Manually validated reference clones mined by their functionality in a large inter-project source-code repository. && \checkmark \scriptsize \ding{58} &  & \checkmark \scriptsize \ding{60} & \checkmark \scriptsize \ding{60} & & \Checkmark\\\midrule
			
			\textit{Kurtz and Le}~\cite{KrutzBenchmark} & 2014 &  Manually validated function pairs within and between randomly selected source files to build a reference corpus. && \checkmark & & & && \Checkmark \\\midrule
			
			\textit{Yuki et el.}~\cite{yukibenchmark} & 2016 &  Automatically validated merged clones mined from development history && \checkmark & & & && \\
			
			
			
			\bottomrule
		\end{longtable}
\end{landscape}
			
	}
	
	\subsection{Detailed Results}
	\label{sec:benchmark_results}
	
	In this section, we describe each of the benchmark in detail, and provide some discussion and critique on their attributes.
	
		\subsubsection{Bellon's Benchmark}
		\label{sec:benchmark_results_bellon}
		
		Bellon's Benchmark.~\cite{bellon,bellonbenchmark} is the product of a clone detection tool evaluation study performed by Bellon et al.~\cite{bellon} in 2002.  The benchmark was built by manually validating 2\% of the 325,935 clone pairs detected by six contemporary clone detection tools in four C and four Java software systems, and adding those found to be true positives (possibly with modifications) to the reference corpus as true clones.  In total, 77 hours were spent validating 6,528 detected clone pairs to build a reference corpus containing 4,319 known true clones.  Bellon et al.~\cite{bellon} propose two clone matching metrics, one strict and one lenient, for measuring recall and precision (lower bound only).
		
		Tools using a variety of detection techniques were used to build the benchmark, including: Dup~\cite{baker,baker:si:92,baker:514697} (token-based), Duplix~\cite{957835} (PDG-based), Duploc~\cite{Ducasse_SMR:SMR317,Ducasse_792593} (text-based), CCFinder~\cite{ccfinderx} (token-based), CLAN~\cite{Mayrand_565012,bellon} (metric-based), and CloneDr~\cite{Baxter_1317484,Baxter_738528} (AST-based).
		
		The utility used to build the benchmark and evaluate the tools was released, and can be used to evaluate the recall of new clone detection tools.  Precision can only be measured for the original tools used to build the benchmark, unless the user extends the benchmark with new clones (although this has some complications).  While Bellon's benchmark does not measure the execution performance of clone detection tools, its collection of subject software systems (of various sizes) can be used as a standard set for performance benchmarking.
		
		Bellon's Benchmark has been a very popular benchmark in the clone community, and multiple extensions have been proposed (e.g,~\cite{gap,moderntools}).  However, the benchmark has also been the subject of criticism.  Baker~\cite{baker} found inconsistencies in the clone types and validation procedure used by Bellon.  Svajlenko et al.~\cite{moderntools} found significant anomalies in the recall measurements by Bellon's Benchmark for modern clone detectors, and suggests that an update of its corpus is warranted.  Charpentier et al.~\cite{Charpentier:2015} performed an empirical assessment of the reference clones in Bellon's Benchmark, and found disagreement when these clone sare re-validated by multiple judges.  In another work, Chapentier et al.~\cite{Charpentier2017} show that non-experts in a particular software system can be unreliable in validating the clones in that software system.  In general, it has been shown that manual clone validation to build benchmarks can be unreliable~\cite{1287259,Charpentier:2015,Charpentier2017}.
		
		\subsubsection{Lavoie and Merlo}
		
		Lavoie and Merlo use the Levenshtein distance metric to automatically build Type-3 clone benchmarks.  They use the Levenstein metric as the definitive oracle for differentiating true and false clones, and implement an algorithm for computing a clone benchmark given a configuration of the metric.  They introduce this methodology as a way of avoiding the extensive efforts and subjectivity of building a clone benchmark by manual validation.  Their methodology can build a comprehensive benchmark for systems up to even millions of lines of code without human efforts.  Since their methodology can investigate all potential clones of a given granularity within a software system, it can measure both recall and precision.
		
		Levenstein distance is the minimal number of edits to transform one string into another.  To measure this for code fragments, the authors transform the code fragments into strings of their token types, which also removes any Type-1 and Type-2 differences between the code fragments.  Edit distance is measured at the token level, and then normalized by the length of the larger code fragment, resulting in a difference ratio between 0.0 (exact match) and 1.0 (total mismatch).  A threshold is used to judge if the pair of code fragments is a true clone or a false clone.  For example, the authors accept pairs of code fragments with a normalized Levenstein distance less than 0.3 to be true clones.  The authors use an algorithm based on metric trees to efficiently identify all pairs of code fragments in a software system that satisfy the chosen threshold.
		
		The authors used their technique to build a benchmark of clones in the TomCat (130KLOC) and Eclipse (1.3MLOC) software systems  They consider only the code fragments that are 70 tokens or larger, and correspond to the block, method or class granularity.  They do not consider the statement-granularity code fragments, although though their technique would support them, likely due to performance constraints.  They accepted as clones the pairs of code fragments with a normalized Levenstein distance lower than 0.30.  For TomCat, they processed 5084 code fragments (13 million pairs) and found 6534 true clones.  For Eclipse, they processed 129 thousand code fragments (8.4 billion pairs) and found 115 thousand true clones.  The Eclipse experiment requires 6.2 days of execution time on a cluster with 32 Operton CPU cores at 2.0GHz and 5GB of memory.  In comparison, Bellon~\cite{bellon} required 3.2 days (or 10.3 normal workdays) to build a benchmark of just 4,319 clones.  There is definitely an advantage of building benchmarks in an automated way.
		
		The advantage of this approach is a complete and objective benchmark can be built without manual efforts.  The primary threat to this benchmarking strategy is it accepts the Levenshtein metric and chosen threshold as the definition of a true Type-3 clone.  Essentially, the benchmark is built using a clone detection tool based on the Levenshtein distance, meaning the benchmark could itself contain false clones incorrectly identified as true clones (false positives) and vice-versa (false negatives).  While a human constructed benchmark can also contain errors (human judges can be subjective, they make inconsistent judgments, and make errors), the authors do not compare the accuracy of the Levenshtein to human judges.  The Levenshtein metric also cannot recognize false clones that are only coincidentally similar.  Researchers do not agree if a purely syntactically similar Type-3 clone (no functional similarity) is a true or false clone.  
		
		However, despite these threats, the benchmark is still useful to measure the recall and precision of a clone detector relative to this Levenshtein approach. The methodology appears to be designed under the premise that the Levenshtein metric is an accurate but expensive way to detect clones, and that clone detectors typically use less accurate but more efficient metrics and algorithms to detect clones.  This benchmark creation approach is too expensive to be used for general clone detection, but the performance of the faster but less accurate clone detectors can be evaluated with respect to it.  However, the authors do not specifically demonstrate this.  At the very least, if this benchmark creation approach is implemented in a rigorous way, it could be used to detect potential bugs or flaws in new clone detection tools.
		
		While the algorithms of this approach have been published~\cite{LavoieBenchmark}, it does not appear that the authors have released the implementation of the algorithms, or the TomCat and Eclipse benchmark.  Therefore usage of this benchmark is difficult as it requires re-implementation of the authors efforts.
		
		\subsubsection{Mutation and Injection Framework}

		The Mutation and Injection Framework~\cite{Roy:2008:TMA:1370256.1370279,4976382,Svajlenko:2013:MAB:2662708.2662710,moderntools} is a fully automatic synthetic benchmark for evaluating the recall of clone detection tools.  It synthesizes realistic copy and paste clones in a mutation-analysis procedure, and automates the evaluation of the subject clone detection tools for the synthetic clones.  The idea was originally presented by Roy et al.~\cite{Roy:2008:TMA:1370256.1370279} including a proof of concept implementation and demonstrating experiment~\cite{4976382} with their clone detector, NiCad~\cite{nicad}.  The framework was then generalized and refined by Svajlenko et al.~\cite{Svajlenko:2013:MAB:2662708.2662710} who have used it in a number of tool evaluation studies~\cite{moderntools,bigclonebench_evaluation,sourcerercc}.
		
		The Mutation and Injection Framework was created in response to difficulties in using benchmarks or benchmarking methodologies, such as Bellon's benchmark~\cite{bellon}, that rely on manual clone validation.  Manual clone validation requires significant time investment~\cite{bellon}, is error-prone~\cite{baker,Charpentier2017,Charpentier:2015} and subjective~\cite{Charpentier2017,Charpentier:2015}, so a clone benchmark that can avoid these efforts is highly valuable.  The framework was designed to support multiple programming languages and granularities, and measure recall more precisely and at a finer granularity than previously possible.
		
		The framework evaluates tools for all the different kinds of edits developers make on copy and pasted code.  The framework takes particular efforts to eliminate or reduce biases in the reference corpus and tool evaluation~\cite{Svajlenko:2013:MAB:2662708.2662710} that exist in other benchmarks.  The framework has two distinct phases: the generation phase and the evaluation phase.  During the generation phase, a reference corpus of synthetic true clones is constructed, and those clones are hidden in copies of a subject software system.  During the evaluation phase, the subject clone detection tools are executed for the reference corpus, and their recall and precision measured specifically for the reference clones.  
		
		No manual clone validation efforts are required during the process, and the framework's implementation fully automates the evaluation through an easy to use menu-based interface.  The framework has been publicly released~\cite{mutationframeworkdl}, along with the reference corpora the authors have generated for their evaluation studies~\cite{moderntools,bigclonebench_evaluation,sourcerercc}.
		
		The advantage of this benchmark is it measures a precise fine-grained recall for the different kinds of clones that are possible. Since it generates clones, it can test the tools against a significant variety of clones.  However, to ensure the artificial clones are true clones, it is conservative in the amount of differences introduced to the cloned code.  Also, the clones may not reflect real clones in practice.  However, the Mutation Framework can perform a much more controlled experiment than possible with a real-world benchmark.  The authors suggest combining the Mutation Framework with a real-world benchmark like BigCloneBench to get the best of both synthetic and real-world clone benchmarking.
		
		\subsubsection{ForkSim}

		ForkSim is a framework for generating sets of artificial forked software systems with known similarities and differences.  It can be used to measure the effectiveness of clone detection tools for migrating software variants towards a software product line.  The framework takes a software system as input, and produces a number of forks by simulating forked software development.  The authors proposed a taxonomy of six forked development activities, which are implemented in the framework, including both independent and shared development activities between the artificial forks.  For simulating differences, the authors use a taxonomy of the types of edits developers make on cloned code.  The development history of the artificial forks is logged such that recall can be measured.  The authors have publicly released the framework, and have demonstrated its use by evaluating NiCad for software variant analysis.
		
		The advantage of this benchmark is it can evaluate tools for a specific use-case of clone detection: analysis of software variants and their migration towards a software product line, which is an important application in industry.  Since the forks are artificially constructed, the generated forks have known similarities and differences, needed for benchmarking.  However, as a synthetic technique, there are a few disadvantages.  The generated forks may not accurately model how real forks are produced.  Additionally, synthetic development is conservatively applied to ensure that the reference data is correct.  Of course, the advantage is a gold standard is produced without human efforts.  While ForkSim creates benchmarks with known similarities and differences, it is not a perfect oracle, so it cannot measure true precision.
		
		\subsubsection{BigCloneBench}
		
		BigCloneBench is a very large benchmark of manually validated clones in the inter-project Java source repository IJaDataset-2.0 (25 thousand systems encompassing 250MLOC).  BigCloneBench was built by mining IJaDataset for function clones implementing commonly needed functionalities.  The clones were mined using a novel and efficient approach designed to minimize subjectivity in the manual clone validation.  The benchmark contains intra-project and inter-project semantic clones (share functionality) spanning the four primary clone types and the entire spectrum of syntactical similarity.  It was designed to evaluate tools for syntactic, semantic, intra and inter-project and large-scale/big-data clone detection.
		
		BigCloneBench was released as a live benchmark, meaning it is under continuous expansion and improvement.  As of this writing, the benchmark is on its second release version, which includes 8 million clones across 43 distinct functionalities.  The benchmark has been used in a number of tool evaluation studies~\cite{bigclonebench_evaluation,sourcerercc,cloneworks}, including evaluations on large-scale clone detection~\cite{sourcerercc,cloneworks}.  The authors have released an evaluation framework, BigCloneEval~\cite{bcbeval}, which automates the evaluation of clone detection tool recall with BigCloneBench.  They have also released experiment artifacts that enable the comparison of execution time and scalability~\cite{sourcerercc,cloneworks,BigCloneBenchdl}.
		
		The authors built BigCloneBench by mining IJaDataset for functions implementing common functionalities.  Automatic search heuristics are used to identify candidate functions implementing a functionality, with manual validation to remove the false positives.  This identifies a large clone class of functions similar by their functionality, which captures a polynomial number of reference clone pairs.  Automatic processing is used to typify these clone pairs as one of the four primary clone types, label them as intra or inter-project clones, and measure their syntactical similarity.  This process is repeated across a number of diverse functionalities to build a large reference corpus of intra-project and inter-project clones of the four primary clone types and spanning the entire spectrum of syntactical similarity.  Their process also identifies reference false clones -- functions that implement different functionalities -- that can be used in a limited measurement of precision with respect to semantic similarity.
		
		The advantage of this benchmark is its breadth, including all four clone types, syntactic and semantic clones, intra-project and inter-project, and clones across 25,000 software systems.  Its validation process is indirect and guided by strict specifications, reducing subjectivity in the benchmark's clones.  A possible disadvantage is the clones are validated only by functionality.  The authors identify large clone classes of code fragments implementing a functionality, but do not inspect each clone pair individually.
		
		\subsubsection{Yuki et al.}
		
		Yuki et al.~\cite{yukibenchmark} in their paper, ``Generating Clone References with Less Human Subjectivity'', propose to build a reference corpus with less subjectivity by automatically mining reference clones from the revision history of a software system.  Specifically, they target cloned methods that have been removed by the software developers by a method merging pattern.  The authors suggest that these are real code clones, since they have been identified and refactored by the authors to remove the duplicate code.  At the very least, this process builds a benchmark of clones that are demonstrably important to the software developers. 
		
		Yuki et al.~\cite{yukibenchmark} introduced this benchmark creation strategy in response to the difficulty in creating reference corpora due to the subjectivity of clone validation.  Subjectivity in clone validation has been a significant challenge in benchmark creation~\cite{baker,Charpentier:2015}, particularly when the judges are not an expert of the chosen subject systems~\cite{Charpentier2017}.  Yuki et al. claim that by automatically mining for cloned methods that have been refactored into a single method, they are able to objectively build a reference corpora without human subjectivity or biases due to a particular researcher's preferred clone definitions.
		
		The authors propose two heuristics for detecting candidate instances of method merging in the development history, which can indicate the removal of a clone by refactoring.  The first pattern is when two methods that exist in one revision are removed in the next while simultaneously a new method is created.  The second pattern is when two methods exist in one revision, one is removed in the next revision while the other is changed.  These heuristics detect cloned method merging instances, but also many false positives.  The false positives are filtered by removing those instances where the merged method is not at least 70\% similar to the methods before merging, and those where method calls to the original methods are not replaced by method calls to the merged method.
		
		The authors tested their procedure for three Java systems stored in subversion software repositories, including: Ant (6022 revision), ArgoUML (3925 revisions) and jEdit (5168 revisions).  In total, they identified 19 clones: seven from Ant, ten from ArgoUML and two from jEdit.  The authors manually checked these clones and found that all were true positive methods clones that were refactored into a common method.
		
		The strength of this approach is that the reference clones have been indirectly validated by the software developers themselves.  Clone validation has been identified as one of the significant challenges in building clone benchmarks~\cite{baker,Charpentier:2015,Roy07asurvey,1287259} as it can be subjective, as evidenced by disagreement amongst experts~\cite{1287259}.  One suggestion has been that only developers of the software system can reliable judge the clones~\cite{Charpentier:2015,Charpentier2017}, however the developers are often not available or busy.  Yuki et al.'s approach captures the opinion of the software developer without requiring their time or presence.
		
		A problem with this technique is that it builds a very small benchmark.  Even after analyzing 15,000 revisions across three software systems the authors are only able to identify nineteen true clones, which is too few to build a reliable clone benchmark.  Possibly a large benchmark could be built by executing this process across the revisions of many software systems.  However, there might be challenges in doing so.  The authors do not comment on the difficulty of finding subject system repositories that are compatible with their process, the execution time and resources required to execute the process, and if they need to tune the process for each subject system individually.  If this process could be executed for public GitHub repositories automatically and without human intervention, the opportunity for a large and diverse benchmark is possible.  Since the authors chose to target only three subject systems, likely there is challenges in the process to solve before it can be scaled to public collections of software repositories.
		
		Another problem is the benchmark has limited scope.  Specifically, it only contains method clones that have been refactored by method merging by the developers.  These systems likely contain many method clones that the developers simply haven't refactored by merging, including clones the developers are not aware of.  The developers may even know of some clones but choose not to refactor them either because the costs outweigh the risks of keeping them, or because they are intentional clones.  A problem with considering only the clones the developers have refactored, is this either captures only those clones that developers already knew about (and didn't need a clone detector to detect), or it is capturing the clones that the developers found using clone detectors (undesirable when building a benchmark).  Refactoring is not the only solution to clone management, and clone detectors are needed to track clones that are undesirable to remove but need to be maintained appropriately.
		
		A threat to the benchmark is the procedure validates the detected candidate merged clone instances using a similarity metric with a threshold of 70\%.  This is itself a sort of clone detection amongst the mined candidates.  Therefore there is a risk of what is the recall and precision of this automatic validation process.  While they found it had 100\% precision in their experiment, this measurement relied on subjective validation by non-developers of the subject system.  The recall of this validation process is not reported, so possibly some true clones are lost from the benchmark.  The authors chose a similarity threshold of 70\% because it is the most common used by clone detectors, which may further bias the benchmark towards the clone detectors.
		
		
		
		\subsubsection{Krutz and Le}

		In their paper, ``A Code Clone Oracle'', Krutz and Le~\cite{KrutzBenchmark} created a small reference corpora of function clones with high confidence using rigorous validation.  The authors manually validating randomly selected function pairs from three open-source C software systems -- including Apache, Python and PostgreSQL.  In total, 1536 function pairs were validated yielding 66 reference true clones, including 43 Type-2 clones, 14 Type-3 and 9 Type-4, but no Type-1.
		
		To select their function pairs (candidate clones) for manual inspection, the authors selected 3-6 files per subject system, and enumerated all possible function pairs, yielding a total of 45,109 functions.  From these functions, they randomly sampled 1536 function pairs to examine if they are clones (a statistically significant sample with 99\% confidence level and a confidence interval of 5).  This includes 357 from Apache, 545 from Python and 634 from PostgreSQL.
		
		Each function pair was examined by three expert judges, who had experience in clone detection.  The three judges were allowed to discuss their results, and the final validation results considered the consensus of the judges.  The judges were also given the detection results of four clone detection tools -- including SimCad~\cite{simcad}, NiCad~\cite{nicad}, MeCC and CCCD -- to help them make their decision.  The authors also validated the clones using four student judges, and compared these results against the expert consensus.  They choose to use only the expert validation results for the final reference corpus.
		
		The authors use a procedure that can build a high quality reference corpus.  By selecting candidate clones at random from the system, they can build a benchmark that exceeds the capabilities of existing clone detection tools, which is good to motivate further development in detection techniques.  By validating the candidate clones by the consensus of three expert judges, they lower the subjectivity in the results, and build high confidence in the reference clones.  However, their benchmark is extremely small -- just 66 clone pairs.  The benchmark lacks variety in the clones, and does not even represent all of the clone types.  While they could increase the size and variety of the reference clones with further validation efforts, their procedure has a number of limitations.
		
		An issue with their approach is scalability.  They investigated 1536 function pairs to identify just 66 true clones for the benchmark (4.2\% efficiency).  In comparison, Bellon~\cite{bellon} investigated 6,528 candidate clone pairs to identify 4,319 true positive clones (66\% efficiency), which required 66 hours of effort.  Krutz and Le propose an approach that is much less efficient, while requiring significantly more validation efforts by requiring consensus amongst multiple judges.  Achieving a large benchmark using this technique is infeasible.
		
		Another concern is the authors involved clone detectors as part of their validation process.  While the clone detectors were not used to make the decision, the authors used the results of the clone detectors to help them make their decision.  This adds a bias to the benchmark due to the influence of the clone detectors themselves.  Ideally, the validation results should be completely independent of any clone detector.
	
	\subsection{Evaluating the Benchmarks}
	\label{sec:benchmarks_evaluation}
	In this section we evaluate the quality of the benchmarks.  To do this, we propose a set of requirements for clone benchmarks, and rank how well each benchmark satisfies them on a four point scale.  We propose desirable benchmark properties across a number of facets, including: general properties, subject systems, reference corpus, measuring recall, and other metrics.
	
	\afterpage{%
		\begin{landscape}
	\footnotesize
		\begin{longtable}{>{\centering\raggedright\arraybackslash}m{3cm}>{\centering}m{1.5cm}>{\centering\raggedright\arraybackslash}m{5cm}>{\centering\raggedright\arraybackslash}m{11cm}}
			\caption{\normalsize Clone Detection Benchmarks Requirements} \label{tab:benchmarks_generalrequirements} \\
			\toprule
			\textbf{Category} & \textbf{Label} & \textbf{Name} & \textbf{Requirement}  \\
			\midrule 
			\endfirsthead
			\toprule
			\textbf{Facet} & \textbf{Label} & \textbf{Name} & \textbf{Requirement}  \\
			\midrule 
			\endhead
			\endfoot
			\endlastfoot
\multirow{7}{*}{General}
& GE1 & Availability    & The benchmark should be free and publicly available. \\
& GE2 & Ease of Use	    & The benchmark should be easy to use. \\
& GE3 & Extensibility	& The benchmark should be open for extension. \\
& GE4 & Demonstrated    & The use of the benchmark should be demonstrated by the authors. \\
& GE5 & Evaluated       & The accuracy of the benchmark should be evaluated by the authors. \\
& GE6 & Community Usage & The benchmark has been used by the community. \\
& GE6 & Maturity        & The benchmark is ready for use and can produce reliable results. \\
\midrule
\multirow{5}{*}{Subject Systems}
& SS1 & Multiple Systems               & The benchmark should consider multiple subject systems. \\
& SS2 & Multiple Languages & The benchmark should consider subject systems of multiple languages. \\
& SS3 & Varying System Sizes           & The benchmark should consider subject systems of varying sizes. \\
& SS4 & Varying Application Domains    & The benchmark should consider subject systems from multiple domains. \\
\midrule
\multirow{8}{*}{Reference Corpus} 
& RC1 & Range of Clone Types        & The reference corpus should contain clones of each clone type. \\
& RC2 & Validated Clones            & The reference corpus should be convincingly validated. \\
& RC3 & Clone Detector Independence & The reference corpus should be independent of the clone detection tools themselves. \\
& RC4 & Strong Context              & The reference corpus should be built with respect to a context. \\
& RC5 & Variety of Clones           & The reference corpus should significant variety of clones. \\
& RC6 & Size of Corpus              & The reference corpus should be large. \\
& RC7 & Real Clones                 & The reference corpus should contain real clones. \\
\midrule
\multirow{6}{*}{Measuring Recall} 
& RE1 & Procedure                 & The benchmark should recommend a procedure for measuring recall with its data/framework. \\
& RE2 & Clone-Matching Algorithms & The benchmark should propose clone-matching algorithms. \\
& RE3 & Tool Supported            & The benchmark should provide a tool that assists with or automates the measure of recall. \\
& RE4 & Repeatable                & The benchmark should allow recall measurements to be repeatable by different experimenters. \\
\midrule
\multirow{5}{*}{Other Metrics}
& OM1 & Measures Precision      & The benchmark should provide or support the measurement of precision.      \\
& OM2 & Measures Execution Time & The benchmark should provide or support the measurement of execution time. \\
& OM3 & Measures Scalability    & The benchmark should provide or support the measurement of scalability.    \\
			\bottomrule
		\end{longtable}
\end{landscape}
	}
	
	The requirements are summarized per facet in Table~\ref{tab:benchmarks_generalrequirements}.  We rank the benchmarks for the requirements on a four point scale, as shown in Table~\ref{tab:benchmarks_fourpoint}.  This is the general guideline of the four point scale, and for each requirement we provide details on how we ranked the benchmarks specifically for that requirement. We also define two additional symbols to denote when a requirement is not applicable to a given benchmark, and when it is unknown if a requirement is satisfied due to lack of details provided by the benchmark authors.
	
	\begin{table}
		\centering
		\caption{Four-Point Scale for Evaluating Benchmarks}\label{tab:benchmarks_fourpoint}
		\begin{tabular}{cc>{\centering\raggedright\arraybackslash}m{10cm}}
			\toprule
			Numerical Rank & Symbol & Meaning \\
			\midrule
			0 & - -   	& Does not satisfy the requirement at all. \\
			1 & - 		& Minimally satisfies the requirement. \\ 
			2 & + 		& Satisfies the requirement. \\
			3 & ++ 		& Excellently satisfies the requirement. \\
			\midrule
			- & x       & No rank (not applicable to the benchmark). \\
			- & ?       & Unknown rank (information unavailable). \\
			\bottomrule
		\end{tabular}
	\end{table}
	
	\afterpage{%
		\begin{landscape}
	\footnotesize
		\begin{longtable}{>{\centering\raggedright\arraybackslash}m{6cm}
						  >{\centering\arraybackslash}m{1cm}
						  >{\centering\arraybackslash}m{1cm}
						  >{\centering\arraybackslash}m{1cm}
						  >{\centering\arraybackslash}m{1cm}
						  >{\centering\arraybackslash}m{1cm}
						  >{\centering\arraybackslash}m{1cm}
						  >{\centering\arraybackslash}m{1cm}
						  >{\centering\arraybackslash}m{1cm}
						 }
			\caption{\normalsize Benchmark Requirement Results} \label{tab:benchmarks_requirementsresults} \\
			\toprule
			\textbf{Requirements}  
			& \rotatebox[origin=c]{90}{\textbf{Bellon's Benchmark}}
			& \rotatebox[origin=c]{90}{\textbf{Mutation Framework}}
			& \rotatebox[origin=c]{90}{\textbf{Lavoie and Merlo}}  
			& \rotatebox[origin=c]{90}{\textbf{ForkSim}}
			& \rotatebox[origin=c]{90}{\textbf{BigCloneBench}}
			& \rotatebox[origin=c]{90}{\textbf{Kurtz and Le}}
			& \rotatebox[origin=c]{90}{\textbf{Yuki et al.}}
			\\
			\midrule 
			\endfirsthead
			
			\caption{\normalsize Benchmark Requirement Results (cont.)} \\ 
			\toprule

			\textbf{Requirements}  
				& \rotatebox[origin=c]{90}{\textbf{Bellon's Benchmark}}
				& \rotatebox[origin=c]{90}{\textbf{Mutation Framework}}
				& \rotatebox[origin=c]{90}{\textbf{Lavoie and Merlo}}  
				& \rotatebox[origin=c]{90}{\textbf{ForkSim}}
				& \rotatebox[origin=c]{90}{\textbf{BigCloneBench}}
				& \rotatebox[origin=c]{90}{\textbf{Krutz and Le}}
				& \rotatebox[origin=c]{90}{\textbf{Yuki et al.}}
				\\
			\midrule 
			
			\endhead
			
			\endfoot
			
			\endlastfoot
GE1 - Availability     				& + + & + + & +   & + + & + + & + + & +   \\
GE2 - Ease of Use      				& + + & + + & - - & +   & + + & -   & - - \\
GE3 - Extensibility    				& -   & + + & + + & + + & + + & -   & + + \\
GE4 - Demonstrated     				& + + & + + & - - & -   & + + & -   & - - \\
GE5 - Evaluated        				& -   & + + & - - & - - & + + & - - & - - \\
GE6 - Community Usage  				& + + & + + & - - & -   & + + & +   & - - \\
GE7 - Maturity        			 	& + + & + + & ++  & + + & + + & +   & -   \\
\midrule
SS1 - Multiple Systems   			& +   & + + & ?   & + + & + + & +   & ?   \\
SS2 - Multiple Languages          	& -   & +   & ?   & +   & - - & -   & ? \\
SS3 - Various Sizes      			& -   & + + & ?   & + + & + + & ?   & ?   \\
SS4 - Various Application Domains 	& + + & + + & ?   & + + & + + & +   & ?   \\
\midrule
RC1 - Range of Clone Types          & +   & +   & +   & +   & + + & + + & ?   \\
RC2 - Validated Clones              & -   & + + & - - & + + & +   & +   & +   \\
RC3 - Clone Detector Independence   & - - & + + & -   & + + & +   & + + & -   \\
RC4 - Strong Context                & -   & + + & +   & + + & + + & - - & + + \\
RC5 - Variety of Clones             & +   & +   & +   & +   & + + & - - & - - \\
RC6 - Size of Corpus                & -   & +   & + + & +   & + + & - - & - - \\
RC7 - Real Clones                   & +   & -   & +   & -   & +   & +   & + + \\
\midrule
RE1 - Procedure                 	& + + & + + & - - & +   & + + & - - & - - \\
RE2 - Clone-Matching Algorithms 	& + + & + + & - - & -   & + + & - - & - - \\
RE3 - Tool Supported           		& +   & + + & - - & - - & + + & - - & - - \\
RE4 - Repeatable                	& +   & + + & -   & -   & +   & -   & - - \\
\midrule
OM1 - Measures Precision      		& -   & +   & + + & - - & -   & - - & - - \\
OM2 - Measures Execution Time 		& +   & - - & - - & - - & + + & +   & - - \\ 
OM3 - Measures Scalability    		& +   & - - & - - & - - & + + & - - & - - \\
\bottomrule
		\end{longtable}
\end{landscape}
			
	}

	In the remainder of this section, we describe each of the requirements in detail, and how we evaluated the benchmarks.  The final results are summarized in Table~\ref{tab:benchmarks_requirementsresults}.

		\subsubsection{General Requirements}
		The general requirements are summarized in Table~\ref{tab:benchmarks_generalrequirements}.  We discuss each in detail below, and their evaluation guidelines.  The evaluation results for the tools are shown in Table~\ref{tab:benchmarks_requirementsresults}, and we discuss below how we assigned the evaluation scores.
		
			\paragraph{GE1 - Availability}
			The benchmark should be free and publicly available.  This is important for adoption of the benchmark, and the repeatability and extension of evaluation experiments.  Benchmarks either include reference corpora, and/or procedures that generate reference corpora.  Raw reference data should be made publicly available for download.  Generation procedures should at least be published as algorithms, although ideally a reference implementation is also provided.  At the very least, all benchmark publications should at least describe the strategy used to produce the benchmark so that others could re-use that strategy to produce similar (if not the same) benchmarks. If the benchmark is not available, then it cannot be used and examined/evaluated by the community.
			
			For availability, we use the following four-point evaluation:
			
			\begin{description}[style=standard,align=right,itemindent=-0.05cm]
				\item[- -]	The benchmark is not released.
				\item[-]	The benchmark is not released, but the benchmark creation strategy is sufficiently described to be reused to produce a similar benchmark.
				\item[+]	The benchmark is available, but requires implementation or other additional efforts from the user to access it.
				\item[++]	The benchmark is fully available, and ready for immediate use.  Any data is released, and any procedures are implemented as public tools or frameworks.
			\end{description}
			
			Almost all of the benchmarks are fully released (++).  While Lavoie \& Merlo and Yuki et al. fully describe their benchmark generation procedures, they do not provide an implementation, so they receive only a (+) rank.
			
			\paragraph{GE2 - Ease of Use}
			The benchmark should be easy to use.  Any reference data should be in an accessible format, and the evaluation procedure should at least be documented or ideally implemented as a tool.  The less burden placed on the user to measure the performance metrics, the more likely the benchmark will be used.  By providing (and ideally implementing) a standard procedure for tool evaluation, evaluations performed by different users are comparable, which is important in benchmarking. 
			
			For ease of use, we use the following four-point evaluation:
			
			\begin{description}[style=standard,align=right,itemindent=-0.05cm]
				\item[- -]  The reference data is not available.
				\item[-]  	The reference data is available, but the user is given no guidance on how to use it to evaluate their tool.
				\item[+]    The reference data is accessible, and the evaluation procedure is well-documented, but not implemented for the user.
				\item[++] The reference data or procedure is easily accessible, and a tool is provided that implements the evaluation procedure for the user.
			\end{description}
			
			A number of the benchmarks are easy to use, with both the reference data released as well as frameworks for evaluating tools using default procedures.  Both Bellon's Benchmark and BigCloneBench have their refrence data and evaluation frameworks/tools publicly released, (+ +).  Lavoie \& Merlo and Yuki et al. are generative benchmarks, but neither example reference data is released, nor are the generation algorithms implemented as tools. This requires significant effort on the user to use these benchmarks in their current state, so we assign (- -).  In contrast, the Mutation Framework includes an easy to use framework implementing its generation technique, and distributes example datasets, so we assign (+ +).  While data from Krutz and Le is available, the authors provide no guidance on how to evaluate tools with their data.  ForkSim provides an evaluation strategy, but it is not implemented as a tool.

			\paragraph{GE3 - Extensibility}
			The benchmark should be extensible by the community.  While ideally the benchmark creators will continue to maintain and improve their benchmark, realistically the authors often move on to new projects.  It is therefore important that the clone community be able to maintain and improve its benchmarks.  For the community to be able to extend a benchmark, the original authors must have sufficiently explained how they built the benchmark.  A major threat to extensibility is if the clone validation process is sufficiently rigorous and documented such that the community can extend the benchmark in a consistent manor, without weakening the integrity of the reference data. 
			
			For extensibility, we use the following four-point evaluation:
				
			\begin{description}[style=standard,align=right,itemindent=-0.05cm]
				\item[- -] The benchmark is closed for extension.  The creation process is not sufficiently documented for repeatability, or the original data is unavailable.
				\item[-] The benchmark creation process is sufficiently described to be repeated, but not detailed enough for newly created reference data to be compatible with the original data.
				\item[+] The benchmark is open for extension, and the process is sufficiently described for the new data to be mixed with the original reference data, but there are some limitations in validating new reference data in a consistent way.
				\item[++] The benchmark is open for extension, and the process and validation procedure is sufficiently described for new reference data to be kept consistent with the original data.
			\end{description}
			
			The generative benchmarks (Mutation Framework, ForkSim, Lavoie and Merlo, and Yuki et al.) are very extensible, as users can use these procedures to generate new reference corpora for any subject systems, so we rank these as (+ +).  The remaining benchmarks rely on manual clone validation.  While extension can be done by validating additional clones, it can be difficult to do this in a consistent way without the original authors. We rank Bellon's Benchmark and Krutz and Le as (-) as their creation strategy is well-documented, but they suffer form the consistent validation issue.  In contrast, we give BigCloneBench a (+ +) rank.  BigCloneBench validates clones indirectly with a strictly defined specification.  While BigCloneBench validation can still contain subjectivity, it is much more controlled compared to the other benchmarks relying on manual validation.
			
			\paragraph{GE4 - Demonstration}
			The use of the benchmark should be demonstrated by its creators.  At the very least, a case study evaluating a single tool, which demonstrates the evaluation procedure.  Ideally, a multi-tool case study, as this demonstrates the benchmark is compatible with multiple clone detection tools.  The best benchmarks are demonstrated in formal tool evaluation studies, which demonstrate how to resolve common benchmarking issues such as appropriate tool configuration, and normalizing for differences in how clone detectors choose to report clones. The authors of the benchmarks are experts in demonstrating its usage.
			
			For demonstration, we use the following four-point evaluation:
			
			\begin{description}[style=standard,align=right,itemindent=-0.05cm]
				\item[- -] The benchmark authors have not used their benchmark to evaluate any tools.
				\item[-]   The benchmark authors have conducted a case study considering a single tool.
				\item[+]   The benchmark authors have conducted a case study considering multiple tools.
				\item[++]  The benchmark authors have conducted a formal tool evaluation study considering multiple tools, and demonstrating how to resolve issues in tool configuration and normalizing for tool reporting differences.
			\end{description}
			
			Ranking in this case was rather straight-forward.  We investigated the benchmark publications and the papers that cite them to evaluate how the authors have demonstrated their benchmarks in their works.
			
			\paragraph{GE5 - Evaluated}
			Benchmarks should themselves be evaluated to build confidence their accuracy.  At minimum, the properties of the benchmark should be measured to demonstrate that it is comprehensive.  Benchmarks can be evaluated by comparing their measurement of performance against reasoned expectations for the tools.  Confidence in accuracy can be built by demonstrating that there is no unexplained anomalies in the benchmarking results.  In the best case, multiple benchmarks can be compared, and their similarities and differences considered.  Similarities build confidence in both benchmarks, while differences should be explainable by the differences in the benchmarking methodologies.
			
			For evaluated, we use the following four-point evaluation:
			
			\begin{description}[style=standard,align=right,itemindent=-0.05cm]
				\item[- -]	The benchmark has not been evaluated.
				\item[-]    The benchmark is evaluated by studying its properties.
				\item[+]	The benchmark is evaluated by conducting a tool evaluation experiment, and comparing the results against the expectations for the tools to check for anomalies.
				\item[+ +]	The benchmark is evaluated by conducting a tool evaluation experiment, and comparing the results against results from other benchmarks, with agreement and disagreement analyzed.
			\end{description}
			
			The Mutation Framework~\cite{moderntools} and BigCloneBench~\cite{bigclonebench_evaluation} were compared against other benchmarks for evaluation, and any anomalies examined and explained, (++).  Bellon's Benchmark's properties were extensively evaluated, and the authors performed tool evaluations with the full benchmark and the mid-point of its creation to show that performance evaluations were stable for the reference data~\cite{bellon}, (-).  The other benchmarks do not try to evaluate the accuracy of their evaluations, or haven't been used in any tool evaluations, (- -).
			
			\paragraph{GE6 - Community Usage}
			Benchmarks are valuable when they are accepted by the community.  Benchmarks are accepted when evaluation studies using them are accepted into the literature.  In the best case, benchmarks are adopted by other researchers and used in their studies.
			
			For community usage , we use the following four-point evaluation:
			
			\begin{description}[style=standard,align=right,itemindent=-0.05cm]
				\item[- -] The benchmark has not been used.
				\item[-]   The benchmark has only been used by its authors in a case study with the benchmark publication.    
				\item[+]   The benchmark has been used in studies published by its authors.
				\item[++]  The benchmark has been used in studies published by authors other than the benchmark's creators.
			\end{description}
			
			The evaluation results are shown in Table~\ref{tab:benchmarks_requirementsresults}.  To evaluate community usage we looked at the citations for the benchmarks.  Bellon's Benchmark has been significantly adopted by the community.  While newer, Mutation Framework and BigCloneBench have been used in numerous studies both by the benchmark authors and by other researchers.  The other benchmarks have either not been used or by only their authors.
			
			\paragraph{GE7 - Maturity}
			Important is the maturity of the benchmark.  New benchmark publications may only present a creation strategy with no reference data or just a small prototype.  It can take time for benchmark reference data to grow as more validation efforts are applied.  Ideally, a benchmark is feature complete and contains a large number of reference clones, at least on the order of one thousand clones.
			
			For maturity, we use the following four-point evaluation:
			
			\begin{description}[style=standard,align=right,itemindent=-0.05cm]
				\item[- -]	The benchmark has only been proposed.
				\item[-]	A benchmark prototype has been produced, but it is incomplete and not yet usable and/or reliable.
				\item[+]	The benchmark is usable, but lacks features or the reference data is still small.
				\item[++]	The benchmark is feature-complete, and the reference data is of significant size (on order of one thousand reference clones or larger).
			\end{description}
			
			BigCloneBench, The Mutation Framework, Lavoie and Merlo, ForkSim and BigCloneBench are mature benchmarks that are ready for general use.  The Kurtz and Le benchmark is usable, but the reference corpus is rather small.  Yuki et al. is in the prototype phase, with its publication focusing on demonstrating its benchmark creation technique.
						
		\subsubsection{Subject System Requirements}
		The performance of a clone detector can vary depending on the subject software system being searched for clones.  Different subject systems may contain different number and different types of clones, or have a different clone density, all of which can affect recall, precision, scalability and execution time.  It is therefore important that benchmarks consider multiple subject systems, and particularly subject systems with different properties.  Most importantly, the benchmark should consider multiple subject systems that vary by: (1) programming language, (2) size in lines of code, and (3) application domain.
		
		To capture the important of subject system variance, we specify four requirements, which are summarized below with their four-point rankings.  Evaluating the benchmarks featuring a reference corpus was straightforward.  Evaluating the generative benchmarks, which trake a subject system as input and produce a clone benchmark, was more challenging.  This includes the Mutation Framework, Lavoie and Merlo, and Yuki et al.  The Mutation Framework is a mature benchmark that is publicly released, so we rate it on the kinds of subject systems it can be executed for.  Lavoie \& Merlo and Yuki et al. could theoretically be executed for various subject systems, but they have no public implementation, so we cannot judge the extent of their subject system support.  The results are found in Table~\ref{tab:benchmarks_requirementsresults}.
		
		For the subject system requirements, we use the following four-point evaluations:
			\paragraph{SS1 - Multiple Systems:}
				\begin{description}[style=standard,align=right,itemindent=-0.05cm]
				\item[- -] The benchmark uses only a single subject system.
				\item[-]   The benchmark uses only a small number of subject systems (2-5).
				\item[+]   The benchmark uses a large number of subject systems (2-10).
				\item[++]  The benchmark considers a lot of subject systems ($>$10).
				\end{description}
			\paragraph{SS2 - Multiple Languages:}
				\begin{description}[style=standard,align=right,itemindent=-0.05cm]
				\item[- -] The benchmark considers a single programming language.
				\item[-]   The benchmark considers two programming languages.
				\item[+]   The benchmark considers three programming languages.
				\item[++]  The benchmark considers four or more programming languages.
				\end{description}
			\paragraph{SS3 - Various System Sizes:}
				\begin{description}[style=standard,align=right,itemindent=-0.05cm]
				\item[- -]	The subject systems are of the same order of magnitude.
				\item[-]	The subject systems are of two different orders of magnitude.
				\item[+]	The subject systems are of three different orders of magnitude.
				\item[++]	The subject systems are of four or more different orders of magnitude.
				\end{description}
			\paragraph{SS4 - Various Application Domains:}
				\begin{description}[style=standard,align=right,itemindent=-0.05cm]
				\item[- -]	The subject systems are of a single application domain.
				\item[-]	The subject systems are of two different application domains.
				\item[+]	The subject systems are of three different application domains.
				\item[++]	The subject systems are of four or more different application domains.
				\end{description}
		
		\subsubsection{Reference Corpus Requirements}
		
		Perhaps the most important aspect of a benchmark is the quality of its reference corpus (or the reference corpora it can generate).  The quality and accuracy of the recall measurement is dependent on the reference corpus.  Here we evaluate the benchmarks for the requirements of a good reference corpus.
			
			\paragraph{RC1 - Range of Clone Types}
			A good reference corpus contains clones of multiple clone types.  Most important are the Type-3 and Type-4 clones, which are difficult to detect with high accuracy.  We evaluate the benchmarks for the highest clone type they contain.
			
			For clone types, we use the following four-point evaluation:
			
				\begin{description}[style=standard,align=right,itemindent=-0.05cm]
					\item[- -] Contains only Type-1 clones.
					\item[-]   Contains clones up to Type-2.
					\item[+]   Contains clones up to Type-3.
					\item[++]  Contains clones up to Type-4.
				\end{description}
			
			Only BigCloneBench and Krutz and Le include Type-4 clones, although Krutz and Le is a very small benchmark.  The remaining include or support up to Type-3 clones, with exception of Yuki et al. which does not document their clone types.
			
			\paragraph{RC2 - Validated Clones}
			
			Very important is how well the the reference data of a clone benchmark is validated.  Poorly validated reference clones are untrustworthy, and weaken the measure of recall.  We rank the benchmarks on a four-point scale of our confidence in their validation from low confidence to very high confidence.
			
			We use the following four-point scale, and justify our decisions below:
			
				\begin{description}[style=standard,align=right,itemindent=-0.05cm]
					\item[- -] The clones are validated with low confidence.
					\item[-]   The clones are validated with medium confidence.
					\item[+]   The clones are validated with high confidence.
					\item[++]  The clones are validated with very high confidence.
				\end{description}
			
			We rank Lavoie and Merlo as low confidence, as the benchmark is automatically constructed using a clone similarity metric (Levenstein distance) and a minimum threshold.  This is essentially building the benchmark using a clone detector, with no validation.
			
			We rank Bellon's Benchmark, where the reference clones were manually validated by a single expert judge, with medium confidence.  Various studies have shown that manual clone validation is very subjective~\cite{1287259,Charpentier2017}, so the reference clones are somewhat suspect.  Indeed, various studies have shown this to be the case with Bellon's Benchmark~\cite{moderntools,Charpentier:2015,baker}.
			
			While Kurtz and Le is also built with manual validation by experts, each reference clone was considered by multiple judges and discussed before the final decision, so we rank it with high confidence.
			
			BigCloneBench avoids direct manual validation of the reference clones.  Instead, judges validate if code fragments implement specific functionalities, given a clear specification of that functionality.  The reference clone pairs are pairs of code fragments that implement the same functionality.  In this way, subjectivity in the decision is greatly reduced, and we therefore rank this benchmark to have high confidence in clone validation.
					
			We rank the Mutation Framework and ForkSim as building reference corpora with very high validation confidence.  These benchmarks produce benchmarks of artificial clones that are carefully constructed.  They focus on building simple clones with well-defined differences to ensure all generated clones are true clones.
			
			\paragraph{RC3 - Clone Detector Independence}
			
			A good reference corpus is built independent of the clone detectors themselves.  Using clone detectors to build the reference corpus creates one that is biased to kind of clones the clone detectors are good at detecting, and fails to evaluate for the clones the clone detectors are missing.  Some benchmarks may not use existing clone detectors, but similarity metrics that are essentially clone detection, or mining techniques that are tangential to clone detection.  We rank the benchmarks here by how rooted in clone detection their reference corpora are:
			
				\begin{description}[style=standard,align=right,itemindent=-0.05cm]
					\item[- -] The benchmark was built using clone detectors.
					\item[-]   The benchmark was built using clone similarity metrics.
					\item[+]   The benchmark was built using mining approaches tangential to clone detection.
					\item[++]  The benchmark was built independently of any clone detection technology or similar.
				\end{description}
			
			Bellon's Benchmark was built using clone detectors (filtered by manual validation), and suffers from becoming out of date as clone detectors have evolved.  Lavoie and Merlo relies upon the Levenstein metric, which has been used for clone detection, but the authors do not rely on a specific clone detector.  While Yuki et al. find clones by mining the revision history of a software system for refactored clones, they remove false positives using a clone similarity metric.  BigCloneBench uses regular expressions to mine for code fragments implementing specific functionalities, with false positives removed by manual inspection, to find large clone classes.  This is a code search technique, which is tangential to clone detection.  The Mutation Framework and ForkSim synthesizes clones, and are completely independent of clone detectors.  Krutz and Le manually search for clones, which is independent of clone detection tools.  While they do consider clone detection output to help them validate the clones, they claim to make their decisions independent of the clone detectors' response, so we give them the benefit of the doubt here.
			
			\paragraph{RC4 - Strong Context}
			
			It is important that the reference corpus be built with respect to a context, so that the measure of recall can be appropriately interpreted.  Specifically, the reference clones should be selected, validated or constructed with some goal in mind.  For example, reference clones relevant to software maintenance, reference clones that are refactoring candidates, and so on.
			
			We rank the benchmarks by how strong their context is, using the following four-point evaluation:
			
				\begin{description}[style=standard,align=right,itemindent=-0.05cm]
					\item[- -] The benchmark lacks a context.
					\item[-]   The benchmark has a poor context.
					\item[+]   The benchmark has a context.
					\item[++]  The benchmark has a strong context.
				\end{description}
			
			We find only Krutz and Le lacks a context, which consists of clones manually identified between a few randomly selected source files in three subject systems.  Bellon's Benchmark consists of clones detectable by the contemporary tools used in its experiment, which is a poor context.  Lavoie and Merlo builds a complete corpus using similarity metric, which we consider to be a context, but not a strong one.  The remainder of the benchmarks have a strong context.  The Mutation Framework constructs clones based on an editing taxonomy, while ForkSim constructs clones based on a forking taxonomy.  BigCloneBench satisfies multiple contexts, including: clones with functional similarity, intra-project vs inter-project clones and large-scale clone detection.  Yuki et al. build a reference corpus of clones that should be refactored, as evidenced by their development history.  
			
			\paragraph{RC5 - Variety of Clones}
			
			It is important for the reference corpus to contain a variety of clones, so that the recall measurement is generalized.  Otherwise, the measurement will be specific to the benchmark.  The aim of a benchmark is to measure recall in a way that estimates performance for any generic subject system.  We rank the benchmarks based on the variety of clones they capture.  We consider variety within the contexts of the benchmarks.  We use the four-point system:
			
				\begin{description}[style=standard,align=right,itemindent=-0.05cm]
					\item[- -] The benchmark has a low variety of clones.
					\item[-]   The benchmark has a medium variety of clones.
					\item[+]   The benchmark has a high variety of clones.
					\item[++]  The benchmark has a significant variety of clones.
				\end{description}
			
			BigCloneBench considers clones of almost 50 distinct functionalities, and captures millions of clones across these functionalities, so we consider it to have significant variety.  Bellon's Benchmark contains clones from a variety of systems and across the three primary clone types, but it is limited to the clones that contemporary detectors could detect, so we reduce its ranking to high variety.  The Mutation Framework and ForkSim synthesize large volumes of artificial clones using a comprehensive taxonomy of clone types, however they generate only simple clones to ensure high confidence, so we reduce ranking to high variety.  Lavoie and Merlo automatically identify all Type-3 clones within a subject system, but are limited by the Levenstein metric, so we reduced to high variety.  Krutz and Le contain only a small number of clones found between a small (3 to 6) random select of source files in three subject systems, so we rank low variety.  Similarly, Yuki et al. find very few clones, so has low variety.
			
			\paragraph{RC6 - Size of Corpus}
			
			It is important that a reference corpus contain a large number of clones in order to build confidence in the measured recall.  A small corpus contains too few data points, which can easily cause recall to swing high or low based on a small number of successes/failures.  Ideally, a reference corpus should contain at least 1000 reference clone pairs.  Each order of magnitude increase builds significant confidence in the results.
			
			For corpus size, we use the following four-point evaluation:
			
				\begin{description}[style=standard,align=right,itemindent=-0.05cm]
					\item[- -] The corpus contains less than 1000 reference clone pairs.
					\item[-]   The corpus contains 1000-9999 reference clone pairs.
					\item[+]   The corpus contains 10,000-99,999 reference clone pairs
					\item[++]  The corpus contains 100,000+ reference clone pairs.
				\end{description}
			
			We rank the benchmarks as per the size of their published reference corpora.  For the benchmarks that generate corpora (Lavoie and Merlo, Mutation Framework and ForkSim), we evaluate them on the size of the example corpora built by the authors.
			
			\paragraph{RC7 - Real Clones}
			
			It is important that a benchmark contain real clones.  If the benchmark synthesizes clones, it is important that they should be realistic.  If the benchmark was built by mining software systems, it is important that it contain clones that are relevant to the developers, and not cases of coincidental similarity.  Obviously, in terms of realism, synthetic clone benchmarks are at a disadvantage.
			
			For real clones, we use the following four-point evaluation:
			
				\begin{description}[style=standard,align=right,itemindent=-0.05cm]
					\item[- -] The clones are completely artificial and arbitrary.
					\item[-]   The clones are artificial, but convincingly created.
					\item[+]   The clones were created by real developers.
					\item[++]  The clones were created by real developers, and are validated as useful to developers.
				\end{description}
			
			The Mutation Framework and ForkSim are synthetic benchmarks, but they build their clones based on clone taxonomies for realism.  Bellon's Benchmark, Lavoie and Merlo, and Kurtz and Le identify clones produced by real developers, but do not necessarily verify these are useful clones to the developers.  BigCloneBench identifies real clones that share functionality, meaning they are not instances of coincidental syntactical similarity, and therefore more likely useful to the developers.  Yuki et al. mine for real clones that have been refactored by the developers, demonstrating that they are useful clones.
		
		\subsubsection{Measuring Recall Requirements}
		
		Beyond providing reference data, it is important that a benchmark assist in the measurement of recall.  The authors should suggest a procedure for measuring recall appropriate for their benchmark, as well as optimal clone-matching algorithms.  Ideally, the author should provide a tool or customizable framework for conducting recall experiments with their benchmark.  At the very least, the procedure should be repeatable, so that tool evaluation results can be verified by multiple researchers, and so that measurements by different researchers are comparable.
		
			\paragraph{RE1 - Procedure}
			
			For procedure, we use the following four-point evaluation:
			
				\begin{description}[style=standard,align=right,itemindent=-0.05cm]
				\item[- -] The authors provide no guidance on how to measure recall with their benchmark.
				\item[-]   The authors provide a partial procedure for measuring recall with their benchmark.
				\item[+]   The authors provide a full procedure for measuring recall with their benchmark.
				\item[+ +] The authors provide procedures for measuring recall with their benchmark from multiple perspectives.
				\end{description}
			
			Bellon's Benchmark, the Mutation Framework, and BigCloneBench have multiple procedures proposed by their authors, and demonstrated in published works.  ForkSim describes an example procedure for evaluating a clone detector with its data.  The other benchmarks provide no suggestions on how to conduct recall experiments with their data.
			
			\paragraph{RE2 - Clone-Matching Algorithms}
			
			For clone-matching algorithms, we use the following four-point evaluation:
			
			\begin{description}[style=standard,align=right,itemindent=-0.05cm]
				\item[- -] The authors do not suggest any clone-matching algorithms.
				\item[-]   The authors partially suggest a clone-matching algorithm (not well explained/justified).
				\item[+]   The authors propose and justify a clone-matching algorithm to be used with their benchmark.
				\item[+ +] The authors propose and justify multiple clone-matching algorithms to be used with their benchmark.
			\end{description}
		
			Bellon's Benchmark, the Mutation Framework and BigCloneBench suggest multiple and customizable clone-matching algorithms that are appropriate for their data.  ForkSim suggests only an exact-match algorithm, which may not be appropriate for all tools.  The other benchmarks provide no clone-matching algorithms appropriate for their data.
		
			\paragraph{RE3 - Tool Support}
			
			For tool support, we use the following four-point evaluation:
			
			\begin{description}[style=standard,align=right,itemindent=-0.05cm]
				\item[- -] The benchmark lacks tool support for measuring recall.
				\item[-]   The benchmark has a tool which partially supports the measurement of recall.
				\item[+]   The benchmark has a tool which measures recall.
				\item[++]  The benchmark has a flexible framework for custom recall measurement experiments.
			\end{description}
		
			Bellon's Benchmark has a tool for measuring recall, after importing the tools' results.  The Mutation Framework and BigCloneBench have flexible evaluation frameworks, that allow significant customizations on the experiment, and can automate the execution of the tools for the benchmark as well.  The other benchmarks do not provide tool support.
		
			\paragraph{RE4 - Repeatable}
			
			For repeatable, we use the following four-point evaluation:
			
			\begin{description}[style=standard,align=right,itemindent=-0.05cm]
				\item[- -] Recall experiments are not repeatable.
				\item[-]   Recall experiments are repeatable, but requires efforts to be fully re-produced by others.
				\item[+]   The benchmark enables researchers to repeat/verify recall experiments.
				\item[++]  The benchmark allows recall experiments to be exported and reviewed by others.
			\end{description}
			
			Researchers can easily repeat experiments with Bellon's Benchmark and BigCloneBench by sharing the clone detector output files.  The Mutation Framework is unique in that it allows recall experiments to be exported and shared, allowing other researchers to view, modify and extend the experiments.  Lavoie and Merlo, ForkSim and Krutz and Le are repeatable, so long as researchers share their procedure.  Yuki et al. is not repeatable as the authors neither release their data nor fully document their approach.
			
		\subsubsection{Other Metrics Requirements}
	
		The goal of a clone benchmark is to provide reference data for measuring recall.  However, it is also valuable if the benchmark supports the measurement of the other performance metrics: precision, execution time and scalability.  While these metrics can be measured without a benchmark, the benchmarks can standardize these measurements (for comparability) and reduce evaluation efforts.
	
			\paragraph{OM2 - Measures Precision}
			
			It is valuable if the benchmark can also measure precision.  A complete oracle is required to measure precision, which is rare in a benchmark.  Incomplete reference corpora can measure a lower and upper bound on precision, which can be useful for comparative analysis.  Measuring precision using reference data is limited by the data's original validation, and by the clone-matching algorithm used, so it is still best to compliment the measurement with manual validation of the tool's output.  Some of the novel clone benchmarks measure precision in a unique way, which provides an alternate perspective from the standard measurement.
			
			For measuring precision, we use the following four-point evaluation:
			
				\begin{description}[style=standard,align=right,itemindent=-0.05cm]
					\item[- -] Does not measure precision.
					\item[-]   Measures a lower/upper bound on precision using a reference corpus.
					\item[+]   Estimates precision using reference corpus and/or automated validation.
					\item[++]  Measures full precision using an oracle.
				\end{description}
		
			Only Lavoie and Merlo provide a full oracle, which can measure true precision, but this is limited by the validation quality of the benchmark (which was not high confidence).  Bellon's Benchmark's and BigCloneBench's incomplete reference corpora enable the measurement of a lower and upper bound on precision.  While FokrSim, Krutz and Le, and Yuki et al. provide reference data, their publications do not comment on the measurement of precision.  Krutz and Le's and Yuki et al.'s reference corpora are perhaps too small for reliable measure of upper/lower-bound precision.  The Mutation Framework provides a unique measure of precision, by automatically validating clones tangential to the reference clones, which is meant to compliment, not replace, a manual measure of precision.
		
			\paragraph{OM2 - Measures Execution Time}
			
			It is desirable for a benchmark to feature a set of subject systems appropriate for measuring execution time.  While it is not challenging for a researcher to collect open-source subject systems from the web, benchmarks can encourage standard sets of subject systems for comparability.
			
			For measuring execution time, we use the following four-point evaluation:
				
				\begin{description}[style=standard,align=right,itemindent=-0.05cm]
					\item[- -] Does not measure execution time.
					\item[-]   Provides a subject system to measure execution time against.
					\item[+]   Provides a variety of subject systems to measure execution time against.
					\item[++]  Provides a systematic collection of subject systems to measure execution time against.
				\end{description}
			
			Bellon's Benchmark provides a variety of subject systems to measure execution time against, and these systems have been adopted as a standard by the community.  BigCloneBench provides thousands of subject systems, as well artificially constructed subject systems that systematically increase in difficulty.  Krutz and Le uses three subject systems as part of their benchmark.  The other benchmarks either don't provide a subject system, or are not publicly released.
		
			\paragraph{OM2 - Measures Scalability}
			
			It is also desirable for the benchmark to provide a set of subject systems with increasing difficulty, so the scalability of the tools can be evaluated.
			
			For measuring scalability, we use the following four-point evaluation:
			
				\begin{description}[style=standard,align=right,itemindent=-0.05cm]
					\item[- -] Does not measure scalability.
					\item[-]   Provides a large subject system to test scalability against.
					\item[+]   Provides a set of subject systems of increasing size to test scalability against.
					\item[++]  Provides a collection of subject systems with systematically increasing difficulty to test scalability against.
				\end{description}
			
			Bellon's Benchmark provides a set of systems of different sizes, enabling a measure of scalability.  BigCloneBench provides a set of subject systems with systematically increasing difficult in terms of lines of code, while normalizing for other factors that affect scalability.  BigCloneBench also provides IJaDataset, an example target for big data clone detection.  The other benchmarks do not enable the measurement of scalability.
	
		\subsubsection{Summary}
		Our multi-faceted evaluation of the benchmarks is summarized in Table~\ref{tab:benchmarks_requirementsresults}.  As can be seen, no benchmark is able to satisfy all of the requirements of a good benchmark.  This does not mean they are bad benchmarks, but rather it shows how difficult it is to build a good clone benchmark.  To overcome limitations in the benchmarks, it is best if researchers and tool developers use multiple benchmarks.  From the evaluation, we see that there are some strong benchmarks.  The benchmarks that best satisfy the requirements are the Mutation and Injection Framework and BigCloneBench.  What is great about this pair of benchmarks is that they cover the weakness of the other.  Some of the benchmarks, such as Krutz and Le and Yuki et al. have poorer evaluations because they are still in the early phases, and require significant expansion.  In particular, Yuki et al. has only been published as a benchmarking concept.  This evaluation shows how these benchmarks need to evolve to become high-quality benchmarks.
	
	\subsection{Threats to Validity}
	\label{sec:benchmarks_threats}
	There are a few threats to the validity of this survey.  While we thoroughly examined the literature to find the benchmarks, we may have missed some publications.  For our evaluation of the benchmarks, some of the requirements have some subjectivity.  In most cases we were able to create a clear and objective ranking, but some of the requirements are more broad concepts that required comparative ranking of the existing works.
	
\section{A Survey of Clone Detector Evaluations by Authors}
\label{sec:toolEvaluations}
In this section, we survey the existing clone detection tools and techniques, and examine how their authors have evaluated the performance of their tools and techniques.  We search the literature for the clone detection tool and technique/algorithm publications, and examine the evaluations performed by their authors.  In particular, we examine if the authors have done a case study, if they compare their tool/technique to others, if they measure the clone detection performance metrics including recall and precision, and if they measure execution performance metrics such as execution time and scalability.  We then rank the works by how well they measured each of the performance metrics, and highlight the tools with the best overall evaluation strategies.  When the authors measure recall, we examine if they do so using an existing clone benchmark, or if they built a reference corpus themselves.  We then measure the overall statistics of how tool/technique authors typically perform their evaluation experiments.

For this survey, we considered only the tool and technique publications.  We did not consider tool evaluation studies, where the authors evaluate and compare multiple tools in an experiment designed to be unbiased.  We examine the tool evaluation papers in a separate survey in Section~\ref{sec:surveyToolComparisons}.  While it is common for tool authors to compare their tools against the competition, these are not the same as tool evaluation papers.  Evaluation by the authors are meant to highlight the performance of their tool, so they may not be completely unbiased.  Tool authors are also experts in their techniques, and best able to demonstrate its performance.  Tool comparison studies typically consider only the most popular and publicly available tools, and the authors of these studies may not be experts in the tools themselves.  It is therefore very important that tool/technique authors evaluate their work.

This section is organized as follows.  In Section~\ref{sec:toolEvaluations_surveyprocedure} we overview our survey procedure, and in Section~\ref{sec:toolEvaluations_evaluationCriteria} we describe our evaluation criteria.  We present the results of our survey in Section~\ref{sec:toolEvaluations_results}, including the ranking of the works by the quality of their evaluation, and an investigation into how the authors measure recall.  Then in Section~\ref{sec:toolEvaluations_standards} we examine the standards for tool evaluation by the tool/technique authors by computing various statistics about their evaluation, including: how frequently a metric is measured by the tool/technique authors, how frequently sets of metrics are measured, and correlations on which metrics are measured together.  We then discuss the threats to our survey in Section~\ref{sec:toolEvaluations_threats}.

	\subsection{Survey Procedure}
	\label{sec:toolEvaluations_surveyprocedure}
	To examine how clone detection tool and technique authors evaluate their tools/techniques, we needed to identify the relevant published works in the literature.  We performed our search by examining the papers in the existing clone detection surveys~\cite{Roy07asurvey,Roy:2009:CEC:1530898.1531101,rattan}, and by searching the leading academic publication databases, including: IEEE, ACM, Springer and Wiley.
	
	There are three excellent clone detection tool surveys in the literature, including: ``A Survey on Software Clone Detection Research'' by Roy and Cordy~\cite{Roy07asurvey} published in 2007, ``Comparison and Evaluation of Code Clone Detection Techniques and Tools: A Qualitative Approach'' by Roy et al.~\cite{Roy:2009:CEC:1530898.1531101} published in 2009, and ``Software clone detection: A systematic review'' by Rattan et al.~\cite{rattan} published in 2013.  These works summarize the clone detection tool and technique publications up to at least 2011.
	
	We trust that the existing surveys are comprehensive up to 2011, so we focused our search of the academic databases for publications in the period of January 2011 to March 2017.  The databases searched, and our search criteria, are summarized in Table~\ref{tab:toolEvaluations_databasesearch}. We determined the relevancy of the papers by first examining their title, then their abstract, and then their complete texts.
	
	\begin{table}
		\centering
		\caption{Survey Databases and Search Criteria}\label{tab:toolEvaluations_databasesearch}
		\begin{tabular}{l>{\arraybackslash}p{11cm}c}
			\toprule
			Database & Search Criteria & \#Papers \\
			\midrule
			ACM &  Year: 2011-2016, Abstract Contains: clone & 367 \vspace{0.5em}\\
			IEEExplore & Years: 2011-2017, Index Term: Cloning, Refine: code clone &  411 \vspace{0.5em}\\
			IEEExplore & Years: 2011-2017, Index Term: Cloning, Refine: clone detection &  388 \vspace{0.5em}\\
			ScienceDirect & Years: 2011-2017, Abstract: clone, Sources: Computer Science & 177 \vspace{0.5em}\\
			Springer-Link & Years: 2011-2017, Title Contains: clone, Text Contains: (code OR software). & 206 \vspace{0.5em}\\
			Wiley & Years: 2011-2017, Title Contains: clone, Text Contains: clone detector & 348\\
			\bottomrule
		\end{tabular}
	\end{table}

	Where possible, we clustered papers regarding the same clone detection tool or technique together, as these should be considered as a single tool/technique.  It is not uncommon for a research group to publish their algorithm in one paper and their tool version in another. We tried to cluster the papers by considering paper authors. We then examined the papers to determine if there was a link between them, such as the same algorithm or tool name, or citations to each other.  However, it was not always clear if two papers were about the same tool or technique, or distinct works.  We erred on the side of caution, and only grouped works that were definitely about same technique/tool.  However, we may have failed to cluster some works that are the same tool/technique.
	
	For the reasons stated previously, we only considered papers which present a clone detection tool or technique, or where the authors evaluate their own tool/technique.  We do not include the tool comparison studies, as these are different kinds of evaluations.  It is important that tool/technique authors provide their own evaluations, as they are the experts in their clone detection works, so we are interested in how they conduct these evaluations.
	
	\subsection{Evaluation Criteria}
	\label{sec:toolEvaluations_evaluationCriteria}
	For each examined paper, we determined how the author(s) evaluated their clone detection tool or technique. We determined whether a case study was performed, if the tool was compared to other competing tools in a meaningful way beyond a simple summary of the related work, if recall and precision were measured, and if execution time and scalability were recorded.  If recall was measured, we determined which standard benchmark was used, or what technique the authors used to build a reference corpus for the measurement.  Using this collected data, we can measure the trends of tool/technique evaluation strategies by the authors.
	
	When judging if an author evaluated their tool or technique in a particular way (e.g., scalability), we were generous, as there is no standard procedure for evaluating clone detection tools/techniques.  We consider the authors to have ran a case study if they demonstrated the execution of their clone detection tool or technique for at least one subject system. We consider the authors to have compared their tool or technique against others if they include a quantitative comparison study, at minimum comparing the detection results of the competing tools to their own.  We did not consider a related work section to be a tool comparison study.
	
	We consider the authors to have measured recall if they evaluate their detection performance for some set of reference clones.  If the reference corpora is very small, such as only tens of clones, we consider this a measurement of recall only if the selected clones are systematically chosen in some way, such as based on taxonomy of clone types or cloning scenarios.  We consider precision to have been measured if the authors have manually or automatically validated some proportion of the output of their tool/technique.  We consider execution time to have been measured if it is reported for at least one subject system.  We consider scalability to be measured if execution time is reported for a collection of software systems of different sizes (lines of code) or complexities, or if the complexity of the algorithm is sufficiently explored and demonstrated.
	
	When documenting the evaluation of a particular tool or technique, we considered only evaluations performed by the tool/technique authors themselves for this survey.  In Section~\ref{sec:toolEvaluations} we survey comparison studies performed by both authors and non-authors of the tools/techniques.
	
	Additionally, we categorized the works into tools and techniques.  We consider a clone detection technique that is given a name by its authors to be a clone detection tool.  We then consider those not given a name to be clone detection techniques.  In both cases, it is expected that the author evaluate the performance of their tool/technique, in particular recall and precision.  However, there is a greater expectation on the evaluation of execution time and scalability for tools.  Authors may not be expected to report execution time and scalability for clone detection techniques, as they may be only partially prototyped and unoptimized.
	
	It is reasonable to assume that if an author gives a name to their clone detection work that they intend to present it as a clone detection tool.  Those that do not name their technique are communicating that they are presenting an algorithm.  This is not a formal distinction or necessarily a very accurate way to differentiate the works.  Most clone detection tools/prototypes are not released publicly, and those that are can be difficult to find or eventually lost.  Still, we see some interesting statistics when we differentiate the named tools and unnamed techniques.
	
	\subsection{Results}
	\label{sec:toolEvaluations_results}
	The results of our survey are shown in Appendix~\ref{app:oversize} in Table~\ref{tab:toolevals}.  The works are separated into clone detection tools (named) and clone detection techniques (unnamed), and are otherwise presented in alphabetical order.  For each tool/technique, we indicate their publications, and their dimensions of evaluation. An `x' denotes that a particular evaluation was performed in one of the publications for that tool/technique.  If recall was measured, we indicate the benchmark used, or the method used to build a reference corpus.  In total we found 198 works, including 90 (45\%) named tools, and 108 (55\%) unnamed techniques.  Most of the works have a single publication, although some have two or more publications.
	
	Our primary interest in this survey is to examine how the authors evaluate the performance of their clone detection tools and techniques.  In Table~\ref{tab:tooleval_standards} we organize the tools and techniques by how they are evaluated.  Specifically, we group the works by which combination of performance metrics (recall, precision, execution time and scalability) the authors evaluated.  For the works where none of the metrics were evaluated, we show which at least had a case-study performed, and those which have no meaningful demonstration or evaluation of their approach.  As can be seen, very few authors evaluate all four of performance metrics for their tool/technique, and nearly a quarter of the publications have no evaluation whatsoever.
		
	\begin{table}
		\centering
		\caption{Survey of Clone Detection Tools/Techniques and their Evaluation by their Authors}\label{tab:tooleval_standards}
		\scriptsize
		\begin{tabular}{ccccccp{13cm}}
			\toprule
			\rot{90}{1em}{Recall} & \rot{90}{1em}{Precision} & \rot{90}{1em}{Execution Time} & \rot{90}{1em}{Scalability} && \rot{90}{1em}{Number of Works} & Tools and Techniques \\
			\midrule
			x & x & x & x && 18 &
			\scriptsize
				BinClone~\cite{6895418},
				CBCD~\cite{6227183},
				CD-Form~\cite{6227879,Cuomo2014390},
				CloneWorks~\cite{cloneworks},
				CMGA~\cite{Wang2014},
				DuDe~\cite{1595830},
				DupLoc~\cite{Ducasse_SMR:SMR317,Ducasse_792593},
				MeCC~\cite{Kim:2011:MMC:1985793.1985835},
				NiCad~\cite{Cordy2014158,5970189,SMR:SMR416,5306301,4556129},
				ScalClone~\cite{Farhadi201546},
				SeByte~\cite{6240495,Keivanloo:2012:JBC:2664398.2664404,6227864,Keivanloo2014426},
				SimCad~\cite{6079770,6613857},
				SourcererCC~\cite{7886988,7883349,6240500,6613042,SMR:SMR1707},
				\textit{Corazza et al.}~\cite{5609715},
				\textit{Falke et al.}~\cite{Falke2008},
				\textit{Kam1n0}~\cite{Ding:2016:KMA:2939672.2939719},
				\textit{Karus and Kilgi}~\cite{7069883},
				\textit{Qu et al.}~\cite{Qu2014544}
			\\
			\midrule
			x & x & x &   && 13 &
			\scriptsize
				BAT~\cite{Hemel:2011:FSL:1985441.1985453},
				CDSW~\cite{6613837},
				clones/cscope~\cite{4023995},
				CodeBlast~\cite{Bhattacharjee:2013:CTA:2480362.2480525},
				cpdetector~\cite{723194,4023995},
				Dup~\cite{baker,baker:si:92,baker:514697},
				JPlag~\cite{Prechelt00findingplagiarisms},
				$MQ_{lone}$~\cite{Storrle2013},
				SHINOBI~\cite{5328752,shinobiTR},
				Simon~\cite{Zilberstein:2016:LCN:2986012.2986013},
				\textit{Abdelkader and Mimoun}~\cite{7483299},
				\textit{Lavoie and Merlo}~\cite{6227861},
				\textit{St{\"o}rrle}~\cite{Storrle2015}
			\\
			x & x &   & x && 2 &
			\scriptsize
				Clonewise~\cite{Cesare2013},
				\textit{Zibran and Roy}~\cite{Zibran:2012:IRF:2245276.2231970,Zibran:2011:TFC:1985404.1985423}
			\\
			x &   & x & x && 5 &
			\scriptsize
				Asta~\cite{4400161,Evans2009},
				HeapAbsCC~\cite{6319249},
				LLVM-Based Framework~\cite{Sargsyan2016,7358259},
				Shuffling Framework~\cite{SMR:SMR1662,6227875,6613037},
				sif~\cite{Manber:1994:FSF:1267074.1267076}
			\\
			  & x & x & x && 12 &
			\scriptsize
				Boreas~\cite{6494937, Yuan:2012:SAA:2162110.2162126},
				CP-Miner~\cite{li_1610609,Li:2004:CTF:1251254.1251274},
				Duplix~\cite{957835},
				FCFinder~\cite{5463293},
				HitoshiIO~\cite{7503720},
				ModelCD~\cite{5070528},
				SaCD~\cite{6615249},
				XIAO~\cite{Dang:2012:XTC:2420950.2421004,Dang:2011:CCD:1985404.1985417},
				\textit{Ishihara et al.}~\cite{6385134},
				\textit{Koschke}~\cite{6178897},
				\textit{Koschke}~\cite{SMR:SMR1592},
				\textit{Saebjornsen et al.}~\cite{Saebjornsen:2009:DCC:1572272.1572287}
			\\
			\midrule
			x & x &   &  && 23 &
			\scriptsize
				AuDeNTES~\cite{Mariani:2012:AAD:2133797.2133799},
				CCCD~\cite{Krutz:2015:EEU:2695664.2695929,6671332},
				CLCMiner~\cite{7582804},
				CPDP~\cite{6613041},
				DroidClone~\cite{7544005},
				DuplicationDetector~\cite{4137427},
				eMetrics~\cite{Lanubile_1192447,Calefato:2004:FCD:2011138.2011140},
				FRISC~\cite{6392103},
				JSCD~\cite{Cheung2016},
				\textit{Abd-El-Hafiz}~\cite{6340252},
				\textit{Al-Omari et al.}~\cite{6385136},
				\textit{Bansal and Tekchandani}~\cite{6897221},
				\textit{Bauer et al.}~\cite{6976124},
				\textit{DL-Clone}~\cite{Schugerl:2011:SCD:1985404.1985413},
				\textit{Ekanayake et al.}~\cite{Ekanayake2012},
				\textit{Iwamoto et al.}~\cite{6363131},
				\textit{Iwamoto et al.}~\cite{6690950},
				\textit{Keivanloo et al.}~\cite{7081830},
				\textit{Li and Sun}~\cite{5478099},
				\textit{Lucia et al.}~\cite{4380246},
				\textit{Sheneamer and Kalita}~\cite{7397263},
				\textit{Sheneamer and Kalita}~\cite{7838289},
				\textit{Stojanovic\' et al}~\cite{Stojanovic2015259}
			\\
			x &   & x &   && 1 &
			\scriptsize
				\textit{Perumal et al.}~\cite{5640465}
			\\
			x &   &   & x && 1 &
			\scriptsize
				CMCD~\cite{6130694}
			\\
			  & x & x &   && 8 &
			\scriptsize
				ClemenX~\cite{5071050},
				Clone Detective (ConQat)~\cite{conqat,conqat_models},
				EqMiner~\cite{Jiang:2009:AMF:1572272.1572283},
				MiLoCo~\cite{6676875},
				RTF~\cite{Basit:2007:ETB:1295014.1295029},
				\textit{Chen et al.}~\cite{Chen2016},
				\textit{Kontogiannis et al.}~\cite{Kontogiannis1996},
				\textit{White et al.}~\cite{7582748}
			\scriptsize
			\\
			  & x &   & x && 0 &
			\scriptsize 
			-
			\\
			  &   & x & x && 11 &
			\scriptsize
				CCFinder(X)~\cite{ccfinderx},
				CtCompare~\cite{6227881},
				Decrescendo~\cite{7880510},
				DyCLINK~\cite{Su:2016:CRD:2950290.2950321},
				HaRe~\cite{Brown:2010:CDE:1706356.1706378},
				SeClone~\cite{6079771,6225474},
				SSD~\cite{Lee:2005:SHP:1094855.1094903},
				\textit{Barbour et al.}~\cite{5521760},
				\textit{Chilowicz et al.}~\cite{CHILOWICZ200947},
				\textit{Hummel et al.}~\cite{5609665},
				\textit{Tekin and Buzluca}~\cite{Tekin2014406}
			\\
			\midrule
			x &   &   &   && 4 &
			\scriptsize
				AST-CC~\cite{6631651},
				CodeCompare~\cite{6631708},
				Scorpio~\cite{5741248},
				\textit{Kong et al.}~\cite{6463128}
			\\
			  & x &   &   && 12 &
			\scriptsize
				Deckard~\cite{deckard},
				Exas~\cite{Nguyen2009},
				R2D2~\cite{Leitao2004},
				\textit{Chen et al.}~\cite{Chen:2014:RRC:2667473.2667486},
				\textit{Dou et al.}~\cite{Dou:2016:DTC:2950290.2950359},
				\textit{Hermans et al.}~\cite{Hermans:2013:DCD:2486788.2486827},
				\textit{Jadon}~\cite{7813733},
				\textit{Joshi et al.}~\cite{Joshi2015},
				\textit{Patil et al.}~\cite{7087126},
				\textit{Priyambadha and Rochimah}~\cite{7062689},
				\textit{Ragkhitwetsagul and Krinke}~\cite{7880502},
				\textit{Uemura et al.}~\cite{7880501}
			\\
			  &   & x &   && 34 &
			\scriptsize
				Agec~\cite{6613854},
				C2D2~\cite{kraft2008cross},
				Covet/CLAN~\cite{Mayrand_565012,bellon},
				D-CCFinder~\cite{4222573},
				GPLAG~\cite{Liu:2006:GDS:1150402.1150522},
				Hanni~\cite{Lillack:2014:DCC:2660190.2662116},
				iClones~\cite{iclones,iclonesthesis},
				JCCD~\cite{5645564},
				LSC Miner~\cite{6377848},
				PDG-DUP~\cite{Komondoor2001,Komondoor2001-2},
				Sim~\cite{Gitchell:1999:SUD:299649.299783},
				Simian~\cite{simian},
				Wrangler~\cite{Li2010},
				\textit{Ahkin and Itsykson}~\cite{Akhin2013},
				\textit{Ali et al.}~\cite{6140712},
				\textit{Cordy et al.}~\cite{Cordy:2004:PLD:1034914.1034915},
				\textit{Dandois and Vanhoof}~\cite{Dandois2012},
				\textit{Di Lucca et al.}~\cite{Lucca01cloneanalysis,1134103},
				\textit{Dumas et al.}~\cite{Dumas2013619,Uba2011},
				\textit{Grant and Cordy}~\cite{5090048},
				\textit{Higo et al.}~\cite{6079769},
				\textit{Huang and Li}~\cite{5376091},
				\textit{Hummel et al.}~\cite{Hummel:2011:IMC:1985404.1985409},
				\textit{Johnson}~\cite{Johnson:1993:IRS:962289.962305,Johnson:1994:VTR:782185.782217,336783},
				\textit{Kamiya}~\cite{7069882},
				\textit{Lavoie et al.}~\cite{Lavoie:2010:CCR:1808901.1808905},
				\textit{Lazar and Banias}~\cite{6840038},
				\textit{Li et al.}~\cite{6821712},
				\textit{Pulkkinen et al.}~\cite{Pulkkinen:2015:RBP:2812428.2812471},
				\textit{Str{\"u}ber}~\cite{Struber2016},
				\textit{Udagawa}~\cite{Udagawa:2014:ESR:2684200.2684290},
				\textit{Udagawa}~\cite{Udagawa:2016:MFS:3011141.3011160},
				\textit{Wahler et al.}~\cite{1386166},
				\textit{Zhang et al.}~\cite{6982615}
			\\
			  &   &   & x && 0 &
			\scriptsize
			-
			\\
			\midrule
			\multicolumn{4}{c}{\scriptsize Case Study Only}   && 11 &
			\scriptsize 
				K-Clone~\cite{kclone},
				Simone~\cite{6405285,7447218,6227873},
				\textit{Agrawal and Yadav}~\cite{6726700},
				\textit{Kamalpriya and Singh}~\cite{7880511},
				\textit{Kaur and Singh}~\cite{Kaur:2014:CDS:2597716.2597723},
				\textit{Keivanloo and Rilling}~\cite{6227963},
				\textit{Kodhai et al.}~\cite{5460547},
				\textit{Matsushita and Sasano}~\cite{Matsushita:2017:DCC:3018882.3018892},
				\textit{Petrik et al.}~\cite{7880355},
				\textit{Sabi et al.}~\cite{7880503},
				\textit{Tekchandani et al.}~\cite{Tekchandani2016}
			\\
			\multicolumn{4}{c}{\scriptsize No Evaluation}   && 43 &
			\scriptsize
				Bauhaus~\cite{Bauhaus,Raza2006},
				clone-digger~\cite{bulychev2008duplicate},
				CloneDr~\cite{Baxter_1317484,Baxter_738528},
				Clone Miner~\cite{Basit:2005:DHS:1081706.1081733,4658086,4796208,Basit:2011:VSC:1985404.1985406},
				Coogle~\cite{Sager:2006:DSJ:1137983.1138000},
				CSeR~\cite{Jacob:2010:ACC:1808901.1808903},
				PC Detector~\cite{6779537},
				SimScan~N/A,
				SMAT~\cite{Yamamoto2005},
				STVsm~\cite{Li2012},
				\textit{Al-Batran et al.}~\cite{Al-Batran:2011:SCD:2050655.2050681},
				\textit{Antony et al.}~\cite{6671325},
				\textit{Ballarin et al.}~\cite{Ballarin2016},
				\textit{Chodarev et al.}~\cite{7158423},
				\textit{Choi et al.}~\cite{Choi2009862},
				\textit{Davey et al.}~\cite{daveycd},
				\textit{Deissenboeck et al.}~\cite{6178896},
				\textit{Devi and Punithavalli}~\cite{5941574},
				\textit{Elva and Leavens}~\cite{6227874},
				\textit{He}~\cite{6308805},
				\textit{Higo and Kusumoto}~\cite{Higo:2014:WMF:2635868.2635886},
				\textit{Ito et al.}~\cite{7880504},
				\textit{Juillerat and Hirsbrunner}~\cite{juillerat2007},
				\textit{Keivanloo and Rilling}~\cite{6649853},
				\textit{Kumar}~\cite{6976163},
				\textit{Kumar et al.}~\cite{Kumar:2014:DDS:2660252.2660397},
				\textit{Lee and Doh}~\cite{Lee:2009:TDC:1651309.1651312},
				\textit{Maeda}~\cite{maeda},
				\textit{Marcus and Maletic}~\cite{989796},
				\textit{Min et al.}~\cite{6478318},
				\textit{Raheja and Tekchandani}~\cite{6832302},
				\textit{Rattan et al.}~\cite{6420770},
				\textit{Rodrigues and Vila{\~a}a}~\cite{Rodrigues2010},
				\textit{Santone}~\cite{Santone:2011:CDT:1985404.1985422},
				\textit{Singh and Raminder}~\cite{Singh:2014:CDU:2597716.2597726},
				\textit{Singh and Sharma}~\cite{SINGH2015915},
				\textit{Sudhamani and Rangarajan}~\cite{SUDHAMANI2015892},
				\textit{Surendran et al.}~\cite{6141393},
				\textit{Sutton et al.}~\cite{Sutton:2005:HEA:1068009.1068191},
				\textit{Tairas and Gray}~\cite{Tairas:2006:PCD:1185448.1185597},
				\textit{Tairas}~\cite{Tairas2012},
				\textit{Tekchandani et al.}~\cite{6832306},
				\textit{Tekin et al.}~\cite{Tekin:2012:MOD:2664398.2664405}
			\\
			\bottomrule
		\end{tabular}
	\end{table}
		
	In the remainder of this section, we rank the works by how well they measure the four performance metrics.  We then extract the works with the best ranking across all four metrics to serve as exemplars for how authors should evaluate their tools.  We conclude this section by examining how the authors measured recall, whether by standardized benchmarks by building new reference corpora.
		
		\subsubsection{Results Ranked by their Measurement of Recall}
		\label{sec:toolEvaluations_results_recall}
		In this section we rank the clone detection tool and technique publications by how well the authors measured recall.  We propose a seven tier ranking, which is summarized in Table~\ref{tab:cluster_recall_ranks}.  We rank highest those works that use one or more standardized clone benchmarks, and below this those whose authors have developed new non-standard reference corpora to evaluate their tools.  We rank lowest those who evaluated against only example clones, or those who use other tools as a baseline.  The ranked works are shown in Table~\ref{tab:cluster_recall_results}.  As can be seen, most works which measure recall either use standard benchmarks, or self-created reference corpora.
	
		\begin{table}
		\small
		\caption{Measurement of Recall Ranks} \label{tab:cluster_recall_ranks}
		\centering
		\begin{tabular}{cl>{\raggedright\arraybackslash}m{9.5cm}}		
			\toprule
			\textbf{Rank}  & \textbf{Name}              & \textbf{Description} \\
			\midrule
			a              & Multiple Benchmarks        & Evaluated using multiple mature and community-accepted benchmarks. \\
			\rule{0pt}{5ex}
			b              & Single Benchmark           & Evaluated using a mature and community-accepted benchmark. \\
			\rule{0pt}{5ex}
			c              & Quality Reference Corpus   & Evaluated using a well-constructed reference corpus built by the authors to evaluate their tool.\\
			\rule{0pt}{5ex}
			d              & Simple Reference Corpus    & Evaluated using a simple reference corpus built by the authors to evaluate their tool. \\
			\rule{0pt}{5ex}
			e              & Clones from the Literature & Evaluated using example clones from the literature. \\
			\rule{0pt}{5ex}
			f              & Clone Detector Baseline    & Evaluated using detection result of other tool(s) without validation. \\
			\rule{0pt}{5ex}
			g              & Unknown                    & Reported, but evaluation procedure is unknown. \\
			\bottomrule
		\end{tabular}
	\end{table}
	
		\begin{table}
		\small
		\centering
		\caption{Clone Detection Tools/Techniques Ranked by their Recall Measurement}\label{tab:cluster_recall_results}
		\begin{tabular}{clc>{\raggedright\arraybackslash}m{8cm}}
			\toprule
			\textbf{Rank} & \textbf{Name} & \textbf{Num} & \textbf{Tools/Techniques} \\
			\midrule
			a & Multiple Benchmarks & 2 &
			\scriptsize
			SourcererCC	\cite{7886988,7883349,6240500,6613042,SMR:SMR1707},
			CloneWorks	\cite{cloneworks}
			\\\rule{0pt}{10ex}
			
			b & Single Benchmark           & 20 & 
			\scriptsize
			Asta	\cite{4400161,Evans2009},
			CCCD	\cite{Krutz:2015:EEU:2695664.2695929,6671332},
			clones/cscope	\cite{4023995},
			cpdetector	\cite{723194,4023995},
			DuDe	\cite{1595830},
			Dup	\cite{baker,baker:si:92,baker:514697},
			DupLoc	\cite{Ducasse_SMR:SMR317,Ducasse_792593},
			FRISC	\cite{6392103},
			Scorpio	\cite{5741248},
			\textit{Abdelkader and Mimoun}	\cite{7483299},
			\textit{Bansal and Tekchandani}	\cite{6897221},
			\textit{Falke et al.}	\cite{Falke2008},
			\textit{Sheneamer and Kalita}	\cite{7397263},
			CDSW	\cite{6613837},
			\textit{Perumal et al.}	\cite{5640465},
			\textit{Keivanloo et al.}	\cite{7081830},
			\textit{Sheneamer and Kalita}	\cite{7838289},
			NiCad	\cite{Cordy2014158,5970189,SMR:SMR416,5306301,4556129},
			SimCad	\cite{6079770,6613857},
			\textit{Zibran and Roy}	\cite{Zibran:2012:IRF:2245276.2231970,Zibran:2011:TFC:1985404.1985423},
			JSCD	\cite{Cheung2016}
			\\\rule{0pt}{6ex}
			
			c & Quality Reference Corpus        & 12 &
			\scriptsize
			\textit{Al-Omari et al.}	\cite{6385136},
			\textit{Qu et al.}	\cite{Qu2014544},
			Simon	\cite{Zilberstein:2016:LCN:2986012.2986013},
			\textit{St{\"o}rrle}	\cite{Storrle2015},
			CBCD	\cite{6227183},
			Clonewise	\cite{Cesare2013},
			DroidClone	\cite{7544005},
			DuplicationDetector	\cite{4137427},
			eMetrics	\cite{Lanubile_1192447,Calefato:2004:FCD:2011138.2011140},
			SeByte	\cite{6240495,Keivanloo:2012:JBC:2664398.2664404,6227864,Keivanloo2014426},
			\textit{Lucia et al.}	\cite{4380246}
			\\\rule{0pt}{8ex}
			
			d & Simple Reference Corpus    & 20 &
			\scriptsize
			\textit{Stojanovi\'c et al}	\cite{Stojanovic2015259},
			BAT	\cite{Hemel:2011:FSL:1985441.1985453},
			CPDP	\cite{6613041},
			CMGA	\cite{Wang2014},
			CodeBlast	\cite{Bhattacharjee:2013:CTA:2480362.2480525},
			CodeCompare	\cite{6631708},
			HeapAbsCC	\cite{6319249},
			\textit{Bauer et al.}	\cite{6976124},
			\textit{Ekanayake et al.}	\cite{Ekanayake2012},
			\textit{Kong et al.}	\cite{6463128},
			\textit{Corazza et al.}	\cite{5609715},
			AuDeNTES	\cite{Mariani:2012:AAD:2133797.2133799},
			BinClone	\cite{6895418},
			JPlag	\cite{Prechelt00findingplagiarisms},
			\textit{Iwamoto et al.}	\cite{6363131},
			\textit{Iwamoto et al.}	\cite{6690950},
			\textit{Li and Sun}	\cite{5478099},
			ScalClone	\cite{Farhadi201546},
			$MQ_{lone}$	\cite{Storrle2013},
			sif	\cite{Manber:1994:FSF:1267074.1267076}
			\\\rule{0pt}{4ex}
			
			e & Clones from the Literature & 5 &
			\scriptsize
			MeCC	\cite{Kim:2011:MMC:1985793.1985835},
			AST-CC	\cite{6631651},
			CMCD	\cite{6130694},
			LLVM-Based Framework	\cite{Sargsyan2016,7358259},
			\textit{Karus and Kilgi}	\cite{7069883}
			\\ \rule{0pt}{5ex}
			
			f & Clone Detector Baseline    & 7 &
			\scriptsize
			CD-Form	\cite{6227879,Cuomo2014390},
			Shuffling Framework	\cite{SMR:SMR1662,6227875,6613037},
			\textit{Abd-El-Hafiz}	\cite{6340252},
			\textit{Kam1n0}	\cite{Ding:2016:KMA:2939672.2939719},
			\textit{Lavoie and Merlo}	\cite{6227861},
			CLCMiner	\cite{7582804},
			\textit{DL-Clone}	\cite{Schugerl:2011:SCD:1985404.1985413}
			\\ \rule{0pt}{3ex}
			
			g & Unknown                    & 1 &
			\scriptsize 
			SHINOBI	\cite{5328752,shinobiTR}
			\\\bottomrule
		\end{tabular}
	\end{table}
	
		In the remainder of this section, we describe and justify the ranks in detail.
	
		\begin{adjustwidth}{1cm}{}
			\textbf{Rank A - Multiple Benchmarks:}  The recall of the tool/technique was evaluated using multiple published clone benchmarks.  By using multiple benchmarks, and comparing their results, the authors get a more complete view of their tool/technique's recall.  Individual benchmarks contain limitations, so using multiple benchmarks built by different means can overcome these limitations.  
			
			\noindent \textbf{Rank B - A Single Benchmark:} The recall of the tool/technique was evaluated using a single benchmark.  However, the benchmark may have limitations in measuring recall, and by using only a single benchmark these limitations cannot be overcome.\\
			
			\noindent \textbf{Rank C - Quality Reference Corpus:}  The recall of the tool/technique was evaluated using a well-constructed reference corpus built by the tool/technique authors themselves.  To achieve this rank, the authors must have taken care to build a quality reference corpus.  They may have built a large corpus, taken extra steps in clone validation, used an innovative mining process, or have a well-defined scope or context, and so on.  The work done is a good start towards building a formal benchmark.\\
			
			\noindent \textbf{Rank D - Simple Reference Corpus:} The recall of the tool/technique was evaluated using a simple reference corpus built by the tool/technique authors themselves.  The reference corpus may be small, lack scope or context, have unclear validation, or be poorly described.\\
			
			\noindent \textbf{Rank E - Clones from the Literature:} The recall of the tool/technique was evaluated against clones taken from the literature.  Typically, this is examples of different types of clones from clone taxonomies, such as from Roy et al.~\cite{Roy:2009:CEC:1530898.1531101}.  This can show what types of clones the tool supports, but does not show how well it works in practice for large variety of these clone types.\\
			
			\noindent \textbf{Rank F - Clone Detector Baseline:} The recall of the tool/technique was evaluated against clones detected by another clone detector(s), without validation.  This approach is often used to show a tool performs as well as an existing well-received tool, but maybe has additional features or faster performance.  This is a poor measure of recall as it assume the baseline tool(s) have correct output.  This technique can easily under-estimate true recall (baseline tool reported false positives not found by the evaluated tool) or over-estimate true recall (baseline tool reported false positives also found by the evaluated tool). \\
			
			\noindent \textbf{Rank G - Unknown:} The recall of the tool/technique is reported, but the evaluation procedure is not sufficiently described.  The measured result is therefore sort of useless as it can't be understood by the reader.\\
		\end{adjustwidth}
		
		With our rankings, we differentiate between standard clone benchmarks and reference corpora built by the tool/technique authors specifically for their publication.  While clone benchmarks consist of reference corpora, the distinction is that the standard clone benchmarks were designed, published and peer-reviewed independently of a particular clone detection tool.  Reference corpora introduced in tool/technique publication papers do not have the same confidence.  These corpora may be improvised, designed to highlight the benefits of the published tool, are typically not well described in the paper, and so on.  They are also not the main contribution of the paper, so not as stringently evaluated by peer-review.  We identify the benchmarks using our survey on the clone benchmarks from Section~\ref{sec:benchmarks}.
		
		Still, many tool/technique authors have built high quality reference corpora, and we differentiate these papers in our ranking as quality vs simple reference corpora.  These authors have introduced techniques/methodologies that could be further explored to build standard clone benchmarks.  However, their description is still only a minor aspect of the tool/technique paper, so are not well-understood.
		
		The other rankings consider how recall is tangentially measured without the use of a proper reference corpora.  
		
		\subsubsection{Results Ranked by their Measurement of Precision}
		\label{sec:toolEvaluations_results_precision}
		In this section, we rank the clone detection tool and technique publications by how well the authors measured precision.  We propose a six tier ranking, which is summarized in Table~\ref{tab:cluster_precision_ranks}.  The best method of measuring precision is manual validation of the output, and we create a number of ranks based on the size of the validation effort.  Below this we consider works that automatically measure precision using an independently built reference corpus or validation tools, and authors who measure precision only qualitatively.  The ranked works are summarized in Table~\ref{tab:cluster_precision_ranks}.  As can be seen, a number of works satisfy our top tier of evaluation.  Measuring precision is not particular difficult, but requires a time investment by the authors.  A large number of works have measured precision against a reference corpus, which is generally limited to a lower or upper bound.  A few works have reported precision without clearly stating how the measurement was made.
		
		\begin{table}
			\small
			\caption{Measurement of Precision Ranking} \label{tab:cluster_precision_ranks}
			\centering
			\begin{tabular}{cl>{\raggedright\arraybackslash}m{7cm}}		
				\toprule
				\textbf{Rank}  & \textbf{Name}              & \textbf{Description} \\
				\midrule
				a              & Significant Manual Validation & Precision is measured by manually validating a significant number of clones (300+). \\
				\rule{0pt}{6ex}
				b              & Large Manual Validation & Precision is measured by manually validating a large number of clones (100-300)\\
				\rule{0pt}{6ex}
				c              & Medium Manual Validation  & Precision is measured by manually validating a medium number of detected clones (50-100).\\
				\rule{0pt}{6ex}
				d              & Small Manual Validation  & Precision is measured by manually validating a small number of detected clones (10-50). \\
				\rule{0pt}{9ex}
				e              & Automatic Validation & Precision is measured automatically using a incomplete or complete reference corpus with a clone-matching algorithm, or some automatic validation technique/metric. \\
				\rule{0pt}{8ex}
				f              & Unknown    & Precision is reported, but its measurement is unclear. At least, the number of clones investigated is not reported. \\
				\bottomrule
			\end{tabular}
		\end{table}
		
		\begin{table}
			\small
			\centering
			\caption{Clone Detection Tools/Techniques Ranked by their Precision Measurement}\label{tab:cluster_precision_results}
			\begin{tabular}{clc>{\raggedright\arraybackslash}m{8cm}}
				\toprule
				\textbf{Rank} & \textbf{Name} & \textbf{Num} & \textbf{Tools/Techniques} \\
				\midrule
				a & Significant Manual Validation & 21 &
				\scriptsize
AuDeNTES	\cite{Mariani:2012:AAD:2133797.2133799},
CLCMiner	\cite{7582804},
clones/cscope	\cite{4023995},
CloneWorks	\cite{cloneworks},
cpdetector	\cite{723194,4023995},
Dup	\cite{baker,baker:si:92,baker:514697},
MeCC	\cite{Kim:2011:MMC:1985793.1985835},
ModelCD	\cite{5070528},
NiCad	\cite{Cordy2014158,5970189,SMR:SMR416,5306301,4556129},
SaCD	\cite{6615249},
SeByte	\cite{6240495,Keivanloo:2012:JBC:2664398.2664404,6227864,Keivanloo2014426},
SourcererCC	\cite{7886988,7883349,6240500,6613042,SMR:SMR1707},
\textit{Abd-El-Hafiz}	\cite{6340252},
\textit{Chen et al.}	\cite{Chen:2014:RRC:2667473.2667486},
\textit{Dou et al.}	\cite{Dou:2016:DTC:2950290.2950359},
\textit{Falke et al.}	\cite{Falke2008},
\textit{Karus and Kilgi}	\cite{7069883},
\textit{Ragkhitwetsagul and Krinke}	\cite{7880502},
\textit{Saebjornsen et al.}	\cite{Saebjornsen:2009:DCC:1572272.1572287},
\textit{Uemura et al.}	\cite{7880501},
\textit{White et al.}	\cite{7582748}
				\\\rule{0pt}{7ex}
				b & Large Manual Validation      & 15 & 
				\scriptsize
Boreas	\cite{6494937, Yuan:2012:SAA:2162110.2162126},
ClemenX	\cite{5071050},
CP-Miner	\cite{li_1610609,Li:2004:CTF:1251254.1251274},
Deckard	\cite{deckard},
eMetrics	\cite{Lanubile_1192447,Calefato:2004:FCD:2011138.2011140},
HitoshiIO	\cite{7503720},
MiLoCo	\cite{6676875},
RTF	\cite{Basit:2007:ETB:1295014.1295029},
SimCad	\cite{6079770,6613857},
\textit{Bauer et al.}	\cite{6976124},
\textit{Chen et al.}	\cite{Chen2016},
\textit{Hermans et al.}	\cite{Hermans:2013:DCD:2486788.2486827},
\textit{Ishihara et al.}	\cite{6385134},
\textit{Priyambadha and Rochimah}	\cite{7062689},
\textit{St{\"o}rrle}	\cite{Storrle2015}
				\\\rule{0pt}{4ex}
				c & Medium Manual Validation   & 5 &
				\scriptsize
BAT	\cite{Hemel:2011:FSL:1985441.1985453}
DuDe	\cite{1595830}
XIAO	\cite{Dang:2012:XTC:2420950.2421004,Dang:2011:CCD:1985404.1985417}
\textit{Jadon}	\cite{7813733}
\textit{Qu et al.}	\cite{Qu2014544}
				\\\rule{0pt}{3ex}
				d & Small Manual Validation & 3 &
				\scriptsize
FCFinder	\cite{5463293},
R2D2	\cite{Leitao2004},
\textit{Kontogiannis et al.}	\cite{Kontogiannis1996}
				\\\rule{0pt}{14ex}
				e & Automatic Validation & 36 &
				\scriptsize
BinClone	\cite{6895418},
CBCD	\cite{6227183},
CCCD	\cite{Krutz:2015:EEU:2695664.2695929,6671332},
CD-Form	\cite{6227879,Cuomo2014390},
CDSW	\cite{6613837},
Clonewise	\cite{Cesare2013},
CMGA	\cite{Wang2014},
CodeBlast	\cite{Bhattacharjee:2013:CTA:2480362.2480525},
CPDP	\cite{6613041},
DroidClone	\cite{7544005},
DuplicationDetector	\cite{4137427},
DupLoc	\cite{Ducasse_SMR:SMR317,Ducasse_792593},
EqMiner	\cite{Jiang:2009:AMF:1572272.1572283},
Exas	\cite{Nguyen2009},
FRISC	\cite{6392103},
JPlag	\cite{Prechelt00findingplagiarisms},
JSCD	\cite{Cheung2016},
$MQ_{lone}$	\cite{Storrle2013},
Simon	\cite{Zilberstein:2016:LCN:2986012.2986013},
\textit{Al-Omari et al.}	\cite{6385136},
\textit{Bansal and Tekchandani}	\cite{6897221},
\textit{Corazza et al.}	\cite{5609715},
\textit{DL-Clone}	\cite{Schugerl:2011:SCD:1985404.1985413},
\textit{Ekanayake et al.}	\cite{Ekanayake2012},
\textit{Iwamoto et al.}	\cite{6363131},
\textit{Iwamoto et al.}	\cite{6690950},
Kam1n0	\cite{Ding:2016:KMA:2939672.2939719},
\textit{Koschke}	\cite{6178897},
\textit{Koschke}	\cite{SMR:SMR1592},
\textit{Lavoie and Merlo}	\cite{6227861},
\textit{Li and Sun}	\cite{5478099},
\textit{Lucia et al.}	\cite{4380246},
\textit{Sheneamer and Kalita}	\cite{7397263},
\textit{Sheneamer and Kalita}	\cite{7838289},
\textit{Stojanovic\' et al}	\cite{Stojanovic2015259},
\textit{Zibran and Roy}	\cite{Zibran:2012:IRF:2245276.2231970,Zibran:2011:TFC:1985404.1985423}
				\\ \rule{0pt}{6ex}
				f & Unknown    & 8 &
				\scriptsize
Clone Detective (ConQat)	\cite{conqat,conqat_models},
Duplix	\cite{957835},
ScalClone	\cite{Farhadi201546},
SHINOBI	\cite{5328752,shinobiTR},
\textit{Abdelkader and Mimoun}	\cite{7483299},
\textit{Joshi et al.}	\cite{Joshi2015},
\textit{Keivanloo et al.}	\cite{7081830},
\textit{Patil et al.}	\cite{7087126}
				\\
\bottomrule
			\end{tabular}
		\end{table}
		
		In the remainder of this section, we describe and justify the ranks in detail.
		
		\begin{adjustwidth}{1cm}{}
			\textbf{Ranks A-D} The most reliable and standard procedure for measuring precision is to manually validate the output of the clone detector, so we make this method the top tiers of our ranking.  As clone validation is very time consuming, only a random sample is validated.  The quality of the results depends on the number of clones validated, so we split the manual validation tier into multiple levels based on the number of clones validated.  For \textbf{Rank A}, we expect at least 300 clones to be manually validated.  Regardless of the population size, a random sample of 300-400 can give a 90-95\% confidence level with a 5\% margin of level.  So we consider works that validate at least 300 clones to have measured precision with significance.  We then create tiers for different common break-points below this: \textbf{Rank B} requires at least 100 clones to be validated, \textbf{Rank C} 50 to 100, and \textbf{Rank D} 10-50.\\
			
			\textbf{Rank E - Automatic Validation} Precision can be measured by comparison to a reference corpora.  If the reference corpora is incomplete, this measures just a lower bound on precision.  If the reference corpora is complete, this measures true precision.  However, complete reference corpora are only possible for very small inputs, which limits the relevance of the precision measure.  Tools or metrics may be designed to automatically validate clones, but these are themselves clone detectors, and will have limits in precision as well.  For these reasons, we rank automatic validation below manual validation. \\
			
			\textbf{Rank F - Unknown} Sometimes authors report precision, but do not clearly indicate how it was measured, or how many clones were validated.  This makes it difficult to understand and trust the measurement, so we rank these works lowest.
		\end{adjustwidth}
		
		\subsubsection{Results Ranked by their Measurement of Execution Time}
		\label{sec:toolEvaluations_results_executionTime}
		
		In this section, we rank the clone detection tool and technique publications by how well the authors measured execution time.  We propose a simple four tier ranking, which is summarized in Table~\ref{tab:cluster_executiontime_ranks}.  Ideally, execution time is measured using multiple subject systems.  In the best case, execution time is also reported for different configurations of the clone detector.  At the very least execution time should be reported for a single subject system, although in some cases we find that execution time is only partially report, such as for only a specific stage of the clone detection algorithm.  The ranked works are summarized in Table~\ref{tab:cluster_executiontime_results}.  As can be seen, most of the works that measure execution time do so using multiple subject systems, with fewer works only considering a single subject system.  A good number of works go the extra step and also report execution time for different configurations of their tools.  Very few works only report execution time partially.
		
		\begin{table}
			\small
			\caption{Measurement of Execution Time Ranks} \label{tab:cluster_executiontime_ranks}
			\centering
			\begin{tabular}{cl>{\raggedright\arraybackslash}m{7cm}}
				\toprule
				\textbf{Rank} & \textbf{Name}              & \textbf{Description} \\
				\midrule
				a & Various Configurations &  Extensive evaluation of execution time, including multiple configurations of the tool.\\
				\rule{0pt}{5ex}
				b & Multiple Subject Systems & Execution time reported for multiple subject systems. \\ \rule{0pt}{5ex}
				c & Single Subject System & Execution time reported for a single subject system. \\ \rule{0pt}{6ex}
				d & Partial Results & Execution time only partially reported. \\
				\bottomrule
			\end{tabular}
		\end{table}
		
		\begin{table}
			\small
			\centering
			\caption{Clone Detection Tools/Techniques Ranked by their Execution Time Measurement}\label{tab:cluster_executiontime_results}
			\begin{tabular}{clc>{\raggedright\arraybackslash}m{6cm}}
				\toprule
				\textbf{Rank} & \textbf{Name} & \textbf{Num} & \textbf{Tools/Techniques} \\
				\midrule
				a & Various Configurations & 17  &
				\scriptsize
					BinClone	\cite{6895418},
					Boreas	\cite{6494937, Yuan:2012:SAA:2162110.2162126},
					CP-Miner	\cite{li_1610609,Li:2004:CTF:1251254.1251274},
					Duplix	\cite{957835},
					DupLoc	\cite{Ducasse_SMR:SMR317,Ducasse_792593},
					GPLAG	\cite{Liu:2006:GDS:1150402.1150522},
					iClones	\cite{iclones,iclonesthesis},
					NiCad	\cite{Cordy2014158,5970189,SMR:SMR416,5306301,4556129},
					ScalClone	\cite{Farhadi201546},
					Shuffling Framework	\cite{SMR:SMR1662,6227875,6613037},
					SimCad	\cite{6079770,6613857},
					\textit{Barbour et al.}	\cite{5521760},
					\textit{Chen et al.}	\cite{Chen2016},
					\textit{Karus and Kilgi}	\cite{7069883},
					\textit{Lavoie and Merlo}	\cite{6227861},
					\textit{Saebjornsen et al.}	\cite{Saebjornsen:2009:DCC:1572272.1572287},
					\textit{Udagawa}	\cite{Udagawa:2014:ESR:2684200.2684290}
				\\ \rule{0pt}{31ex}
				b & Multiple Subject Systems & 62 &
				\scriptsize
					Asta	\cite{4400161,Evans2009},
					BAT	\cite{Hemel:2011:FSL:1985441.1985453},
					C2D2	\cite{kraft2008cross},
					CBCD	\cite{6227183},
					CCFinder(X)	\cite{ccfinderx},
					CD-Form	\cite{6227879,Cuomo2014390},
					CDSW	\cite{6613837},
					ClemenX	\cite{5071050},
					clones/cscope	\cite{4023995},
					CloneWorks	\cite{cloneworks} ,
					CMGA	\cite{Wang2014},
					CodeBlast	\cite{Bhattacharjee:2013:CTA:2480362.2480525},
					Covet/CLAN	\cite{Mayrand_565012,bellon},
					cpdetector	\cite{723194,4023995},
					CtCompare	\cite{6227881},
					D-CCFinder	\cite{4222573},
					Decrescendo	\cite{7880510},
					Dup	\cite{baker,baker:si:92,baker:514697},
					DyCLINK	\cite{Su:2016:CRD:2950290.2950321},
					EqMiner	\cite{Jiang:2009:AMF:1572272.1572283},
					HaRe	\cite{Brown:2010:CDE:1706356.1706378},
					HeapAbsCC	\cite{6319249},
					HitoshiIO	\cite{7503720},
					JCCD	\cite{5645564},
					LLVM-Based Framework	\cite{Sargsyan2016,7358259},
					MeCC	\cite{Kim:2011:MMC:1985793.1985835},
					ModelCD	\cite{5070528},
					$MQ_{lone}$	\cite{Storrle2013},
					PDG-DUP	\cite{Komondoor2001,Komondoor2001-2},
					SaCD	\cite{6615249},
					SeByte	\cite{6240495,Keivanloo:2012:JBC:2664398.2664404,6227864,Keivanloo2014426},
					SeClone	\cite{6079771,6225474},
					sif	\cite{Manber:1994:FSF:1267074.1267076},
					Sim	\cite{Gitchell:1999:SUD:299649.299783},
					SourcererCC	\cite{7886988,7883349,6240500,6613042,SMR:SMR1707},
					SSD	\cite{Lee:2005:SHP:1094855.1094903},
					\textit{Abdelkader and Mimoun}	\cite{7483299},
					\textit{Chilowicz et al.}	\cite{CHILOWICZ200947},
					\textit{Corazza et al.}	\cite{5609715},
					\textit{Cordy et al.}	\cite{Cordy:2004:PLD:1034914.1034915},
					\textit{Dandois and Vanhoof}	\cite{Dandois2012},
					\textit{Dumas et al.}	\cite{Dumas2013619,Uba2011},
					\textit{Falke et al.}	\cite{Falke2008},
					\textit{Hummel et al.}	\cite{Hummel:2011:IMC:1985404.1985409},
					\textit{Hummel et al.}	\cite{5609665},
					\textit{Ishihara et al.}	\cite{6385134},
					\textit{Johnson}	\cite{Johnson:1993:IRS:962289.962305,Johnson:1994:VTR:782185.782217,336783},
					\textit{Kam1n0}	\cite{Ding:2016:KMA:2939672.2939719},
					\textit{Kamiya}	\cite{7069882},
					\textit{Kontogiannis et al.}	\cite{Kontogiannis1996},
					\textit{Koschke}	\cite{6178897},
					\textit{Koschke}	\cite{SMR:SMR1592},
					\textit{Lavoie et al.}	\cite{Lavoie:2010:CCR:1808901.1808905},
					\textit{Li et al.}	\cite{6821712},
					\textit{Perumal et al.}	\cite{5640465},
					\textit{Qu et al.}	\cite{Qu2014544},
					\textit{St{\"o}rrle}	\cite{Storrle2015},
					\textit{Str{\"u}ber}	\cite{Struber2016},
					\textit{Tekin and Buzluca}	\cite{Tekin2014406},
					\textit{Wahler et al.}	\cite{1386166},
					\textit{White et al.}	\cite{7582748},
					\textit{Zhang et al.}	\cite{6982615}
				\\ \rule{0pt}{11ex}
				c & Single Subject System & 21 &
				\scriptsize
					Agec	\cite{6613854},
					DuDe	\cite{1595830},
					FCFinder	\cite{5463293},
					Hanni	\cite{Lillack:2014:DCC:2660190.2662116},
					JPlag	\cite{Prechelt00findingplagiarisms},
					MiLoCo	\cite{6676875},
					RTF	\cite{Basit:2007:ETB:1295014.1295029},
					SHINOBI	\cite{5328752,shinobiTR},
					Simian	\cite{simian},
					Simon	\cite{Zilberstein:2016:LCN:2986012.2986013},
					Wrangler	\cite{Li2010},
					XIAO	\cite{Dang:2012:XTC:2420950.2421004,Dang:2011:CCD:1985404.1985417},
					\textit{Ahkin and Itsykson}	\cite{Akhin2013},
					\textit{Ali et al.}	\cite{6140712},
					\textit{Di Lucca et al.}	\cite{Lucca01cloneanalysis,1134103},
					\textit{Grant and Cordy}	\cite{5090048},
					\textit{Higo et al.}	\cite{6079769},
					\textit{Huang and Li}	\cite{5376091},
					\textit{Lazar and Banias}	\cite{6840038},
					\textit{Pulkkinen et al.}	\cite{Pulkkinen:2015:RBP:2812428.2812471},
					\textit{Udagawa}	\cite{Udagawa:2016:MFS:3011141.3011160}
				\\ \rule{0pt}{4ex}
				d & Partial Results & 2 &
				\scriptsize
					Clone Detective (ConQat)	\cite{conqat,conqat_models},
					LSC Miner	\cite{6377848}
				\\
				\bottomrule
			\end{tabular}
		\end{table}
		
		\subsubsection{Results Ranked by their Measurement of Scalability}
		\label{sec:toolEvaluations_results_scalability}
		We now rank the clone detection tools and techniques by how well their authors measured scalability.  We propose a five tier ranking, which is summarized in Table~\ref{tab:cluster_scalability_ranks}.  The best evaluations are those that demonstrate the scalability of their tool against a series of increasingly difficult inputs, and we rank highest those that use a systematically chosen/created set of inputs.  We notice that often the papers will report execution time for inputs of various sizes, but not all of them present the results clearly, so we rank higher those authors who clearly report and discuss their results.  We rank lower those who demonstrate scalability using only a large system, and those who show scalability only qualitatively.  The works that measured scalability are ranked in Table~\ref{tab:cluster_scalability_results}.  As can be seen, a number of works measure scalability using a variety of inputs, but few choose the set of input in a systematic way.
		
		\begin{table}
			\small
			\caption{Measurement of Scalability Ranks} \label{tab:cluster_scalability_ranks}
			\centering
			\begin{tabular}{cl>{\raggedright\arraybackslash}m{7cm}}
				\toprule
				\textbf{Rank} & \textbf{Name}              & \textbf{Description} \\
				\midrule
				a & Systematic Inputs &  A systematic and controlled experiment using inputs of uniformly increasing difficulty, with clear presentation of the results.\\
				\rule{0pt}{5ex}
				b & Various Inputs with Clear Presentation & Executed for inputs of various sizes, with clear presentation of the results. \\ \rule{0pt}{5ex}
				c & Various Inputs with Poor Presentation & Executed for inputs of various sizes, but the results are not clearly presented. \\ \rule{0pt}{6ex}
				d & Large System & Executed for a couple large systems to demonstrate scalability, but little variety in input sizes tested. \\ \rule{0pt}{5ex}
				e & Qualitative & Execution time not measured, but scalability described qualitatively. \\
				\bottomrule
			\end{tabular}
		\end{table}
		
		\begin{table}
			\small
			\centering
			\caption{Clone Detection Tools/Techniques Ranked by their Scalability Measurement}\label{tab:cluster_scalability_results}
			\begin{tabular}{clc>{\raggedright\arraybackslash}m{6cm}}
				\toprule
				\textbf{Rank} & \textbf{Name} & \textbf{Num} & \textbf{Tools/Techniques} \\
				\midrule
				a & Systematic Inputs & 3 &
				\scriptsize
				CloneWorks	\cite{cloneworks},
				CtCompare	\cite{6227881},
				SourcererCC	\cite{7886988,7883349,6240500,6613042,SMR:SMR1707}
				\\ \rule{0pt}{10ex}
				b & Various Inputs with Clear Presentation          & 22 &
				\scriptsize
				CD-Form	\cite{6227879,Cuomo2014390},
				CP-Miner	\cite{li_1610609,Li:2004:CTF:1251254.1251274},
				Duplix	\cite{957835},
				DupLoc	\cite{Ducasse_SMR:SMR317,Ducasse_792593},
				HeapAbsCC	\cite{6319249},
				LLVM-Based Framework	\cite{Sargsyan2016,7358259},
				ModelCD	\cite{5070528},
				NiCad	\cite{Cordy2014158,5970189,SMR:SMR416,5306301,4556129},
				SaCD	\cite{6615249},
				ScalClone	\cite{Farhadi201546},
				SeClone	\cite{6079771,6225474},
				SimCad	\cite{6079770,6613857},
				SSD	\cite{Lee:2005:SHP:1094855.1094903},
				\textit{Barbour et al.}	\cite{5521760},
				\textit{Chilowicz et al.}	\cite{CHILOWICZ200947},
				\textit{Falke et al.}	\cite{Falke2008},
				\textit{HaRe}	\cite{Brown:2010:CDE:1706356.1706378},
				\textit{Hummel et al.}	\cite{5609665},
				\textit{Koschke}	\cite{6178897},
				\textit{Koschke}	\cite{SMR:SMR1592},
				\textit{Qu et al.}	\cite{Qu2014544},
				\textit{Tekin and Buzluca}	\cite{Tekin2014406}				
				\\ \rule{0pt}{4ex}
				c & Various Inputs with Poor Presentation      & 5 &
				\scriptsize
				Asta	\cite{4400161,Evans2009},
				BinClone	\cite{6895418},
				CCFinder(X)	\cite{ccfinderx},
				sif	\cite{Manber:1994:FSF:1267074.1267076},
				\textit{Karus and Kilgi}	\cite{7069883}
				\\ \rule{0pt}{8ex}
				d & Large Systems    & 17 &
				\scriptsize
				Boreas	\cite{6494937, Yuan:2012:SAA:2162110.2162126}
				CBCD	\cite{6227183},
				Clonewise	\cite{Cesare2013},
				CMCD	\cite{6130694},
				CMGA	\cite{Wang2014},
				DuDe	\cite{1595830},
				DyCLINK	\cite{Su:2016:CRD:2950290.2950321},
				FCFinder	\cite{5463293},
				HitoshiIO	\cite{7503720},
				MeCC	\cite{Kim:2011:MMC:1985793.1985835},
				SeByte	\cite{6240495,Keivanloo:2012:JBC:2664398.2664404,6227864,Keivanloo2014426},
				Shuffling Framework	\cite{SMR:SMR1662,6227875,6613037},
				XIAO	\cite{Dang:2012:XTC:2420950.2421004,Dang:2011:CCD:1985404.1985417},
				\textit{Corazza et al.}	\cite{5609715},
				\textit{Ishihara et al.}	\cite{6385134},
				\textit{Kam1n0}	\cite{Ding:2016:KMA:2939672.2939719},
				\textit{Saebjornsen et al.}	\cite{Saebjornsen:2009:DCC:1572272.1572287}
				\\ \rule{0pt}{3ex}
				e & Qualitative & 2 &
				\scriptsize
				Decrescendo	\cite{7880510},
				\textit{Zibran and Roy}	\cite{Zibran:2012:IRF:2245276.2231970,Zibran:2011:TFC:1985404.1985423}
				\\
				\bottomrule
			\end{tabular}
		\end{table}
		
		We discuss the ranks in further detail below:
		
		\begin{adjustwidth}{1cm}{}	
			\textbf{Rank A - Systematic Inputs:} The authors have conducted a systematic and controlled experiment across inputs of increasing difficulty, namely size in lines of code.  The authors have taken care to select or create inputs that increase in size but otherwise have similar properties, such as clone density, which could also affect scalability.\\
			
			\noindent \textbf{Rank B - Various Inputs with Clear Presentation:} The authors have executed their tool for various inputs of increasing difficulty, namely size in lines of code.  However, the inputs are not chosen systematically, so they may not exactly create a series of uniformly increasing execution difficulty in terms of execution time and system resource requirements (e.g., memory).  The results are clearly presented, such as in a table or graph, and discussed. \\
			
			\noindent \textbf{Rank C - Various Inputs with Poor Presentation:} The authors have executed their tool for inputs of various sizes and reported their execution performance, but the results are not clearly presented in the paper.  For example, the inputs may be various case studies and the execution performance results are scattered throughout the paper.  Meaning the reader has to piece together the scalability performance. \\
			
			\noindent \textbf{Rank D - Large System:} The authors demonstrate scalability by executing the tool for one or more large software systems, where large depends on the scalability targets of the tool (but at least tens of thousands of lines of code).  The authors have shown the tool can be executed for large inputs, but does not demonstrate how execution performance scales with increasing input size.  Readers therefore can't easily predict how the tool will perform for even larger inputs, or smaller inputs. \\
			
			\noindent \textbf{Rank E - Qualitative:} The authors discuss scalability in a meaningful way, but only qualitatively.  For example they may just discuss algorithmic complexity, but do not demonstrate the scalability for real inputs on an example hardware.\\
		\end{adjustwidth}
	
		\subsubsection{Exemplars in Tool Evaluation}
		\label{sec:toolEvaluations_results_exemplars}
		We have ranked the clone detection tools and techniques on how the authors have measured the four primarily performance metrics.  These rankings can be used by the community to find examples on the best way to evaluate the performance of their clone detection tools and techniques.  We extract from these rankings the works which rank within the top one or two tiers for each of the four metrics.  We propose these works as exemplar works for how authors should evaluate their clone detection tools and techniques.  These publications should be used by authors of new tools/techniques as inspiration for their evaluation experiments, and by peer reviewers to judge new publications in clone detection research.  As we have shown, evaluation by the authors is lacking in the literature.  It is very important that future works are well evaluated.
	
		We propose the following exemplars for how authors should be expected to evaluate their clone detection tools/techniques:
		\begin{itemize}
			\item CloneWorks~\cite{cloneworks}
			\item \textit{Falke et al.}~\cite{Falke2008}
			\item NiCad~\cite{Cordy2014158,5970189,SMR:SMR416,5306301,4556129}
			\item SimCad~\cite{6079770,6613857}
			\item SourcererCC~\cite{sourcerercc}
		\end{itemize}
	
		\subsubsection{How Tool Authors Measure Recall}
		\label{sec:toolEvaluations_results_howrecall}
		Only 67 of the 198 tools and techniques have their authors measuring their recall.  We are interested in how the authors measured recall, specifically how they obtained or created reference clone corpora.  We find that the authors either use one of the standard clone benchmarks (Section~\ref{sec:benchmarks}), or create a new reference data for their evaluation using one of the common methodologies (Section~\ref{sec:buildingReferenceCorpus_methods}).
	
		In Table~\ref{tab:tooleval_recall}, we summarize how the tool authors have evaluated recall.  We split the table into two halves, in the first half we list the standard benchmarks that authors have used to measure the recall of their clone detection tool/technique.  In the second half, we list the methodologies the authors have used when they created their own reference corpora to measure recall.  The frequencies add up to greater than 67, as a number of authors have used multiple benchmarks.
	
		\begin{table}
		\small
		\centering
		\caption{Methods and Benchmarks used by Tool Authors to Measure Recall} \label{tab:tooleval_recall}
		\begin{tabular}{>{\arraybackslash}p{4cm}>{\centering\arraybackslash}p{2cm}>{\centering\arraybackslash}p{2cm}>{\arraybackslash}p{6cm}}
			\toprule
			\textbf{Method} & \textbf{Frequency} & \textbf{Ratio} & \textbf{Method Description} \\
			\midrule
			\multicolumn{4}{c}{\textbf{Using Standard Benchmarks}} \vspace{1em}\\
			
			Bellon's Benchmark                & 14 & 21\% & Bellon's Benchmark's reference corpus (or extension). \vspace{0.5em}\\
			
			Krutz's Benchmark & 1 & 1\% & Krutz's manually validated clone corpus.\vspace{0.5em}\\
			
			Mutation Framework                & 5  & 6\% & Mutation and Injection Framework (or procedure).\vspace{0.5em}\\
			
			BigCloneBench & 4 & 5\% & BigCloneBench's cross-project big data clone corpora. \vspace{0.5em} \\

			\midrule
			\multicolumn{4}{c}{\textbf{Using Self-Built Reference Corpora}} \vspace{1em} \\
			Manual Inspection                 & 17 & 22\% & Manual inspection of a system for clones.\vspace{0.5em}\\ 
			
			Clone Injection                   & 12 & 15\% & Software system manually seeded with known clones, or software system with known clones manually constructed.\vspace{0.5em}\\
			
			Clone Detector(s)                 & 7  & 9\% & Output of one or more clone detectors used as an oracle. \vspace{0.5em}\\
			
			Literature                        & 7  & 9\% & Tested against clones and clone scenarios from academic literature.\vspace{0.5em}\\
			
			Artificial Clones                 & 4  & 5\% & Reference clones or software system with known clones automatically or semi-automatically synthesized. \vspace{0.5em}\\
			
			Clone Detector(s) with Validation & 4  & 5\% & Output of one or more clone detectors used as an oracle after manual removing of false positives. \vspace{0.5em}\\

			Unknown                           & 2  & 3\% & Recall reported without sufficient description of how.\\
			\bottomrule
		\end{tabular}
		\end{table}
	
			\paragraph{Standard Benchmarks} Tool and technique authors have measured recall using various standard benchmarks, including: Bellon's Benchmark, Krutz's Benchmark, the Mutation Framework, and BigCloneBench.  The most popular benchmark has been Bellon's Benchmark, which is also the benchmark that has been available for the longest.
	
			While the easiest way to measure recall is to use an existing reference corpus, most the authors that measured recall crated new reference data.  With the exception of Bellon's Benchmark, most of these benchmarks have only been recently published and released.  We expect that adoption of standard benchmarks by clone detection tool and technique authors will improve with time.
	
			\paragraph{Self-Built} Most of the works that measure recall have done so by the authors creating their own reference corpora.  This has been because standard benchmarks have been historically unavailable or difficult to use.  Sometimes clone detection tools are not compatible with existing benchmarks, especially in the case of emerging detection techniques that require data not captured by the benchmarks (e.g., compiled code, commit logs), or where the detection tool targets particular use-cases that are not represented by the existing benchmarks (e.g., student plagiarism detection).  It is interesting to see how tool/technique authors have coped with this by building their own reference data.
	
			The most popular technique has been \textbf{manual inspection}, where a subject system is manually searched for the clones it contains.  This technique is only possible for very small systems, where a manual inspection is feasible.  Even small subject systems can contain far too many pairs of code fragments to inspect manually.  Unless a systematic manual search process is used, it may not be reliable to find all clones.  This technique produces a reference corpus that lacks variety, as a single or small group of very-small systems is unlikely to contain many kinds of clones.  Therefore, measuring recall in this way can lack generality.
	
			The second most popular approach has been \textbf{clone injection}, where an existing or new subject system is manually seeded with known clones.  This technique gives the author more control over the kinds of clones in the subject system, potentially allowing a more general measure of recall.  However, the reference corpus is still typically small as manually adding clones to a software system is time consuming~\cite{bellon}.  Another concern is the source of the clones to be injected, whether they have been extracted from a real software system, or hand crafted.
	
			Another approach used by authors is to use an existing \textbf{clone detector(s)} to build a reference corpus.  This is a poor approach, as there is no guarantee the existing tool(s) have good recall and precision themselves.  The reference corpus may be incomplete, or may contain false positives.  Typically, this approach has been used when the authors want to show that their tool is as-good as the well-accepted and popular existing tools.  The authors then show that their tool is superior in some other capacity, such as execution time or usability.
	
			A good approach has been to evaluate the clone detector against clones documented in the academic \textbf{literature}.  Most commonly, this has been done using the cloning scenarios proposed by Roy et al.~\cite{4556127,Roy:2009:CEC:1530898.1531101}.  This approach is the basis of the Mutation and Injection Framework, which automates synthesis of many clones using these scenarios.  The advantage of this technique is it can show that the tool handles the various cloning scenarios.  However, doing this by hand means very few examples have been tested, so it has not been tested against the various ways the source syntax may express these scenarios.  This technique is comprehensive in the kinds of types of clones that can exist, but is not comprehensive in the variety of ways they can be expressed.  Authors that have used this technique have typically measured recall for only tens of clones.
	
			Another technique has been using \textbf{artificial clones}, where an existing or new software system is automatically seeded with synthetically produced clones.  Sometimes this is done based on a taxonomy of the types of edits made on copy and pasted or plagiarized code.  This technique is also similar to the Mutation and Injection Framework.  The limitation of this technique is the synthetic clones may not reflect real clones in practice, so it is based to pair this approach with a benchmark of real clones.
	
			Another technique has been to use \textbf{clone detector(s) with validation}, where one or more clone detectors are executed for a subject system(s), and their results are taken as the reference corpus after removing the false positives.  Since clone detection results can be large, sometimes only a random sample of the detection results are considered.  This is most similar to Bellon's Benchmark, which used the same technique.  The problem with this method is it does not contain the kinds of clones the tools are unable to detect, and suffers from the subjectivity of manual clone validation.  It is also a very time consuming approach.
	
			For two of the works, we were unable to identify the methodology used to build the reference corpus. Actually, in most cases the authors have provided only minimal details on how they built their reference corpora.  Also, in most cases their reference clones are not released for public examination.  This makes it impossible to reproduce or validate the results.  It also means that future studies cannot directly compare to their results, as their reference corpora is not available.  This is why standard benchmarks are strongly needed by the clone detection community.

	\subsection{Standards of Evaluation by Authors}
	\label{sec:toolEvaluations_standards}
	Previously in this section, we have ranked how well the authors have evaluated their clone detection tools and techniques.  We are interested in what is the standards of evaluation by the authors.  To understand this, we measure various statistics about the frequency in which the evaluation criteria are satisfied by the authors.  We begin in Section~\ref{sec:toolEvaluations_standards_frequency} by measuring the frequency in which the individual evaluation criteria are satisfied.  Then in Section~\ref{sec:toolEvaluations_standards_freqscopes} we evaluate how frequently the authors have measured different combinations of the performance metrics.  Lastly, in Section~\ref{sec:toolEvaluations_correlation}, we evaluate the correlations between the evaluation criteria, including how often the different performance metrics are measured together.  The results show the state of tool evaluation by the authors.  We show statistics for all works, and individually for the named tools, and unnamed cone detection techniques, as discussed earlier in Section~\ref{sec:toolEvaluations_evaluationCriteria}.
		
		\subsubsection{Frequency of Metric Evaluations by Authors}
		\label{sec:toolEvaluations_standards_frequency}
	
		We measured how frequently each of the evaluation criteria are examined by the tool authors. In Table~\ref{tab:tooleval_criteria}, we report the percentage of the works that evaluate each of the evaluation criteria.  We report our findings across all of the works, and individually for the named clone detection tools and unnamed clone detection techniques.  For each of the evaluation criteria, we found that authors of named tools were more likely to measure that criteria than authors of unnamed techniques.
		
		\begin{table}
			\centering
			\caption{Frequency of Evaluations of Clone Detection Tools by Authors} \label{tab:tooleval_criteria}
			\begin{tabular}{>{\centering\arraybackslash}m{5cm}>{\centering\arraybackslash}m{3cm}>{\centering\arraybackslash}m{3cm}>{\centering\arraybackslash}m{3cm}}
				\toprule
				Evaluation Type & All Tools and techniques & Named Tools & Unnamed Techniques \\
				\midrule
				Case Study      & 93\% & 98\% & 89\% \\
				Tool Comparison & 38\% & 49\% & 29\% \\
				\midrule
				Recall          & 34\% & 47\% & 23\% \\
				Precision       & 44\% & 54\% & 36\% \\
				\midrule
				Execution Time  & 52\% & 68\% & 38\% \\
				Scalability     & 25\% & 39\% & 13\% \\
				\bottomrule
			\end{tabular}
		\end{table}
		
		\paragraph{Case Study} We find that an overwhelming majority of the works, 93\%, are evaluated at least using a case study.  Named clone detection tools are somewhat more likely to have a case study than unnamed techniques, 98\% versus 89\%, but in both cases it is very common. Case studies range from a demonstration of the tool/technique for a single software system, reporting at least the size of the detection result, to large-scale studies examining the kinds of clones detected and exploring its configurations and use-cases.  In some cases, the tool/technique is demonstrated through its use in a real industrial use-case.
		
		\paragraph{Tool Comparison} Comparison to other tools is much more rare, just 38\%. While authors typically discuss other tools and techniques in their related work, rarely are they directly and quantitatively compared against the author's work.  This may be due to lack of access to the competing tools, as few are publicly released.  It may also be due to the difficulty in conducting tool comparison studies, due to the differences between the tools, and historical lack of universal clone benchmarks.  This suggests that works are often accepted for publication based on the novelty of their algorithms, and not because their tool/technique is demonstrably and quantitatively better or distinct from the related work.
		
		Tool comparison is more common for named tools, 49\%, than for unnamed techniques, 29\%.  It may be that there is a higher expectation for tools to be compared against the popular/competing tools, while techniques are accepted on the merit of a novel algorithm or detection metrics/strategy.  Unnamed clone detection techniques may not have a working prototype, which can make comparison to the existing tools more challenging.
		
		\paragraph{Recall and Precision} Precision and recall are important measures of clone detection performance.  However, only 44\% of works measure precision, and only 34\% measure recall.  It is understandable that recall evaluation is low, as it requires a reference corpus.  Historically, there have been few clone benchmarks, and reference corpora are very challenging and time-consuming to build.  However, precision is very easy to measure, only requiring the author to manually validate a sample of their detected clones.  The efforts required to measure precision are small compared to the effort required to research, develop, and publish a new clone detection tool or technique, so it is reasonable to expect authors to at least report precision.
		
		It is much more common for precision and recall to be measured for named tools, 54\% and 47\%, compared to unnamed techniques, 36\% and 23\%. This may be because named tool works have more mature tools/prototypes, or there is a higher expectation for evaluation for clone detection tool papers.  It may be that clone detection techniques are published on the merit of their new algorithms alone, and reviewers do not expect a thorough evaluation of their detection performance.  Works published as clone detection techniques may not have a mature enough prototype (or a prototype at all) for measurement of recall and precision.  However, these are still important measures, and it is hard to see the value in a new detection technique if its detection performance is not evaluated.
		
		\paragraph{Execution Time and Scalability} Execution time is the most commonly measured performance metric, 52\%.  It is more commonly reported for named tools, 68\%, than unnamed techniques, 38\%.  This could be because some unnamed techniques lack a prototype, so it is not possible to measure execution time.  Or the current prototype was implemented without concern for speed, so the authors are hesitant to report it.  Still, all papers featuring at least a prototype should report execution time as it is simple to measure.
		
		It is much more rare for scalability to be measured, just 25\% of the works.  Although it is more commonly evaluated for named tools, 39\%, than unnamed techniques, 13\%. Some have systematically evaluated scalability by reporting execution time (and sometimes memory usage) for various subject system sizes.  Others have also explored scalability for different tool configuration settings, such as similarity thresholds.  Scalability is not challenging to evaluate, requiring the author to pull subject systems of various sizes from open-source repositories, and report the execution time of their tool/prototype.
		
		\paragraph{Summary Comments} While most of the clone detection tools and techniques are published with at least a case study, the other evaluation criteria are more often than not neglected.  Most often execution time is measured, but this is still for only 52\% of the published tools and techniques.
	
		\subsubsection{Frequency of Evaluation Scopes}
		\label{sec:toolEvaluations_standards_freqscopes}
		
		In the previous section, we examined the frequency of different types of evaluations by authors of the clone detection tools.  Now we look at what are the standard ways that authors evaluate their tools.  In other words, which set of metrics do tool/technique authors most frequently evaluate.  In Table~\ref{tab:tooleval_standards2} we show the frequency in which tool and technique authors evaluate certain combinations of performance metrics.  We focus on the four primary metrics of recall, precision, execution-time and scalability.  We show the statistics for all the tools/techniques, as well as individually for named tools and unnamed tools (techniques).
		
		\begin{table}
			\caption{Standards of Clone Detection Tool Evaluation by Authors}\label{tab:tooleval_standards2}
			\begin{tabular}{>{\centering\arraybackslash}m{1.25cm}>{\centering\arraybackslash}m{2.5cm}>{\centering\arraybackslash}m{2cm}>{\centering\arraybackslash}m{1.25cm}>{\centering\arraybackslash}m{1cm}>{\centering\arraybackslash}m{1.25cm}>{\centering\arraybackslash}m{1.25cm}>{\centering\arraybackslash}m{2cm}}
			\toprule
				Recall & Precision & Execution Time & Scalability & & All  & Named Tools & Unnamed Techniques \\
	                 x &         x &              x &           x & &  9\% &        14\% &                5\% \\
			\midrule
	                 x &         x &              x &             & &  7\% &        11\% &                3\% \\
					 x &         x &                &           x & &  1\% &         1\% &                1\% \\
					 x &           &              x &           x & &  3\% &         6\% &                0\% \\
					   &         x &              x &           x & &  6\% &         9\% &                4\% \\
			\midrule
	                 x &         x &                &             & & 12\% &        10\% &               13\% \\
					 x &           &              x &             & &  1\% &         0\% &                1\% \\
					 x &           &                &           x & &  1\% &         1\% &                0\% \\
					   &         x &              x &             & &  4\% &         6\% &                3\% \\
					   &         x &                &           x & &  0\% &         0\% &                0\% \\
					   &           &              x &           x & &  6\% &         8\% &                4\% \\
			\midrule
				     x &           &                &             & &  2\% &         3\% &                1\% \\
					   &         x &                &             & &  6\% &         3\% &                8\% \\
	                   &           &              x &             & & 17\% &        14\% &               19\% \\
	                   &           &                &           x &	&  0\% &         0\% &                0\% \\
	        \midrule
				       &           &                &             & & 27\% &        13\% &               39\% \\
	        \bottomrule
			\end{tabular}
		\end{table}
		
		Most common, in 27\% of the works, the authors have measured none of these performance metrics.  In 25\% of the works the authors measured just one of the metrics, in 24\% of the works the authors measured two of the metrics, in 17\% they measured three metrics, and in 9\% of the works they measured all four.
		
		When the authors only measured one metric, they most commonly measured execution time and then precision.  When they measured two metrics, it is most common they measured recall and precision together. When three metrics are measured, it is most common that recall, precision and execution time are measured.  Overall, it is fairly rare that all four metrics are evaluated.  It is rare that scalability is measured without execution time.  When it is measured without execution time, this is where the authors evaluate scalability by examining complexity and memory usage alone.
		
		Considering all combinations, the top-4 most common cases are: (1) no metrics evaluated, (2) only execution-time is measured, (3) recall and precision are measured, and (4) all four metrics are measured.  Comparing named tools and unnamed tools, it is much more common that authors of unnamed tools do not measure any of the metrics.
		
		Given that precision, execution time and scalability are simple to measure, it is surprisingly that this case is not more common.  When the authors have at least prototyped their tools, execution time and scalability can be measured using open-source systems, for which precision can be estimated by manually validating a sample of the clones detected in these studies.  Only recall is significantly challenging to measure, as it requires a reference corpus (clone benchmark).
		
		These results show that tool and technique authors need to become more proactive in evaluating their tools/techniques.  The availability of clone benchmarks that are accessible and easy to use should hopefully encourage authors to thoroughly evaluate their tools/techniques.
	
		\subsubsection{Correlation Between Evaluations}
		\label{sec:toolEvaluations_correlation}
		
		Also of interested is the correlation between the the evaluation criteria.  For example, given that an author has measured the precision of their clone detection tool/technique, what is the chance they also measured recall, and vice-versa.  This way we can examine how these dimensions of tool evaluation are correlated in the works.
		
		We show these correlations in Table~\ref{tab:tooleval_correlation}.  For each evaluation criteria, we present the number of tools/techniques that were evaluated for that criteria, and then the ratio of those which were also evaluated for the other evaluation criteria.  There are some interesting observations we can make from this data.
		
		\begin{table}
			\small
			\centering
			\caption{Correlation Between Evaluation Criteria} \label{tab:tooleval_correlation}
			\begin{tabular}{>{\centering\arraybackslash}m{3cm}>{\centering\arraybackslash}m{1.25cm}>{\centering\arraybackslash}m{1cm}>{\centering\arraybackslash}m{2cm}>{\centering\arraybackslash}m{1cm}>{\centering\arraybackslash}m{1.5cm}>{\centering\arraybackslash}m{1.5cm}>{\centering\arraybackslash}m{2cm}}
				\toprule
				& & \multicolumn{6}{c}{Probability that this Criteria is also Measured} \\
				Measured        & Number & {Case Study} & {Tool Comparison} & {Recall} & {Precision} & {Execution Time} & {Scalability} \\
				\midrule
				Case Study      & 184 & x     & 41\% & 36\% & 48\% & 55\% & 27\% \\ 
				Tool Comp.      &  75 & 100\% & x    & 56\% & 64\% & 56\% & 36\% \\ 
				Recall          &  67 & 100\% & 63\% & x    & 84\% & 55\% & 39\% \\ 
				Precision       &  88 & 100\% & 55\% & 64\% & x    & 58\% & 36\% \\
				Execution Time  & 102 & 100\% & 41\% & 36\% & 50\% & x    & 45\% \\
				Scalability     &  49 & 100\% & 55\% & 53\% & 65\% & 94\% & x    \\
				\bottomrule
			\end{tabular}
		\end{table}
	
		For all of the criteria, if measured, then there is a 100\% chance a case study was performed.  This is because we count any of the other evaluation criteria as a case study.  In this way, the chance of the other criteria given a case study was done can be considered a baseline chance of the other evaluation criteria being measured, given that we ignore the few papers that have no evaluation of the tool or technique.  In no case do we see the measurement of one criteria reducing the chance another was measured to below the baseline probability.
		
		When recall is measured, there it is high chance (84\%) that precision is also measured.  The reverse is not as strongly true, 64\%, although recall is most likely measured when precision is also measured.  Recall is more challenging to measure, and if the author has taken the time to evaluate recall they likely will make the time to measure precision too.  It is difficult to interpret a measurement of recall without a measurement of precision, and vice-versa.
		
		Given that a work measures the scalability of their tool/technique, it is very likely that execution time is also reported.  This is because scalability is typically measured by presenting execution time for different input sizes, although scalability has been presented by simply indicating successful execution without specifying time.  What is interesting, is that the likelihood execution time is measured does not vary significantly for the other criteria.
		
		A tool comparison is most likely when recall is measured.  Given that the author has produced a corpus or has access to a clone benchmark, they are more likely to take the time to evaluate multiple tools with that reference corpora.

	\subsection{Threats to Validity of this Survey}
	\label{sec:toolEvaluations_threats}

	There are a few threats to the validity of this survey.  It is likely that we have not found all of the tools in the literature.  We relied on the previous surveys being comprehensive on the available works up to 2011.  We then thoroughly searched the major databases for publications between 2011 and march 2017.  However, with some of the databases we had to constraint our search query in order to receive a reasonable number of results to manually examine.  While we may have missed some papers, we are confidant we have a sufficient sample to measure accurate trends on clone detection tool evaluation by the authors.  We validate this assumption by the fact that our search yielded all the major clone detection tools we were already aware of.

	
	
\section{A Survey of Tool Comparison Studies}
\label{sec:surveyToolComparisons}

In this section, we survey the existing clone detection tool comparison studies and experiments.  For this survey we only consider those papers that perform an unbiased tool comparison experiment.  We do not consider publications that introduce a new clone detection tool or technique, and include a tool comparison study as part of its evaluation.  While tool authors often perform high-quality experiments, these studies have been designed to highlight the performance and novelty of their new tool, so there is a potential bias in the experiment design.

This survey is organized as follows.  In Section~\ref{sec:surveyToolComparisons_procedure} we discuss our survey procedure, and in Section~\ref{sec:surveyToolComparisons_results} we summarize the results.  Then in Section~\ref{sec:surveyToolComparisons_qualitative} we summarize and critique the qualitative tool comparison studies, and the same in Section~\ref{sec:surveyToolComparisons_quantitative} for the quantitative studies.  We discuss the threats to this survey in Section~\ref{sec:surveyToolComparisons_threats}, with some concluding remarks in Section~\ref{sec:surveyToolComparisons_conclusions}.

	\subsection{Survey Procedure}
	\label{sec:surveyToolComparisons_procedure}
	
	We identified the existing tool comparison studies by examining the existing clone detection surveys~\cite{Roy07asurvey,Roy:2009:CEC:1530898.1531101,rattan}.  We also searched the relevant academic databases for publications.  We did this as part of our survey of the evaluation of clone detection tools by their authors, as described in Section~\ref{sec:toolEvaluations_surveyprocedure}.  The search terms were broad enough to capture both clone detection tool publications and tool comparison studies.  We searched the related work sections and citations of the benchmarks and tool comparison papers to find any other works.  We also searched the academic databases for papers that cite the important clone detection tool comparison publications.  For example, Bellon et al.'s~\cite{bellon} study was a major milestone in clone detection tool comparison, and most comparison studies at least cite this work.
	
	\subsection{Results}
	\label{sec:surveyToolComparisons_results}
	
	In total we found twelve tool comparison works spread across fifteen publications.  A few of the works included multiple publications, both a conference publication and an extended journal publication.  We split the works into two categories: (1) predominately qualitative tool comparison studies and (2) predominately quantitative tool comparison studies.
	
	The qualitative studies are summarized in Table~\ref{tab:toolstudies_qualatative}.  The papers are listed in chronological order, and where a work has multiple publications we list the publication with the most details.  For each work, we list the number of tools evaluated, the number of subject systems considered, the sizes of the subject systems, and the programming languages considered.  We then provide a short and long description of the qualitative analysis performed.
	
	\begin{sidewaystable}[ht]
		\begin{center}
			\scriptsize
			\begin{longtable}{
					>{	\arraybackslash}m{5cm}																										>{\centering\arraybackslash}m{1cm}	>{\centering\arraybackslash}m{0.5cm}	>{\centering\arraybackslash}m{0.5cm}	>{\centering\arraybackslash}m{0.5cm}		>{\centering\arraybackslash}m{1.5cm}	>{\centering\arraybackslash}m{3cm}		>{\arraybackslash}m{8cm}}
				\caption{Survey of Qualitative Tool Comparison Studies} \label{tab:toolstudies_qualatative} \\
				\toprule
				\textbf{Paper} 																																&	\textbf{Year}					&	\rot{90}{1em}{\textbf{\#Tools}}		&	\rot{90}{1em}{\textbf{\#Systems}}	&	\rot{90}{1em}{\textbf{System Size}}		&	{\textbf{Language}}					&	\textbf{Qualatative Summary}  		& \textbf{Comments}  \\ \midrule
				\endfirsthead 
				\caption{Survey of Qualitative Tool Comparison Studies (cont.)} \\ 
				\toprule 
				\textbf{Paper} 																																&	\textbf{Year}						&	\rot{90}{1em}{\textbf{\#Tools}}		&	\rot{90}{1em}{\textbf{\#Systems}}	&	\rot{90}{1em}{\textbf{System Size}}		&	{\textbf{Language}}					&	\textbf{Qualatative Summary}  		& \textbf{Comments} \\ \midrule
				\endhead 
				\endfoot 
				Evaluating clone detection techniques from a refactoring perspective~\cite{1342759}															&	2004							&	3									&	5									&	537 to 98K								&	Java,C++							&	Refactoring-Based					&	Manually examined clones to judge refactorability of the detected clones based on four criteria: suitable, relevant, confidence, focus. \\ \midrule
				Comparison and Evaluation of Code Clone Detection Techniques and Tools: A Qualitative Approach~\cite{Roy:2009:CEC:1530898.1531101,4556127}	&	2009							&	38									&										&											&	Independent							&	Property-Based Scenario-Based		&	Extensive evaluation of clone detector features and properties.  Evaluation of detection performance for cloning scenarios based on described features. \\ \midrule
				An extended assessment of type-3 clones as detected by state-of-the-art tools~\cite{Tiarks2011,5279980}										&   2011							&	5									&	14									&	19K-235K								&	Java, CEC							&	Validation and Categorization		& 	Manually validated Type-3 clones detected by the tools.  Investigated their sytnactic differences and derived their semantic abstractions.  Investigated characteristics that suggest true-positive from human perspective.\\ \midrule
				Searching for Configurations in Clone Evaluation: A Replication Study~\cite{Ragkhitwetsagul2016}											&	2016							&	4									&	14									&	56K-250K								&	Java								&	Infering Relative Performance		&	Estimating changes to recall and precision based on clone detection size.		Replicates EvaClone study with four tools and different system (Mockito).  Shows that tuning tool configurations by agreement may result in more false positives or false negatives. \\ \midrule	\end{longtable}
		\end{center}
	\end{sidewaystable}
	
	The quantitative studies are summarized in Table~\ref{tab:toolstudies_quantitative}.  The papers are listed in chronological order, and where a work has multiple publications we list its comprehensive (journal) publication.  For each work we list the number of tools compared, the number of subject systems used, the sizes of the subject systems, the programming languages considered, and the clone detection performance metrics measured.  Each of the publications measure recall, so we list the reference corpora used (benchmark name or methodology) and its size.
	
	\begin{sidewaystable}[ht]
		\begin{center}
			\scriptsize
			\begin{longtable}{
					>{\arraybackslash}m{5cm}																							>{\centering\arraybackslash}m{1cm}	>{\centering\arraybackslash}m{0.5cm}	>{\centering\arraybackslash}m{0.5cm}	>{\centering\arraybackslash}m{0.5cm}	>{\centering\arraybackslash}m{1.5cm}	m{0.75cm}								m{0.75cm}								m{0.75cm}									m{0.5cm}								>{\centering\arraybackslash}m{3cm}								>{\centering\arraybackslash}m{2cm}}
				\caption{Survey of Quantitative Tool Comparison Studies} \label{tab:toolstudies_quantitative} \\
				\toprule
				\textbf{Paper} 																												&	\textbf{Year}					&	\rot{90}{1em}{\textbf{\#Tools}}		&	\rot{90}{1em}{\textbf{\#Systems}}	&	\rot{90}{1em}{\textbf{System Size}}	&	{\textbf{Language}}					&	\rot{90}{1em}{\textbf{Recall}}		&	\rot{90}{1em}{\textbf{Precision}}	&	\rot{90}{1em}{\textbf{Execution Time}}	&	\rot{90}{1em}{\textbf{Scalability}}	&	\textbf{Reference Corpus} 									&	\textbf{Corpus Size} 	\\ \midrule
				\endfirsthead 
				\caption{Survey of QuantitativeTool Comparison Studies (cont.)} \\ 
				\toprule 
				\textbf{Paper} 																												&	\textbf{Year}					&	\rot{90}{1em}{\textbf{\#Tools}}		&	\rot{90}{1em}{\textbf{\#Systems}}	&	\rot{90}{1em}{\textbf{System Size}}	&	{\textbf{Language}}					&	\rot{90}{1em}{\textbf{Recall}}		&	\rot{90}{1em}{\textbf{Precision}}	&	\rot{90}{1em}{\textbf{Execution Time}}	&	\rot{90}{1em}{\textbf{Scalability}}	&	\textbf{Reference Corpus} 									&	\textbf{Corpus Size} 	\\ \midrule
				\endhead 
				\endfoot 
				Evaluating clone detection tools for use during preventative maintenance~\cite{1134103}										&	2002							&	5									&	1									&	16k									&	Java,C++							&	x									&	x									&											&										&	Clone Detectors with Validation								&	1463 clone pairs 					 \\ \midrule
				On the use of clone detection for identifying crosscutting concern code~\cite{1542064}										&	2005							&	3									&	1									&	16K									&	C									&	x									&	x									&											&										&	Manual Inspection											&	16KLOC of concerns					 \\ \midrule
				Comparison and Evaluation of Clone Detection Tools~\cite{bellon}															&	2006							&	6									&	8									&	11K-235K							&	Java, C								&	x									&	x									&	x										&	x									&	Bellon's Benchmark											&	4319 clone pairs					 \\ \midrule
				Searching for Better Configurations: A Rigorous Approach to Clone Evaluation~\cite{Wang:2013:SBC:2491411.2491420}			&	2013							&	6									&	8									&	11K-235K							&	Java, C								&	x									&	x									&											&										&	Bellon's Benchmark											&	4319 clone pairs					 \\ \midrule
				On the robustness of clone detection to code obfuscation~\cite{6613045}														&	2013							&	3									&	1									&	2400								&	Java								&	x									&										&											&										&	Artificial Clones											&	336 clone pairs						 \\ \midrule
				Evaluating Modern Clone Detection Tools~\cite{moderntools}																	&	2014							&	11									&	9									&	11K-235K							&	Java, C								&	x									&										&											&										&	Bellon's Benchmark (and derivatives) and Mutation Framework	&	4319 and  75,000 clone pairs		 \\ \midrule
				Evaluating clone detection tools with BigCloneBench~\cite{7332459}															&	2015							&	10									&	2									&	250M								&	Java								&	x									&										&											&										&	BigCloneBench and Mutation Framework						&	8 million and 37,500 clone pairs	 \\ \midrule
				Similarity of Source Code in the Presence of Pervasive Modifications~\cite{7816530,7781805}									&	2016							&	30									&	1									&	50 classes							&	Java								&	x									&	x									&											&										&	Artificial Clones											&	2500 clone pairs					 \\ \midrule
			\end{longtable}
		\end{center}	
	\end{sidewaystable}

	\subsection{Qualitative Studies}
	\label{sec:surveyToolComparisons_qualitative}
	
	In this section we summarize and discuss each of the qualitative tool comparison studies.\newline
	
	\noindent \textbf{Evaluating Clone Detection Techniques from a Refactoring perspective}\newline
	\textit{Filip Van Rysselberghe and Serge Demeyer}\newline
		\indent Rysselberghe and Demeyer~\cite{1342759} implemented three clone detection techniques from the lierature and executed them for five subject systems to evaluate their detection results from a refactoring perspective.  For clone detection, the implemented techniques based on string matching, token-based with parameterized matches by suffix trees, and one based on metric fingerprints.  They used four Java software systems (537LOC to 100KLOC) and one C++ system (12KLOC).  They evaluated the techniques for four criteria relating to refactoring: (1) suitable (for a refactoring tool), (2) relevance (prioritizes larger clone classes), (3) confidence (precision), and (4) focus (within a class versus project-wide).
	
		For each category, the authors discuss which of the three investigated clone detection techniques performed best.  However, their analysis is very short, and lacks any examples.  Some of their criticisms of the clone detection techniques could be solved by pre-processing or post-processing.  For example, they indicate that metric fingerprinting is the most suitable for refactoring as it detects method-level clones, while the token-based and string-based are not because they ignore the boundaries of logical code units.  The authors implemented the techniques, so they could have unified their compatibility by implementing the string-based and token-based techniques over methods as well.
	
	However, examining clone detectors for refactoring is an important study, and the authors highlighted various ways in which clone detectors could be improved for this use-case.\newline
	
	\noindent \textbf{Comparison and Evaluation of Code Clone Detection Techniques and Tools: A Qualitative Approach}\newline
	\textit{Chanchal K. Roy, James R. Cordy and Rainer Koschke} \newline
		\indent Roy et al.~\cite{Roy:2009:CEC:1530898.1531101,4556127} performed a large-scale property-based and scenario-based evaluation of 38 clone detection tools.  The authors recommend that clone detection users identify the properties and cloning scenarios relevant to their task, and then use the author's evaluations to select the set of tools best for their task.
	
		In their property-based evaluation, they categorized the tools for twenty-two facets across ten categories, including: usage (platform, dependencies, availability), interaction (UI, output format, IDE support), language (paradigm and support), clone (relation, granularity, types), technical (comparison algorithm, comparison granularity, complexity), adjustment (configurable pre/post-processing, configurable heuristics/thresholds), transformation/normalization support, code representation (text, metrics, vector, trees, PDG, etc), program analysis (parsing type/complexity), and evaluation (empirical, availability of results, subject systems).  This evaluation exhaustively annotates the properties of their tools, and makes it easy for the reader to select a tool with their desired features.
	
		For their scenario-based evaluation, the authors created an empirically validated taxonomy of the sixteen types of edits developers makes on copy and pasted code, for clones of the four primary clone types.  Using their knowledge of the tools' features from the property-based evaluation, they judged how well each of the tools' support these sixteen cloning scenarios.  They used a seven point scale that captures the certainty or uncertainty of a tools' support based on the underlying algorithm, features, empirical evaluations, etc.
	
		The authors have provided a very strong qualitative evaluation of the existing tools.  The advantage of this methodology is it does not require the clone detection tool to be publicly available, which is often the case with clone detection publications.  The problem with this methodology is it is time-consuming to classify the tools, and there is the threat of miss-classification.  This would be a good methodology to be adopted as a standard in the community, with tool authors classifying their own tools, perhaps with feedback from the community.  The community could then update the evaluation facets and categories as new clone detection techniques are created. \newline
	
	\noindent \textbf{An extended assessment of type-3 clones as detected by state-of-the-art tools} \newline
	\textit{Rebecca Tiarks, Rainer Koschke and Raimar Falke} \newline
		\indent Tiarks et al.~\cite{Tiarks2011,5279980} evaluated the Type-3 clones reported by six state-of-the-art tools.  They executed the tools for the six subject systems from Bellon's Benchmark, as well as six industry systems.  In total, 381,628 Type-3 clones were detected, for which 751 were carefully validated over 40 hours of manual validation.  They found only 189 of these clones were true positives from human perspective.  They studied these clones to answer three research questions.  They determined what kind of syntactical differences occur between Type-3 clones, and classify the clones by these kinds of differences.  They then derive from these syntactical differences the semantic abstractions that exist in the Type-3 clones.  In total they identify fourteen categories of abstractions.  They then study fifteen kinds of similarity metrics to determine if they can be used to differentiate true and false Type-3 clones from the human perspective.  They specifically look to see how well these metrics separate the true and false positives into separable distributions (e.g., using a threshold).  They also investigate the use of metric combinations and decision trees (machine learning).
	
		The important findings of this evaluation study is that most (approximately 75\%) of the detected Type-3 clones were rejected by human inspection.  Their identification of common syntactical differences and semantic abstractions are helpful for developing new Type-3 clone detectors and designing automated or semi-automated Type-3 clone refactoring patterns and tools.  Their study of metrics for distinguishing true and false positive Type-3 clones will help with designing better clone detectors.
	
		However, there is a few flaws in this study.  Four of the five tools were produced by the author's research lab, which could bias the results.  The validation is not reported per tool, so it is unknown if the significant 75\% false positives during validation is due to poor precision of all the tools, or poor precision of just a few tools. \newline
	
	\noindent \textbf{Searching for Configurations in Clone Evaluation – A Replication Study} \newline
	\textit{Chaiyong Ragkhitwetsagul, Matheus Paixao, Manal Adham, Saheed Busari, Jens Krinke and John H. Drake} \newline
		\indent Ragkhitwetsagul et al.~\cite{Ragkhitwetsagul2016} replicate a study by Wang et al.~\cite{Wang:2013} where clone detector configurations are optimized for a subject system(s) by searching for the configurations that maximize agreement between the tools.  The search is performed using a genetic algorithm and framework called EvaClone.  They found optimal configurations of four clone detection tools for 14 individual revisions of the Mockito software system.  They compared the detection results for the optimized configurations against the tools' default configurations in terms of detection result agreement, and studied the stability of the configurations across the revisions.
	
		The authors found that the EvaClone does indeed find configurations that produce higher agreement amongst the tools compared to the default configurations.  Also, the authors found that no configuration was optimal across all revisions of the same software systems, with instability in the configurations between subsequent revisions.  However, the authors notice that EvaClone maximizes agreement between the tools by significantly increasing or decreasing the number of cloned lines detected by the tools compared to their default settings, which produces more false positives or false negatives, respectively.
	
		This is a very good study that shows the dangers of searching for tool configurations by maximizing the agreement between clone detection tools.  However, the study would have been stronger if precision and recall had been measured for the tools optimized and default configurations to confirm the suspicion that EvaClone is either increasing the number of false positives or false negatives detected by the tools to force agreement on the cloned lines within Mockito.  At the very least, precision could have easily been measured to strengthen their findings.
		
		While this publication relies upon measurable facts, we do not include it as a quantitative evaluation as it does not measure any of the common clone detection performance metrics (recall, precision, execution time and scalability).  The results in this paper are extrapolated from the clone detection sizes, so the conclusions are also qualitative.  \newline	
	
	\subsection{Quantitative Studies}
	\label{sec:surveyToolComparisons_quantitative}
	
	In this section we summarize and discuss the quantitative studies, which are listed in Table~\ref{tab:toolstudies_quantitative}.  As can be seen, recall is the most commonly measured performance metric, which is the opposite of our findings with the tool evaluations performed by the authors themselves (Section~\ref{sec:toolEvaluations}).  The sizes of the reference corpora used range from just hundreds of clones to million of clone pairs.  Precision is commonly measured, but execution performance is often omitted.
	
	\noindent \textbf{Evaluating clone detection tools for use during preventative maintenance} \newline
	\textit{Elzabeth Burd and John Bailey}\newline
		\indent Burd and Bailey~\cite{1134103} measured the recall and precision of five clone detection tools (including CCFinder, CloneDr, Covet, JPlag and Moss) from the perspective of preventative maintenance.  The tools were executed for a small subject system, GraphTool, locally developed by a post-graduate student, and therefore familiar to the authors.  In total 1,463 unique clones were detected and manually validated by the authors.  The clones were validated as true positives if they would be useful for preventative maintenance.  Using this oracle, recall and precision were measured for the tools.  Additionally, the authors investigated the properties of the detected clones, and the overlap and differences between the tools' outputs.  The authors recommended the best tools in terms of various concerns of preventative maintenance, including: (1) high recall, (2) high precision, (3) replication frequency (clone class size), (4) ease of refactorability, (5) tool usability.
	
		As one of the earliest tool comparison studies, this work is very commendable.  However, there are a few limitations.  Only a single subject system and programming language are considered, so the results may not be generalizable.  The results also depend upon the validation of the clones, and this process is not sufficiently described.  The reference corpus was built using the participating clone detection tools themselves, which means it may be missing clones that none of the participating tools are able to detect, possibly over-estimating recall.  However, the authors did investigate all of the detected clones, so their measurement of precision is precise.\newline
	
	\noindent \textbf{On the use of clone detection for identifying crosscutting concern code} \newline
	\textit{Magiel Bruntink, Arie van Deursen, Remco van Engelen and Tom Tourw\'e}\newline
		\indent Bruntink et al.~\cite{1542064} evaluated the performance of clone detection tools specifically for identifying cross-cutting concerns. They manually identified and annotated five cross-cutting in a 16KLOC industrial C system.  The identified cross-cutting concerns included memory error handling, null-value checking, range checking, error handling, and tracing/logging.  In total, 4182 source lines across 45 source files were identified as concerns, including the intersection of 386 functions.
	
		They executed three clone detectors (including ccdmiml, CCFinder and PDG-DUP) for the subject system and evaluated how well they detected the concern code.  They measure recall and precision per concern given a selection of the clone classes in the detection reports.  Recall is measured as the ratio of the concern's code lines detected as clones, and precision as the ratio of the detected clones lines that are in the concern's code.  They propose a greedy algorithm for selecting clone classes from a clone detector's output that aims to maximize the average precision across the concerns.  This selection algorithm is used to measure recall/precision specifically for the concerns, ignoring the other kinds of clones detected.  They then plot recall vs. precision for each clone detector and concern, and discuss the results.  They also investigate the performance of the tools when the detection results are combined.  Overall, they find that clone detectors show promise for detecting cross-cutting concerns.
	
		This is an excellent study that evaluates clone detectors for a new use-case.  The authors take reasonable steps to filter out non-concern related clones so that the clone detectors can be evaluated for the concern code in isolation.  This study motivates research into automatic or semi-automatic methods of categorizing detected clones in concern and non-concern related clones.  The study has a few threats to the validity of its results.  First, only a single and rather small subject system is considered.  It could be that precision would be lower for a larger system where there is more chance of false positives including concern code to be detected.  Additionally, it relied upon manual inspection and validation, which can be error-prone.  \newline
	
	\noindent \textbf{Comparison and Evaluation of Clone Detection Tools} \newline
	\textit{Stefan Bellon, Rainer Koschke, Giuliano Antoniol, Jens Krinke and Ettore Merlo}\newline
		\indent Bellon et al.~\cite{bellon} measured the relative recall and precision of six clone detection tools for four C and four Java software systems.  The reference corpus was built by validating 2\% of the detected clones, randomly and evenly distributed across the clone detectors and subject systems.  The subject tools were configured and executed by their authors.  Validation was performed by Bellon, who was kept unaware of which tool proposed a given clone.  Recall and precision (lower bound) were measured using this reference corpus and two clone matching algorithms.  The experiment was conducted in two phases, with an initial run to identify and resolve the problems with the procedure, and the main run to measure the reported results.  The authors thoroughly explore the detection performance and results of the tools, including the properties of the clones.  Performance, including execution time and memory requirements, are also reported for the clone detectors.  Although the tools were not executed on the same machine, so the results are not perfectly comparable.
		
		This tool comparison experiment was the first large-scale experiment to measure the recall and precision of clone detectors for multiple subject systems and multiple programming languages.  It is a very commendable experiment, and the reference data has been used by many tool authors to evaluate their works.  The limitation of the experiment is it can only measure relative performance, since the clone detectors were used to built the reference corpus.  Missing is clones none of the tools are able to detect, and the corpus is likely biased towards the kinds of clones the tools are best at detecting.  Reliability in the validation of the clones by Bellon has been shown to be suspect by multiple researchers~\cite{baker,Charpentier:2015}.\newline
	
	\noindent \textbf{Searching for Better Configurations: A Rigorous Approach to Clone Evaluation} \newline
	\textit{Tiantian Wang, Mark Harman, Yue Jia and Jens Krinke}\newline
		\indent Wang et al.\cite{Wang:2013:SBC:2491411.2491420} measured the recall and (lower bound) precision of six clone detection tools using Bellon's Benchmark and optimally chosen configurations.  They created a tool, EvaClone, that uses a genetic algorithm to search for configurations for a set of tools that maximizes a fitness function.  For Bellon's Benchmark, they designed a fitness function that maximizes agreement on the cloned lines with a subject system(s).  They searched for configurations for the C and Java systems as sets, and for each individual subject system.  They measure recall and precision for each tool using their default configuration, their optimal configuration for the given subject systems, and their optimal configuration for the set of subject systems.  They find that their search algorithm improves recall in comparison to the default configuration, but hurts precision.
	
		The authors have addressed what they call the ``confounding configuration problem'', that the configuration of the tools greatly affects the outcome of a clone detection experiment.  The authors present an efficient algorithm for exploring the search space of tool configurations given a fitness function.  The strength of this work is they rigorously considered tool configurations when evaluating the tools.  However, their fitness function appears to maximize recall at a significant cost to precision.  They do not suggest an alternate fitness function for balancing both the recall and precision concern.  They also do not investigate their algorithm with multiple sets of tools to see if the presence of a single tool can dominate the re-configuration for agreement.
		
		A missed opportunity is they do not attempt to create targeted configurations of the tools considering the benchmark properties and tool documentation, and then see if their search algorithm can produce better configurations.  The default tool configurations are typically very conservative, and intended for the user to get their toes wet before exploring the configuration choices.  It would be interesting to compare the configurations found by EvaClone against configurations decided by an experienced user.\newline
	
	\noindent \textbf{On the Robustness of Clone Detection to Code Obfuscation} \newline
	\textit{Sandro Schulze and Daniel Meyer} \newline
		\indent Schulze and Meyer~\cite{6613045} measured the robustness of three clone detection tools (including, JPlag, CloneDigger and Scorpio) to obfuscated code.  The authors took a single small C software system and created 15 alternate versions of its source code with fifteen different combinations of five obfuscation techniques, including: renaming, code expansion, code contraction, loop transformation and conditional transformation.  The authors then executed the clone detectors for each pair of original software system and obfuscated software system, and measure how the obfuscations reduce the recall of the clone detectors given the knowledge that the obfuscated systems are semantically identical to the original system.  They then evaluate the different clone detection techniques with respect to these obfuscations.
	
		This is an interesting study for evaluating clone detection tools for detecting plagiarized or stolen code in the presence of source-level obfuscation.  The obfuscations also map well to general Type-2, Type-3 and Type-4 clone differences, so this is also a good evaluation of the tools against common clone edits.  The weakness in the study is the technique is applied to only a single and small subject system.  The results could be improved by repeating this study for multiple subject systems of various source languages and paradigms.\newline
	
	\noindent \textbf{Evaluating Modern Clone Detection Tools}
	\textit{Jeffrey Svajlenko and Chanchal K. Roy}
		\indent Svajlenko and Roy~\cite{moderntools} measured and compared the recall of eleven clone detection tools using three variants of  Bellon's Benchmark and the Mutation and Injection Framework.  They measured recall for both Java and C clones.  To evaluate the accuracy of the benchmarks themselves, the authors compared the benchmarks against each other and against their expectations and knowledge of the tools.  They find significant disagreement between the benchmarks, but find that the Mutation and Injection Framework has better agreement with their expectations of performance.  Additionally, they found anomalies between the results from Bellon's Benchmark and knowledge of the tools' characteristics. The authors suggest that the reference corpus used by Bellon's Benchmark is not relevant for modern clone detection tools.  They demonstrate this by showing that CCFinderX's recall measured by Bellon's Benchmark is significantly lower than its earlier version CCFinder.  The authors suggested that a new real-world benchmark was needed to replace Bellon's Benchmark as a complimenet to synthetic benchmarking by the Mutation and Injection Framework.  The authors response to their finding was to introduce BigCloneBench~\cite{bigclonebench,bcbeval} in later publications.\newline
	
	\noindent \textbf{Evaluating Clone Detection Tools with BigCloneBench}
	\textit{Jeffrey Svajlenko and Chanchal K. Roy} \newline
		\indent Svajlenko and Roy~\cite{7332459} measure the recall of ten clone detection tools using their real-world benchmark, BigCloneBench, and their synthetic benchmark, the Mutation and Injection Framework.  The authors demonstrate the need for both synthetic and real-world benchmarking in tool evaluation.  With the Mutation and Injection Framework, they measure the tools' precise capabilities for each clone type in an controlled and unbiased experiment.  With BigCloneBench, they measure how the tools perform for complex clones from thousands of real software systems.  They measure recall per clone type, and for particular regions of syntactical similarity.  They also measure recall specifically for intra-project versus inter-project clones.  The authors compare real-world vs synthetically measured recall to show how they tell something different about the clone detection tools.  Additionally, they use multiple clone matching metrics to show how well the clone detectors capture the reference clones.
		
		This study provides an extensive look at the recall of the popular and publicly available clone detection tools, using both synthetic and real-world benchmarking methodologies, and measuring recall for clones with different properties.  In their clone detection tool paper introducing SourcererCC (Hitesh et al.~\cite{sourcerercc}), the authors extended this experiment with SourcererCC and a few of the existing tools to also measure precision, execution time and scalability.  However, we do not include this paper in this survey's table, as we are focusing on the papers that are just tool comparison experiments. \newline
	
	\noindent \textbf{Similarity of Source Code in the Presence of Pervasive Modifications} \newline
	\textit{Chaiyong Ragkhitwetsagul, Jens Krinke and David Clark} \newline
		\indent Ragkhitwetsagul et al.~\cite{7816530,7781805} evaluated the recall and precision of 30 tools and metric for clone detection in the presence of pervasive modifications, specifically source-level and bytecode-level obfuscation.  Specifically, they sought to answer three questions: (1) how well do current detection techniques perform for clones with pervasive modifications, (2) what are the optimal configurations of these techniques, and (3) does compilation of the source-code followed by de-compilation improve the detection of pervasively modified clones.  The authors considered tools and metrics from different domains, including five clone detectors, six plagiarism detectors, twelve compression algorithms, and seven other metrics (mostly string similarity/differentiating techniques).
	
		The authors built a benchmark with known ground truth by obfuscating a dataset of five Java source files.  Obfuscation was performed using both source-level and bytecode-level obfuscation tools.  From each source file, nine cloned versions with pervasive changes are produced by a combination of source-level and byte-level obfuscation, and two de-compilation tools (for the byte-code obfuscated files). The final dataset contains 50 source files with 500 true and 2000 false file clones.  As a post-processing step, all the source files in the dataset are pretty-printed. They executed the tools and metrics for these file pairs, and measures recall, precision and f-measure.  The tools were executed for many configurations and similarity thresholds, and the configuration with optimal f-measure reported.
	
		For their base experiment, they execute the tools for the dataset, and choose the optimal configurations that optimize f-measure.  For their second experiment, they normalize the source code by compiling it and then de-compiling it using two major decompilers (Krakatau and Procyon).  They find that the compilation/de-compilation is a good normalization for improving clone detection accuracy in the presence of pervasive modifications.
	
		The message of this paper is that compilation followed by de-compilation is an effective normalization for detecting clones where the copied code has been obfuscated, perhaps to hide license violations, plagiarism or intellectual theft.  They also show which are the best tools/metrics for detecting copied code that has been hidden by source-level and/or byte-level obfuscation.
	
		The strength of this paper is the authors evaluated many detection strategies, and optimized the tool configurations to balance their recall and precision.  However, there are some weaknesses in the reference corpus.  It is very small, built from just five original source files.  Since the reference data is built automatically, the authors could have built a much larger corpus and improved confidence in their results.  Other studies using artificial clones generate on the order of tens of thousands of clones~\cite{moderntools,bigclonebench_evaluation}.
	
		The corpus is also not well described.  The authors only provide one small example, and do not report how obfuscation effected the similarity of the reference clones.  The authors could improve their study by generating multiple datasets with different kinds and levels/amounts of source-level and byte-level obfuscation, to see how this changes detection performance.  While the authors show that compilation/de-compilation as a normalization improves detection in many cases, they do not re-optimize the configurations of the tools with respect to this added normalization.  The tool configurations were only optimized for the version of the dataset without this post-processing.\newline
	
	\subsection{Threats to Validity}
	\label{sec:surveyToolComparisons_threats}
	
	As with all surveys, the threat is that we have not found all of the publications.  However, we were quite extensive in our survey process, and since there is so few tool comparison papers, it is typical that they cite each other.  So we are confident we have found all or most of the tool comparison papers.
	
	\subsection{Conclusions}
	\label{sec:surveyToolComparisons_conclusions}

	We have found that there are very few tool comparison studies, despite the publication of at least 197 clone detection tools and techniques.  This is a significant threat in the clone detection research, as many techniques have been proposed, but very few have been objectively evaluated.  A challenge in tool evaluation studies is the availability of the tools.  While at least 197 exist in the literature, very few of these have been publicly released and maintained.  While many tool authors have evaluated the performance of their own tool (Section~\ref{sec:toolEvaluations}), and some have even compared their tool against other tools and techniques, these evaluations are designed to highlight the unique advantages/performances of the featured tool.  Tool comparison studies are needed to objectively compare the performance of tools, with an experiment designed irrespective of the individual tools.  This is not a criticism of studies implemented by tool authors, readers expect their experiments to highlight what is distinct about their tool compared to the related work.

\section{Conclusion}
\label{sec:conclusion}
In this work, we thoroughly explored and evaluated the literature on clone detection tool and technique evaluation.  We began by surveying the clone detection benchmarks that are in the literature.  We thoroughly evaluated and compared these benchmarks across multiple facets, and recommended the best benchmarks to be used at the time of writing. 

We then surveyed the published clone detection tool and techniques to see how authors evaluate their own tools.  We ranked these works by how well they measured recall, precision, execution time and scalability.  We identified the works that best evaluated all four metrics to act as exemplars for how authors should evaluate their clone detection tools/techniques.  We then report statistics on how authors evaluate their own tools/techniques.  We find that authors often neglect to evaluate their tools/techniques, which is a threat in the literature.  This lack of evaluation is due to the historic lack of clone benchmarks and standard evaluation strategies.  We hope that our study will motivate future authors to evaluate their tools/techniques, and use our exemplars and evaluations as a guide on how to run good evaluation studies.

Lastly, we surveyed the tool comparison studies, to see how researchers (not necessarily tool authors) evaluate and compare clone detection tools.  We summarize these works, and compare their attributes.  We identify both qualitative and quantitative comparisons of the tools.  Overall, there is very few studies comparing the clone detection tools despite so many published works on clone detection.

Our findings is that there is a significant lack of clone detection tool evaluation, despite significant work in introducing new detection tools and techniques.  We find that this has been due to a historic lack of good clone benchmarks.  There is a strong need by the clone detection research community for strong benchmarks and for good exemplars on how to conduct tool evaluation and comparison experiments.

\FloatBarrier 


\appendix

\begin{landscape}
	\section{Oversized Tables} \label{app:oversize}
	\footnotesize
	\begin{center}
\begin{longtable}{>{\arraybackslash}m{4.5cm}>{\centering\arraybackslash}m{3cm}m{0.5cm}m{0.5cm}m{0.5cm}m{0.5cm}m{0.5cm}m{0.5cm}>{\centering\raggedright\arraybackslash}m{6cm}}
			\caption{Clone Detection Tool and Techniques and their Evaluations by their Authors} \label{tab:toolevals} \\
			\toprule
			\textbf{Named Tool or \textit{Authors}}                                                                                                                                                                                                 &  \textbf{Citations}                                                                      &  \rot{90}{1em}{\textbf{Case Study}}    &  \rot{90}{1em}{\textbf{Comparison}}   &  \rot{90}{1em}{\textbf{Recall}}    &  \rot{90}{1em}{\textbf{Precision}}    &  \rot{90}{1em}{\textbf{Execution}}    &  \rot{90}{1em}{\textbf{Scalability}}    &  \textbf{Benchmark/Refernce Corpus}                       \\\midrule
			\endfirsthead 
			\toprule 
			\textbf{Tool}                                                                                                                                                                                                 &  \textbf{Citations}                                                                      &  \rot{90}{1em}{\textbf{Case Study}}    &  \rot{90}{1em}{\textbf{Comparison}}   &  \rot{90}{1em}{\textbf{Recall}}    &  \rot{90}{1em}{\textbf{Precision}}    &  \rot{90}{1em}{\textbf{Execution}}    &  \rot{90}{1em}{\textbf{Scalability}}    &  \textbf{Benchmark}                       \\\midrule
			\endhead 
			\endfoot
			Agec                                                                                                                                                                                                            & \cite{6613854}                                                                             & x                                        &                                         &                                      &                                         & x                                       &                                           &                                                                \\ 
			Asta                                                                                                                                                                                                            &  \cite{4400161,Evans2009}                                                                  & x                                        & x                                       & x                                    &                                         & x                                       & x                                         & Bellon's Benchmark                                    \\ 
			AST-CC                                                                                                                                                                                                          & \cite{6631651}                                                                             & x                                        & x                                       & x                                    &                                         &                                         &                                           & Clones from Literature                                         \\
\bottomrule
		\end{longtable}
	\end{center}
\end{landscape}

\bibliographystyle{IEEEtran}
\bibliography{bib}

\begin{thebibliography}{100}
\providecommand{\url}[1]{#1}
\csname url@samestyle\endcsname
\providecommand{\newblock}{\relax}
\providecommand{\bibinfo}[2]{#2}
\providecommand{\BIBentrySTDinterwordspacing}{\spaceskip=0pt\relax}
\providecommand{\BIBentryALTinterwordstretchfactor}{4}
\providecommand{\BIBentryALTinterwordspacing}{\spaceskip=\fontdimen2\font plus
\BIBentryALTinterwordstretchfactor\fontdimen3\font minus
  \fontdimen4\font\relax}
\providecommand{\BIBforeignlanguage}[2]{{%
\expandafter\ifx\csname l@#1\endcsname\relax
\typeout{** WARNING: IEEEtran.bst: No hyphenation pattern has been}%
\typeout{** loaded for the language `#1'. Using the pattern for}%
\typeout{** the default language instead.}%
\else
\language=\csname l@#1\endcsname
\fi
#2}}
\providecommand{\BIBdecl}{\relax}
\BIBdecl

\bibitem{Roy07asurvey}
C.~K. Roy and J.~R. Cordy, ``A survey on software clone detection research,''
  School of Computing, Queen’s University, Tech. Rep. TR 2007-541, 2007, 115
  pp.

\bibitem{Roy:2009:CEC:1530898.1531101}
\BIBentryALTinterwordspacing
C.~K. Roy, J.~R. Cordy, and R.~Koschke, ``Comparison and evaluation of code
  clone detection techniques and tools: A qualitative approach,'' \emph{Sci.
  Comput. Program.}, vol.~74, no.~7, pp. 470--495, May 2009. [Online].
  Available: \url{http://dx.doi.org/10.1016/j.scico.2009.02.007}
\BIBentrySTDinterwordspacing

\bibitem{rattan}
D.~Rattan, R.~Bhatia, and M.~Singh, ``Software clone detection: A systematic
  review,'' \emph{Information and Software Technology}, vol.~55, no.~7, pp.
  1165 -- 1199, 2013.

\bibitem{ZibranSurvey}
M.~F. Zibran and C.~K. Roy, ``The road to software clone management: A
  survey,'' Department of Computer Science, University of Saskatchewan, Tech.
  Rep. TR 2012-03, 2012, 62 pp.

\bibitem{bellon}
S.~Bellon, R.~Koschke, G.~Antoniol, J.~Krinke, and E.~Merlo, ``Comparison and
  evaluation of clone detection tools,'' \emph{IEEE Transactions on Software
  Engineering}, vol.~33, no.~9, pp. 577--591, Sept 2007.

\bibitem{nicad}
C.~Roy and J.~Cordy, ``Nicad: Accurate detection of near-miss intentional
  clones using flexible pretty-printing and code normalization,'' in
  \emph{Program Comprehension, 2008. ICPC 2008. The 16th IEEE International
  Conference on}, June 2008, pp. 172--181.

\bibitem{moderntools}
J.~Svajlenko and C.~K. Roy, ``Evaluating modern clone detection tools,'' in
  \emph{The 30th International Conference on Software Maintenance and
  Evolution}, ser. ICSME 2014, 2014, p.~10.

\bibitem{bigclonebench_evaluation}
------, ``Evaluating clone detection tools with bigclonebench,'' in
  \emph{Software Maintenance and Evolution (ICSME), 2015 IEEE International
  Conference on}, Sept 2015, pp. 131--140.

\bibitem{iclones}
N.~G{\"o}de and R.~Koschke, ``Incremental clone detection,'' in \emph{Software
  Maintenance and Reengineering, 2009. CSMR '09. 13th European Conference on},
  March 2009, pp. 219--228.

\bibitem{ccfinderx}
T.~Kamiya, S.~Kusumoto, and K.~Inoue, ``Ccfinder: a multilinguistic token-based
  code clone detection system for large scale source code,'' \emph{Software
  Engineering, IEEE Transactions on}, vol.~28, no.~7, pp. 654--670, Jul 2002.

\bibitem{1287259}
A.~Walenstein, N.~Jyoti, J.~Li, Y.~Yang, and A.~Lakhotia, ``Problems creating
  task-relevant clone detection reference data,'' in \emph{WCRE}, 2003, pp.
  285--294.

\bibitem{Charpentier:2015}
\BIBentryALTinterwordspacing
A.~Charpentier, J.-R. Falleri, D.~Lo, and L.~R{\'e}veill\`{e}re, ``An empirical
  assessment of bellon's clone benchmark,'' in \emph{Proceedings of the 19th
  International Conference on Evaluation and Assessment in Software
  Engineering}, ser. EASE '15.\hskip 1em plus 0.5em minus 0.4em\relax New York,
  NY, USA: ACM, 2015, pp. 20:1--20:10. [Online]. Available:
  \url{http://doi.acm.org/10.1145/2745802.2745821}
\BIBentrySTDinterwordspacing

\bibitem{KrutzBenchmark}
\BIBentryALTinterwordspacing
D.~E. Krutz and W.~Le, ``A code clone oracle,'' in \emph{Proceedings of the
  11th Working Conference on Mining Software Repositories}, ser. MSR
  2014.\hskip 1em plus 0.5em minus 0.4em\relax New York, NY, USA: ACM, 2014,
  pp. 388--391. [Online]. Available:
  \url{http://doi.acm.org/10.1145/2597073.2597127}
\BIBentrySTDinterwordspacing

\bibitem{bigclonebench}
\BIBentryALTinterwordspacing
J.~Svajlenko, J.~F. Islam, I.~Keivanloo, C.~K. Roy, and M.~M. Mia, ``Towards a
  big data curated benchmark of inter-project code clones,'' in
  \emph{Proceedings of the 2014 IEEE International Conference on Software
  Maintenance and Evolution}, ser. ICSME '14.\hskip 1em plus 0.5em minus
  0.4em\relax Washington, DC, USA: IEEE Computer Society, 2014, pp. 476--480.
  [Online]. Available: \url{http://dx.doi.org/10.1109/ICSME.2014.77}
\BIBentrySTDinterwordspacing

\bibitem{LavoieBenchmark}
\BIBentryALTinterwordspacing
T.~Lavoie and E.~Merlo, ``Automated type-3 clone oracle using levenshtein
  metric,'' in \emph{Proceedings of the 5th International Workshop on Software
  Clones}, ser. IWSC '11.\hskip 1em plus 0.5em minus 0.4em\relax New York, NY,
  USA: ACM, 2011, pp. 34--40. [Online]. Available:
  \url{http://doi.acm.org/10.1145/1985404.1985411}
\BIBentrySTDinterwordspacing

\bibitem{Roy:2008:TMA:1370256.1370279}
C.~K. Roy and J.~R. Cordy, ``Towards a mutation-based automatic framework for
  evaluating code clone detection tools,'' in \emph{Proceedings of the 2008
  C3S2E Conference}, ser. C3S2E '08.\hskip 1em plus 0.5em minus 0.4em\relax New
  York, NY, USA: ACM, 2008, pp. 137--140.

\bibitem{4976382}
C.~Roy and J.~Cordy, ``A mutation/injection-based automatic framework for
  evaluating code clone detection tools,'' in \emph{Software Testing,
  Verification and Validation Workshops, 2009. ICSTW '09. International
  Conference on}, April 2009, pp. 157--166.

\bibitem{Svajlenko:2013:MAB:2662708.2662710}
J.~Svajlenko, C.~K. Roy, and J.~R. Cordy, ``A mutation analysis based
  benchmarking framework for clone detectors,'' in \emph{Proceedings of the 7th
  International Workshop on Software Clones}, ser. IWSC '13.\hskip 1em plus
  0.5em minus 0.4em\relax Piscataway, NJ, USA: IEEE Press, 2013, pp. 8--9.

\bibitem{clusterprecision}
J.~Svajlenko and C.~K. Roy, ``A machine learning based approach for evaluating
  clone detection tools for a generalized and accurate precision,''
  \emph{International Journal of Software Engineering and Knowledge
  Engineering}, vol.~26, no. 09n10, pp. 1399--1429, 2016.

\bibitem{bellonbenchmark}
S.~Bellon, ``Stefan bellon's clone detector benchmark,''
  \url{http://www.softwareclones.org/research-data.php}.

\bibitem{baker}
B.~Baker, ``Finding clones with dup: Analysis of an experiment,'' \emph{IEEE
  Transactions on Software Engineering}, vol.~33, no.~9, pp. 608--621, 2007.

\bibitem{baker:si:92}
\BIBentryALTinterwordspacing
B.~S. Baker, ``{A program for identifying duplicated code},'' in \emph{Computer
  Science and Statistics: Proc. Symp. on the Interface}, March 1992, pp.
  49--57. [Online]. Available:
  \url{http://citeseer.nj.nec.com/baker92program.html}
\BIBentrySTDinterwordspacing

\bibitem{baker:514697}
------, ``On finding duplication and near-duplication in large software
  systems,'' in \emph{Proceedings of 2nd Working Conference on Reverse
  Engineering}, Jul 1995, pp. 86--95.

\bibitem{957835}
J.~Krinke, ``Identifying similar code with program dependence graphs,'' in
  \emph{Proceedings Eighth Working Conference on Reverse Engineering}, 2001,
  pp. 301--309.

\bibitem{Ducasse_SMR:SMR317}
\BIBentryALTinterwordspacing
S.~Ducasse, O.~Nierstrasz, and M.~Rieger, ``On the effectiveness of clone
  detection by string matching,'' \emph{Journal of Software Maintenance and
  Evolution: Research and Practice}, vol.~18, no.~1, pp. 37--58, 2006.
  [Online]. Available: \url{http://dx.doi.org/10.1002/smr.317}
\BIBentrySTDinterwordspacing

\bibitem{Ducasse_792593}
S.~Ducasse, M.~Rieger, and S.~Demeyer, ``A language independent approach for
  detecting duplicated code,'' in \emph{Software Maintenance, 1999. (ICSM '99)
  Proceedings. IEEE International Conference on}, 1999, pp. 109--118.

\bibitem{Mayrand_565012}
J.~Mayrand, C.~Leblanc, and E.~M. Merlo, ``Experiment on the automatic
  detection of function clones in a software system using metrics,'' in
  \emph{1996 Proceedings of International Conference on Software Maintenance},
  Nov 1996, pp. 244--253.

\bibitem{Baxter_1317484}
I.~D. Baxter, C.~Pidgeon, and M.~Mehlich, ``Dms reg;: program transformations
  for practical scalable software evolution,'' in \emph{Proceedings. 26th
  International Conference on Software Engineering}, May 2004, pp. 625--634.

\bibitem{Baxter_738528}
I.~D. Baxter, A.~Yahin, L.~Moura, M.~Sant'Anna, and L.~Bier, ``Clone detection
  using abstract syntax trees,'' in \emph{Proceedings. International Conference
  on Software Maintenance (Cat. No. 98CB36272)}, Nov 1998, pp. 368--377.

\bibitem{gap}
H.~Murakami, Y.~Higo, and S.~Kusumoto, ``A dataset of clone references with
  gaps,'' in \emph{MSR'14}, 2014, pp. 412--415.

\bibitem{Charpentier2017}
A.~Charpentier, J.-R. Falleri, F.~Morandat, E.~Ben Hadj~Yahia, and
  L.~R{\'e}veill{\`e}re, ``Raters' reliability in clone benchmarks
  construction,'' \emph{Empirical Software Engineering}, vol.~22, no.~1, pp.
  235--258, 2017.

\bibitem{6648182}
J.~Svajlenko, C.~K. Roy, and S.~Duszynski, ``Forksim: Generating software forks
  for evaluating cross-project similarity analysis tools,'' in \emph{2013 IEEE
  13th International Working Conference on Source Code Analysis and
  Manipulation (SCAM)}, Sept 2013, pp. 37--42.

\bibitem{bcbeval}
J.~Svajlenko and C.~K. Roy, ``Bigcloneeval: A clone detection tool evaluation
  framework with bigclonebench,'' in \emph{2016 IEEE International Conference
  on Software Maintenance and Evolution (ICSME)}, 2016.

\bibitem{yukibenchmark}
Y.~Yuki, Y.~Higo, K.~Hotta, and S.~Kusumoto, ``Generating clone references with
  less human subjectivity,'' in \emph{2016 IEEE 24th International Conference
  on Program Comprehension (ICPC)}, May 2016, pp. 1--4.

\bibitem{sourcerercc}
H.~Sajnani, V.~Saini, J.~Svajlenko, C.~K. Roy, and C.~V. Lopes, ``Sourcerercc:
  scaling code clone detection to big-code,'' in \emph{Proceedings of the 38th
  International Conference on Software Engineering}.\hskip 1em plus 0.5em minus
  0.4em\relax ACM, 2016, pp. 1157--1168.

\bibitem{mutationframeworkdl}
J.~Svajlenko and C.~Roy, ``The mutation and injection framework,''
  \url{http://www.jeff.svajlenko.com/mutationframework}.

\bibitem{cloneworks}
\BIBentryALTinterwordspacing
J.~Svajlenko and C.~K. Roy, ``Fast and flexible large-scale clone detection
  with cloneworks,'' in \emph{Proceedings of the 39th International Conference
  on Software Engineering Companion}, ser. ICSE-C '17.\hskip 1em plus 0.5em
  minus 0.4em\relax Piscataway, NJ, USA: IEEE Press, 2017, pp. 27--30.
  [Online]. Available: \url{https://doi.org/10.1109/ICSE-C.2017.3}
\BIBentrySTDinterwordspacing

\bibitem{BigCloneBenchdl}
J.~Svajlenko and C.~Roy, ``Bigclonebench,''
  \url{http://www.jeff.svajlenko.com/bigclonebench.html}.

\bibitem{simcad}
M.~Uddin, C.~Roy, and K.~Schneider, ``Simcad: An extensible and faster clone
  detection tool for large scale software systems,'' in \emph{Program
  Comprehension (ICPC), 2013 IEEE 21st International Conference on}, May 2013,
  pp. 236--238.

\bibitem{6895418}
M.~R. Farhadi, B.~C.~M. Fung, P.~Charland, and M.~Debbabi, ``Binclone:
  Detecting code clones in malware,'' in \emph{2014 Eighth International
  Conference on Software Security and Reliability (SERE)}, June 2014, pp.
  78--87.

\bibitem{6227183}
J.~Li and M.~D. Ernst, ``Cbcd: Cloned buggy code detector,'' in \emph{2012 34th
  International Conference on Software Engineering (ICSE)}, June 2012, pp.
  310--320.

\bibitem{6227879}
A.~Cuomo, A.~Santone, and U.~Villano, ``A novel approach based on formal
  methods for clone detection,'' in \emph{2012 6th International Workshop on
  Software Clones (IWSC)}, June 2012, pp. 8--14.

\bibitem{Cuomo2014390}
\BIBentryALTinterwordspacing
------, ``Cd-form: A clone detector based on formal methods,'' \emph{Science of
  Computer Programming}, vol. 95, Part 4, pp. 390 -- 405, 2014, special Issue
  on Software Clones (IWSC'12). [Online]. Available:
  \url{http://www.sciencedirect.com/science/article/pii/S0167642313003067}
\BIBentrySTDinterwordspacing

\bibitem{Wang2014}
\BIBentryALTinterwordspacing
T.~Wang, K.~Wang, X.~Su, and P.~Ma, ``Detection of semantically similar code,''
  \emph{Frontiers of Computer Science}, vol.~8, no.~6, pp. 996--1011, 2014.
  [Online]. Available: \url{http://dx.doi.org/10.1007/s11704-014-3430-1}
\BIBentrySTDinterwordspacing

\bibitem{1595830}
R.~Wettel and R.~Marinescu, ``Archeology of code duplication: recovering
  duplication chains from small duplication fragments,'' in \emph{Seventh
  International Symposium on Symbolic and Numeric Algorithms for Scientific
  Computing (SYNASC'05)}, Sept 2005, pp. 8 pp.--.

\bibitem{Kim:2011:MMC:1985793.1985835}
\BIBentryALTinterwordspacing
H.~Kim, Y.~Jung, S.~Kim, and K.~Yi, ``Mecc: Memory comparison-based clone
  detector,'' in \emph{Proceedings of the 33rd International Conference on
  Software Engineering}, ser. ICSE '11.\hskip 1em plus 0.5em minus 0.4em\relax
  New York, NY, USA: ACM, 2011, pp. 301--310. [Online]. Available:
  \url{http://doi.acm.org/10.1145/1985793.1985835}
\BIBentrySTDinterwordspacing

\bibitem{Cordy2014158}
\BIBentryALTinterwordspacing
J.~R. Cordy and C.~K. Roy, ``Tuning research tools for scalability and
  performance: The nicad experience,'' \emph{Science of Computer Programming},
  vol.~79, pp. 158 -- 171, 2014, experimental Software and Toolkits (EST 4): A
  special issue of the Workshop on Academic Software Development Tools and
  Techniques (WASDeTT-3 2010). [Online]. Available:
  \url{http://www.sciencedirect.com/science/article/pii/S0167642311002024}
\BIBentrySTDinterwordspacing

\bibitem{5970189}
------, ``The nicad clone detector,'' in \emph{2011 IEEE 19th International
  Conference on Program Comprehension}, June 2011, pp. 219--220.

\bibitem{SMR:SMR416}
\BIBentryALTinterwordspacing
C.~K. Roy and J.~R. Cordy, ``Near-miss function clones in open source software:
  an empirical study,'' \emph{Journal of Software Maintenance and Evolution:
  Research and Practice}, vol.~22, no.~3, pp. 165--189, 2010. [Online].
  Available: \url{http://dx.doi.org/10.1002/smr.416}
\BIBentrySTDinterwordspacing

\bibitem{5306301}
C.~K. Roy, ``Detection and analysis of near-miss software clones,'' in
  \emph{2009 IEEE International Conference on Software Maintenance}, Sept 2009,
  pp. 447--450.

\bibitem{4556129}
C.~K. Roy and J.~R. Cordy, ``Nicad: Accurate detection of near-miss intentional
  clones using flexible pretty-printing and code normalization,'' in \emph{2008
  16th IEEE International Conference on Program Comprehension}, June 2008, pp.
  172--181.

\bibitem{Farhadi201546}
\BIBentryALTinterwordspacing
M.~R. Farhadi, B.~C. Fung, Y.~B. Fung, P.~Charland, S.~Preda, and M.~Debbabi,
  ``Scalable code clone search for malware analysis,'' \emph{Digital
  Investigation}, vol.~15, pp. 46 -- 60, 2015, special Issue: Big Data and
  Intelligent Data Analysis. [Online]. Available:
  \url{http://www.sciencedirect.com/science/article/pii/S1742287615000705}
\BIBentrySTDinterwordspacing

\bibitem{6240495}
I.~Keivanloo, C.~K. Roy, and J.~Rilling, ``Sebyte: A semantic clone detection
  tool for intermediate languages,'' in \emph{2012 20th IEEE International
  Conference on Program Comprehension (ICPC)}, June 2012, pp. 247--249.

\bibitem{Keivanloo:2012:JBC:2664398.2664404}
\BIBentryALTinterwordspacing
------, ``Java bytecode clone detection via relaxation on code fingerprint and
  semantic web reasoning,'' in \emph{Proceedings of the 6th International
  Workshop on Software Clones}, ser. IWSC '12.\hskip 1em plus 0.5em minus
  0.4em\relax Piscataway, NJ, USA: IEEE Press, 2012, pp. 36--42. [Online].
  Available: \url{http://dl.acm.org/citation.cfm?id=2664398.2664404}
\BIBentrySTDinterwordspacing

\bibitem{6227864}
------, ``Java bytecode clone detection via relaxation on code fingerprint and
  semantic web reasoning,'' in \emph{2012 6th International Workshop on
  Software Clones (IWSC)}, June 2012, pp. 36--42.

\bibitem{Keivanloo2014426}
\BIBentryALTinterwordspacing
------, ``Sebyte: Scalable clone and similarity search for bytecode,''
  \emph{Science of Computer Programming}, vol. 95, Part 4, pp. 426 -- 444,
  2014, special Issue on Software Clones (IWSC'12). [Online]. Available:
  \url{http://www.sciencedirect.com/science/article/pii/S0167642313002773}
\BIBentrySTDinterwordspacing

\bibitem{6079770}
M.~S. Uddin, C.~K. Roy, K.~A. Schneider, and A.~Hindle, ``On the effectiveness
  of simhash for detecting near-miss clones in large scale software systems,''
  in \emph{2011 18th Working Conference on Reverse Engineering}, Oct 2011, pp.
  13--22.

\bibitem{6613857}
M.~S. Uddin, C.~K. Roy, and K.~A. Schneider, ``Simcad: An extensible and faster
  clone detection tool for large scale software systems,'' in \emph{2013 21st
  International Conference on Program Comprehension (ICPC)}, May 2013, pp.
  236--238.

\bibitem{7886988}
H.~Sajnani, V.~Saini, J.~Svajlenko, C.~K. Roy, and C.~V. Lopes, ``Sourcerercc:
  Scaling code clone detection to big-code,'' in \emph{2016 IEEE/ACM 38th
  International Conference on Software Engineering (ICSE)}, May 2016, pp.
  1157--1168.

\bibitem{7883349}
V.~Saini, H.~Sajnani, J.~Kim, and C.~Lopes, ``Sourcerercc and sourcerercc-i:
  Tools to detect clones in batch mode and during software development,'' in
  \emph{2016 IEEE/ACM 38th International Conference on Software Engineering
  Companion (ICSE-C)}, May 2016, pp. 597--600.

\bibitem{6240500}
H.~Sajnani, J.~Ossher, and C.~Lopes, ``Parallel code clone detection using
  mapreduce,'' in \emph{2012 20th IEEE International Conference on Program
  Comprehension (ICPC)}, June 2012, pp. 261--262.

\bibitem{6613042}
H.~Sajnani and C.~Lopes, ``A parallel and efficient approach to large scale
  clone detection,'' in \emph{2013 7th International Workshop on Software
  Clones (IWSC)}, May 2013, pp. 46--52.

\bibitem{SMR:SMR1707}
\BIBentryALTinterwordspacing
H.~Sajnani, V.~Saini, and C.~Lopes, ``A parallel and efficient approach to
  large scale clone detection,'' \emph{Journal of Software: Evolution and
  Process}, vol.~27, no.~6, pp. 402--429, 2015, jSME-13-0129.R2. [Online].
  Available: \url{http://dx.doi.org/10.1002/smr.1707}
\BIBentrySTDinterwordspacing

\bibitem{5609715}
A.~Corazza, S.~D. Martino, V.~Maggio, and G.~Scanniello, ``A tree kernel based
  approach for clone detection,'' in \emph{2010 IEEE International Conference
  on Software Maintenance}, Sept 2010, pp. 1--5.

\bibitem{Falke2008}
\BIBentryALTinterwordspacing
R.~Falke, P.~Frenzel, and R.~Koschke, ``Empirical evaluation of clone detection
  using syntax suffix trees,'' \emph{Empirical Software Engineering}, vol.~13,
  no.~6, pp. 601--643, 2008. [Online]. Available:
  \url{http://dx.doi.org/10.1007/s10664-008-9073-9}
\BIBentrySTDinterwordspacing

\bibitem{Ding:2016:KMA:2939672.2939719}
\BIBentryALTinterwordspacing
S.~H. Ding, B.~C. Fung, and P.~Charland, ``Kam1n0: Mapreduce-based assembly
  clone search for reverse engineering,'' in \emph{Proceedings of the 22Nd ACM
  SIGKDD International Conference on Knowledge Discovery and Data Mining}, ser.
  KDD '16.\hskip 1em plus 0.5em minus 0.4em\relax New York, NY, USA: ACM, 2016,
  pp. 461--470. [Online]. Available:
  \url{http://doi.acm.org/10.1145/2939672.2939719}
\BIBentrySTDinterwordspacing

\bibitem{7069883}
S.~Karus and K.~Kilgi, ``Code clone detection using wavelets,'' in \emph{2015
  IEEE 9th International Workshop on Software Clones (IWSC)}, March 2015, pp.
  8--14.

\bibitem{Qu2014544}
\BIBentryALTinterwordspacing
W.~Qu, Y.~Jia, and M.~Jiang, ``Pattern mining of cloned codes in software
  systems,'' \emph{Information Sciences}, vol. 259, pp. 544 -- 554, 2014.
  [Online]. Available:
  \url{http://www.sciencedirect.com/science/article/pii/S0020025510001787}
\BIBentrySTDinterwordspacing

\bibitem{Hemel:2011:FSL:1985441.1985453}
\BIBentryALTinterwordspacing
A.~Hemel, K.~T. Kalleberg, R.~Vermaas, and E.~Dolstra, ``Finding software
  license violations through binary code clone detection,'' in
  \emph{Proceedings of the 8th Working Conference on Mining Software
  Repositories}, ser. MSR '11.\hskip 1em plus 0.5em minus 0.4em\relax New York,
  NY, USA: ACM, 2011, pp. 63--72. [Online]. Available:
  \url{http://doi.acm.org/10.1145/1985441.1985453}
\BIBentrySTDinterwordspacing

\bibitem{6613837}
H.~Murakami, K.~Hotta, Y.~Higo, H.~Igaki, and S.~Kusumoto, ``Gapped code clone
  detection with lightweight source code analysis,'' in \emph{2013 21st
  International Conference on Program Comprehension (ICPC)}, May 2013, pp.
  93--102.

\bibitem{4023995}
R.~Koschke, R.~Falke, and P.~Frenzel, ``Clone detection using abstract syntax
  suffix trees,'' in \emph{2006 13th Working Conference on Reverse
  Engineering}, Oct 2006, pp. 253--262.

\bibitem{Bhattacharjee:2013:CTA:2480362.2480525}
\BIBentryALTinterwordspacing
A.~Bhattacharjee and H.~M. Jamil, ``Codeblast: A two-stage algorithm for
  improved program similarity matching in large software repositories,'' in
  \emph{Proceedings of the 28th Annual ACM Symposium on Applied Computing},
  ser. SAC '13.\hskip 1em plus 0.5em minus 0.4em\relax New York, NY, USA: ACM,
  2013, pp. 846--852. [Online]. Available:
  \url{http://doi.acm.org/10.1145/2480362.2480525}
\BIBentrySTDinterwordspacing

\bibitem{723194}
R.~Koschke, J.~F. Girard, and M.~Wurthner, ``An intermediate representation for
  integrating reverse engineering analyses,'' in \emph{Proceedings Fifth
  Working Conference on Reverse Engineering}, Oct 1998, pp. 241--250.

\bibitem{Prechelt00findingplagiarisms}
L.~Prechelt, G.~Malpohl, and M.~Philippsen, ``Finding plagiarisms among a set
  of programs with jplag,'' \emph{JOURNAL OF UNIVERSAL COMPUTER SCIENCE},
  vol.~8, pp. 1016--1038, 2000.

\bibitem{Storrle2013}
\BIBentryALTinterwordspacing
H.~St{\"o}rrle, ``Towards clone detection in uml domain models,''
  \emph{Software {\&} Systems Modeling}, vol.~12, no.~2, pp. 307--329, 2013.
  [Online]. Available: \url{http://dx.doi.org/10.1007/s10270-011-0217-9}
\BIBentrySTDinterwordspacing

\bibitem{5328752}
S.~Kawaguchi, T.~Yamashina, H.~Uwano, K.~Fushida, Y.~Kamei, M.~Nagura, and
  H.~Iida, ``Shinobi: A tool for automatic code clone detection in the ide,''
  in \emph{2009 16th Working Conference on Reverse Engineering}, Oct 2009, pp.
  313--314.

\bibitem{shinobiTR}
T.~Yamashina, H.~Uwano, K.~Fushida, Y.~Kamei, M.~Nagura, S.~Kawaguchi, and
  H.~Iida, ``Shinobi: A real-time code clone detection frool for software
  maintenance,'' Graduate School for Information Science, Nara Institute of
  Science and Technology, Tech. Rep. NAIST-IS-TR2007011, 2008.

\bibitem{Zilberstein:2016:LCN:2986012.2986013}
\BIBentryALTinterwordspacing
M.~Zilberstein and E.~Yahav, ``Leveraging a corpus of natural language
  descriptions for program similarity,'' in \emph{Proceedings of the 2016 ACM
  International Symposium on New Ideas, New Paradigms, and Reflections on
  Programming and Software}, ser. Onward! 2016.\hskip 1em plus 0.5em minus
  0.4em\relax New York, NY, USA: ACM, 2016, pp. 197--211. [Online]. Available:
  \url{http://doi.acm.org/10.1145/2986012.2986013}
\BIBentrySTDinterwordspacing

\bibitem{7483299}
M.~Abdelkader and M.~Mimoun, ``Clone detection using time series and dynamic
  time warping techniques,'' in \emph{2015 Third World Conference on Complex
  Systems (WCCS)}, Nov 2015, pp. 1--6.

\bibitem{6227861}
T.~Lavoie and E.~Merlo, ``An accurate estimation of the levenshtein distance
  using metric trees and manhattan distance,'' in \emph{2012 6th International
  Workshop on Software Clones (IWSC)}, June 2012, pp. 1--7.

\bibitem{Storrle2015}
\BIBentryALTinterwordspacing
H.~St{\"o}rrle, \emph{Effective and Efficient Model Clone Detection}.\hskip 1em
  plus 0.5em minus 0.4em\relax Cham: Springer International Publishing, 2015,
  pp. 440--457. [Online]. Available:
  \url{http://dx.doi.org/10.1007/978-3-319-15545-6_25}
\BIBentrySTDinterwordspacing

\bibitem{Cesare2013}
\BIBentryALTinterwordspacing
S.~Cesare, Y.~Xiang, and J.~Zhang, \emph{Clonewise -- Detecting Package-Level
  Clones Using Machine Learning}.\hskip 1em plus 0.5em minus 0.4em\relax Cham:
  Springer International Publishing, 2013, pp. 197--215. [Online]. Available:
  \url{http://dx.doi.org/10.1007/978-3-319-04283-1_13}
\BIBentrySTDinterwordspacing

\bibitem{Zibran:2012:IRF:2245276.2231970}
\BIBentryALTinterwordspacing
M.~F. Zibran and C.~K. Roy, ``Ide-based real-time focused search for near-miss
  clones,'' in \emph{Proceedings of the 27th Annual ACM Symposium on Applied
  Computing}, ser. SAC '12.\hskip 1em plus 0.5em minus 0.4em\relax New York,
  NY, USA: ACM, 2012, pp. 1235--1242. [Online]. Available:
  \url{http://doi.acm.org/10.1145/2245276.2231970}
\BIBentrySTDinterwordspacing

\bibitem{Zibran:2011:TFC:1985404.1985423}
\BIBentryALTinterwordspacing
------, ``Towards flexible code clone detection, management, and refactoring in
  ide,'' in \emph{Proceedings of the 5th International Workshop on Software
  Clones}, ser. IWSC '11.\hskip 1em plus 0.5em minus 0.4em\relax New York, NY,
  USA: ACM, 2011, pp. 75--76. [Online]. Available:
  \url{http://doi.acm.org/10.1145/1985404.1985423}
\BIBentrySTDinterwordspacing

\bibitem{4400161}
W.~S. Evans, C.~W. Fraser, and F.~Ma, ``Clone detection via structural
  abstraction,'' in \emph{14th Working Conference on Reverse Engineering (WCRE
  2007)}, Oct 2007, pp. 150--159.

\bibitem{Evans2009}
\BIBentryALTinterwordspacing
------, ``Clone detection via structural abstraction,'' \emph{Software Quality
  Journal}, vol.~17, no.~4, pp. 309--330, 2009. [Online]. Available:
  \url{http://dx.doi.org/10.1007/s11219-009-9074-y}
\BIBentrySTDinterwordspacing

\bibitem{6319249}
L.~Dong, J.~Wang, and L.~Chen, ``Modular heap abstraction-based code clone
  detection for heap-manipulating programs,'' in \emph{2012 12th International
  Conference on Quality Software}, Aug 2012, pp. 197--200.

\bibitem{Sargsyan2016}
\BIBentryALTinterwordspacing
S.~Sargsyan, S.~Kurmangaleev, A.~Belevantsev, and A.~Avetisyan, ``Scalable and
  accurate detection of code clones,'' \emph{Programming and Computer
  Software}, vol.~42, no.~1, pp. 27--33, 2016. [Online]. Available:
  \url{http://dx.doi.org/10.1134/S0361768816010072}
\BIBentrySTDinterwordspacing

\bibitem{7358259}
A.~Avetisyan, S.~Kurmangaleev, S.~Sargsyan, M.~Arutunian, and A.~Belevantsev,
  ``Llvm-based code clone detection framework,'' in \emph{2015 Computer Science
  and Information Technologies (CSIT)}, Sept 2015, pp. 100--104.

\bibitem{SMR:SMR1662}
\BIBentryALTinterwordspacing
J.~Svajlenko, I.~Keivanloo, and C.~K. Roy, ``Big data clone detection using
  classical detectors: an exploratory study,'' \emph{Journal of Software:
  Evolution and Process}, vol.~27, no.~6, pp. 430--464, 2015, jSME-13-0126.R1.
  [Online]. Available: \url{http://dx.doi.org/10.1002/smr.1662}
\BIBentrySTDinterwordspacing

\bibitem{6227875}
I.~Keivanloo, C.~K. Roy, J.~Rilling, and P.~Charland, ``Shuffling and
  randomization for scalable source code clone detection,'' in \emph{2012 6th
  International Workshop on Software Clones (IWSC)}, June 2012, pp. 82--83.

\bibitem{6613037}
J.~Svajlenko, I.~Keivanloo, and C.~K. Roy, ``Scaling classical clone detection
  tools for ultra-large datasets: An exploratory study,'' in \emph{2013 7th
  International Workshop on Software Clones (IWSC)}, May 2013, pp. 16--22.

\bibitem{Manber:1994:FSF:1267074.1267076}
\BIBentryALTinterwordspacing
U.~Manber, ``Finding similar files in a large file system,'' in
  \emph{Proceedings of the USENIX Winter 1994 Technical Conference on USENIX
  Winter 1994 Technical Conference}, ser. WTEC'94.\hskip 1em plus 0.5em minus
  0.4em\relax Berkeley, CA, USA: USENIX Association, 1994, pp. 2--2. [Online].
  Available: \url{http://dl.acm.org/citation.cfm?id=1267074.1267076}
\BIBentrySTDinterwordspacing

\bibitem{6494937}
Y.~Yuan and Y.~Guo, ``Boreas: an accurate and scalable token-based approach to
  code clone detection,'' in \emph{2012 Proceedings of the 27th IEEE/ACM
  International Conference on Automated Software Engineering}, Sept 2012, pp.
  286--289.

\bibitem{Yuan:2012:SAA:2162110.2162126}
\BIBentryALTinterwordspacing
Y.~Yuan, ``A scalable and accurate approach based on count matrix for detecting
  code clones,'' in \emph{Proceedings of the 11th Annual International
  Conference on Aspect-oriented Software Development Companion}, ser. AOSD
  Companion '12.\hskip 1em plus 0.5em minus 0.4em\relax New York, NY, USA: ACM,
  2012, pp. 21--22. [Online]. Available:
  \url{http://doi.acm.org/10.1145/2162110.2162126}
\BIBentrySTDinterwordspacing

\bibitem{li_1610609}
Z.~Li, S.~Lu, S.~Myagmar, and Y.~Zhou, ``Cp-miner: finding copy-paste and
  related bugs in large-scale software code,'' \emph{IEEE Transactions on
  Software Engineering}, vol.~32, no.~3, pp. 176--192, March 2006.

\bibitem{Li:2004:CTF:1251254.1251274}
\BIBentryALTinterwordspacing
------, ``Cp-miner: A tool for finding copy-paste and related bugs in operating
  system code,'' in \emph{Proceedings of the 6th Conference on Symposium on
  Opearting Systems Design \& Implementation - Volume 6}, ser. OSDI'04.\hskip
  1em plus 0.5em minus 0.4em\relax Berkeley, CA, USA: USENIX Association, 2004,
  pp. 20--20. [Online]. Available:
  \url{http://dl.acm.org/citation.cfm?id=1251254.1251274}
\BIBentrySTDinterwordspacing

\bibitem{5463293}
Y.~Sasaki, T.~Yamamoto, Y.~Hayase, and K.~Inoue, ``Finding file clones in
  freebsd ports collection,'' in \emph{2010 7th IEEE Working Conference on
  Mining Software Repositories (MSR 2010)}, May 2010, pp. 102--105.

\bibitem{7503720}
F.-H. Su, J.~Bell, G.~Kaiser, and S.~Sethumadhavan, ``Identifying functionally
  similar code in complex codebases,'' in \emph{2016 IEEE 24th International
  Conference on Program Comprehension (ICPC)}, May 2016, pp. 1--10.

\bibitem{5070528}
N.~H. Pham, H.~A. Nguyen, T.~T. Nguyen, J.~M. Al-Kofahi, and T.~N. Nguyen,
  ``Complete and accurate clone detection in graph-based models,'' in
  \emph{2009 IEEE 31st International Conference on Software Engineering}, May
  2009, pp. 276--286.

\bibitem{6615249}
Q.~Q. Shi, L.~P. Zhang, F.~J. Meng, and D.~S. Liu, ``A novel detection approach
  for statement clones,'' in \emph{2013 IEEE 4th International Conference on
  Software Engineering and Service Science}, May 2013, pp. 27--30.

\bibitem{Dang:2012:XTC:2420950.2421004}
\BIBentryALTinterwordspacing
Y.~Dang, D.~Zhang, S.~Ge, C.~Chu, Y.~Qiu, and T.~Xie, ``Xiao: Tuning code
  clones at hands of engineers in practice,'' in \emph{Proceedings of the 28th
  Annual Computer Security Applications Conference}, ser. ACSAC '12.\hskip 1em
  plus 0.5em minus 0.4em\relax New York, NY, USA: ACM, 2012, pp. 369--378.
  [Online]. Available: \url{http://doi.acm.org/10.1145/2420950.2421004}
\BIBentrySTDinterwordspacing

\bibitem{Dang:2011:CCD:1985404.1985417}
\BIBentryALTinterwordspacing
Y.~Dang, S.~Ge, R.~Huang, and D.~Zhang, ``Code clone detection experience at
  microsoft,'' in \emph{Proceedings of the 5th International Workshop on
  Software Clones}, ser. IWSC '11.\hskip 1em plus 0.5em minus 0.4em\relax New
  York, NY, USA: ACM, 2011, pp. 63--64. [Online]. Available:
  \url{http://doi.acm.org/10.1145/1985404.1985417}
\BIBentrySTDinterwordspacing

\bibitem{6385134}
T.~Ishihara, K.~Hotta, Y.~Higo, H.~Igaki, and S.~Kusumoto, ``Inter-project
  functional clone detection toward building libraries - an empirical study on
  13,000 projects,'' in \emph{2012 19th Working Conference on Reverse
  Engineering}, Oct 2012, pp. 387--391.

\bibitem{6178897}
R.~Koschke, ``Large-scale inter-system clone detection using suffix trees,'' in
  \emph{2012 16th European Conference on Software Maintenance and
  Reengineering}, March 2012, pp. 309--318.

\bibitem{SMR:SMR1592}
\BIBentryALTinterwordspacing
------, ``Large-scale inter-system clone detection using suffix trees and
  hashing,'' \emph{Journal of Software: Evolution and Process}, vol.~26, no.~8,
  pp. 747--769, 2014. [Online]. Available:
  \url{http://dx.doi.org/10.1002/smr.1592}
\BIBentrySTDinterwordspacing

\bibitem{Saebjornsen:2009:DCC:1572272.1572287}
\BIBentryALTinterwordspacing
A.~S{\ae}bj{\o}rnsen, J.~Willcock, T.~Panas, D.~Quinlan, and Z.~Su, ``Detecting
  code clones in binary executables,'' in \emph{Proceedings of the Eighteenth
  International Symposium on Software Testing and Analysis}, ser. ISSTA
  '09.\hskip 1em plus 0.5em minus 0.4em\relax New York, NY, USA: ACM, 2009, pp.
  117--128. [Online]. Available:
  \url{http://doi.acm.org/10.1145/1572272.1572287}
\BIBentrySTDinterwordspacing

\bibitem{Mariani:2012:AAD:2133797.2133799}
\BIBentryALTinterwordspacing
L.~Mariani and D.~Micucci, ``Audentes: Automatic detection of tentative
  plagiarism according to a reference solution,'' \emph{Trans. Comput. Educ.},
  vol.~12, no.~1, pp. 2:1--2:26, Mar. 2012. [Online]. Available:
  \url{http://doi.acm.org/10.1145/2133797.2133799}
\BIBentrySTDinterwordspacing

\bibitem{Krutz:2015:EEU:2695664.2695929}
\BIBentryALTinterwordspacing
D.~E. Krutz, S.~A. Malachowsky, and E.~Shihab, ``Examining the effectiveness of
  using concolic analysis to detect code clones,'' in \emph{Proceedings of the
  30th Annual ACM Symposium on Applied Computing}, ser. SAC '15.\hskip 1em plus
  0.5em minus 0.4em\relax New York, NY, USA: ACM, 2015, pp. 1610--1615.
  [Online]. Available: \url{http://doi.acm.org/10.1145/2695664.2695929}
\BIBentrySTDinterwordspacing

\bibitem{6671332}
D.~E. Krutz and E.~Shihab, ``Cccd: Concolic code clone detection,'' in
  \emph{2013 20th Working Conference on Reverse Engineering (WCRE)}, Oct 2013,
  pp. 489--490.

\bibitem{7582804}
X.~Cheng, Z.~Peng, L.~Jiang, H.~Zhong, H.~Yu, and J.~Zhao, ``Mining revision
  histories to detect cross-language clones without intermediates,'' in
  \emph{2016 31st IEEE/ACM International Conference on Automated Software
  Engineering (ASE)}, Sept 2016, pp. 696--701.

\bibitem{6613041}
B.~Muddu, A.~Asadullah, and V.~Bhat, ``Cpdp: A robust technique for plagiarism
  detection in source code,'' in \emph{2013 7th International Workshop on
  Software Clones (IWSC)}, May 2013, pp. 39--45.

\bibitem{7544005}
S.~Alam, R.~Riley, I.~Sogukpinar, and N.~Carkaci, ``Droidclone: Detecting
  android malware variants by exposing code clones,'' in \emph{2016 Sixth
  International Conference on Digital Information and Communication Technology
  and its Applications (DICTAP)}, July 2016, pp. 79--84.

\bibitem{4137427}
H.~Liu, Z.~Ma, L.~Zhang, and W.~Shao, ``Detecting duplications in sequence
  diagrams based on suffix trees,'' in \emph{2006 13th Asia Pacific Software
  Engineering Conference (APSEC'06)}, Dec 2006, pp. 269--276.

\bibitem{Lanubile_1192447}
F.~Lanubile and T.~Mallardo, ``Finding function clones in web applications,''
  in \emph{Seventh European Conference onSoftware Maintenance and
  Reengineering, 2003. Proceedings.}, March 2003, pp. 379--386.

\bibitem{Calefato:2004:FCD:2011138.2011140}
\BIBentryALTinterwordspacing
F.~Calefato, F.~Lanubile, and T.~Mallardo, ``Function clone detection in web
  applications: A semiautomated approach,'' \emph{J. Web Eng.}, vol.~3, no.~1,
  pp. 3--21, May 2004. [Online]. Available:
  \url{http://dl.acm.org/citation.cfm?id=2011138.2011140}
\BIBentrySTDinterwordspacing

\bibitem{6392103}
H.~Murakami, K.~Hotta, Y.~Higo, H.~Igaki, and S.~Kusumoto, ``Folding repeated
  instructions for improving token-based code clone detection,'' in \emph{2012
  IEEE 12th International Working Conference on Source Code Analysis and
  Manipulation}, Sept 2012, pp. 64--73.

\bibitem{Cheung2016}
\BIBentryALTinterwordspacing
W.~T. Cheung, S.~Ryu, and S.~Kim, ``Development nature matters: An empirical
  study of code clones in javascript applications,'' \emph{Empirical Software
  Engineering}, vol.~21, no.~2, pp. 517--564, 2016. [Online]. Available:
  \url{http://dx.doi.org/10.1007/s10664-015-9368-6}
\BIBentrySTDinterwordspacing

\bibitem{6340252}
S.~K. Abd-El-Hafiz, ``A metrics-based data mining approach for software clone
  detection,'' in \emph{2012 IEEE 36th Annual Computer Software and
  Applications Conference}, July 2012, pp. 35--41.

\bibitem{6385136}
F.~Al-Omari, I.~Keivanloo, C.~K. Roy, and J.~Rilling, ``Detecting clones across
  microsoft .net programming languages,'' in \emph{2012 19th Working Conference
  on Reverse Engineering}, Oct 2012, pp. 405--414.

\bibitem{6897221}
G.~Bansal and R.~Tekchandani, ``Selecting a set of appropriate metrics for
  detecting code clones,'' in \emph{2014 Seventh International Conference on
  Contemporary Computing (IC3)}, Aug 2014, pp. 484--488.

\bibitem{6976124}
V.~Bauer, T.~Völke, and E.~Jürgens, ``A novel approach to detect
  unintentional re-implementations,'' in \emph{2014 IEEE International
  Conference on Software Maintenance and Evolution}, Sept 2014, pp. 491--495.

\bibitem{Schugerl:2011:SCD:1985404.1985413}
\BIBentryALTinterwordspacing
P.~Schugerl, ``Scalable clone detection using description logic,'' in
  \emph{Proceedings of the 5th International Workshop on Software Clones}, ser.
  IWSC '11.\hskip 1em plus 0.5em minus 0.4em\relax New York, NY, USA: ACM,
  2011, pp. 47--53. [Online]. Available:
  \url{http://doi.acm.org/10.1145/1985404.1985413}
\BIBentrySTDinterwordspacing

\bibitem{Ekanayake2012}
\BIBentryALTinterwordspacing
C.~C. Ekanayake, M.~Dumas, L.~Garc{\'i}a-Ba{\~{n}}uelos, M.~La~Rosa, and
  A.~H.~M. ter Hofstede, \emph{Approximate Clone Detection in Repositories of
  Business Process Models}.\hskip 1em plus 0.5em minus 0.4em\relax Berlin,
  Heidelberg: Springer Berlin Heidelberg, 2012, pp. 302--318. [Online].
  Available: \url{http://dx.doi.org/10.1007/978-3-642-32885-5_24}
\BIBentrySTDinterwordspacing

\bibitem{6363131}
M.~Iwamoto, S.~Oshima, and T.~Nakashima, ``Token-based code clone detection
  technique in a student's programming exercise,'' in \emph{2012 Seventh
  International Conference on Broadband, Wireless Computing, Communication and
  Applications}, Nov 2012, pp. 650--655.

\bibitem{6690950}
------, ``A token-based illicit copy detection method using complexity for a
  program exercise,'' in \emph{2013 Eighth International Conference on
  Broadband and Wireless Computing, Communication and Applications}, Oct 2013,
  pp. 575--580.

\bibitem{7081830}
I.~Keivanloo, F.~Zhang, and Y.~Zou, ``Threshold-free code clone detection for a
  large-scale heterogeneous java repository,'' in \emph{2015 IEEE 22nd
  International Conference on Software Analysis, Evolution, and Reengineering
  (SANER)}, March 2015, pp. 201--210.

\bibitem{5478099}
Z.~O. Li and J.~Sun, ``A metric space based software clone detection
  approach,'' in \emph{2010 2nd IEEE International Conference on Information
  Management and Engineering}, April 2010, pp. 393--397.

\bibitem{4380246}
A.~de~Lucia, M.~Risi, G.~Tortora, and G.~Scanniello, ``Clustering algorithms
  and latent semantic indexing to identify similar pages in web applications,''
  in \emph{2007 9th IEEE International Workshop on Web Site Evolution}, Oct
  2007, pp. 65--72.

\bibitem{7397263}
A.~Sheneamer and J.~Kalita, ``Code clone detection using coarse and
  fine-grained hybrid approaches,'' in \emph{2015 IEEE Seventh International
  Conference on Intelligent Computing and Information Systems (ICICIS)}, Dec
  2015, pp. 472--480.

\bibitem{7838289}
------, ``Semantic clone detection using machine learning,'' in \emph{2016 15th
  IEEE International Conference on Machine Learning and Applications (ICMLA)},
  Dec 2016, pp. 1024--1028.

\bibitem{Stojanovic2015259}
\BIBentryALTinterwordspacing
S.~Stojanović, Z.~Radivojević, and M.~Cvetanović, ``Approach for estimating
  similarity between procedures in differently compiled binaries,''
  \emph{Information and Software Technology}, vol.~58, pp. 259 -- 271, 2015.
  [Online]. Available:
  \url{http://www.sciencedirect.com/science/article/pii/S0950584914001517}
\BIBentrySTDinterwordspacing

\bibitem{5640465}
A.~Perumal, S.~Kanmani, and E.~Kodhai, ``Extracting the similarity in detected
  software clones using metrics,'' in \emph{2010 International Conference on
  Computer and Communication Technology (ICCCT)}, Sept 2010, pp. 575--579.

\bibitem{6130694}
Y.~Yuan and Y.~Guo, ``Cmcd: Count matrix based code clone detection,'' in
  \emph{2011 18th Asia-Pacific Software Engineering Conference}, Dec 2011, pp.
  250--257.

\bibitem{5071050}
T.~T. Nguyen, H.~A. Nguyen, N.~H. Pham, J.~M. Al-Kofahi, and T.~N. Nguyen,
  ``Clemanx: Incremental clone detection tool for evolving software,'' in
  \emph{2009 31st International Conference on Software Engineering - Companion
  Volume}, May 2009, pp. 437--438.

\bibitem{conqat}
E.~Juergens, F.~Deissenboeck, and B.~Hummel, ``Clonedetective - a workbench for
  clone detection research,'' in \emph{Proceedings of the 31st International
  Conference on Software Engineering}, ser. ICSE '09.\hskip 1em plus 0.5em
  minus 0.4em\relax Washington, DC, USA: IEEE Computer Society, 2009, pp.
  603--606.

\bibitem{conqat_models}
F.~Deissenboeck, B.~Hummel, E.~Jürgens, B.~Schätz, S.~Wagner, J.~F. Girard,
  and S.~Teuchert, ``Clone detection in automotive model-based development,''
  in \emph{2008 ACM/IEEE 30th International Conference on Software
  Engineering}, May 2008, pp. 603--612.

\bibitem{Jiang:2009:AMF:1572272.1572283}
\BIBentryALTinterwordspacing
L.~Jiang and Z.~Su, ``Automatic mining of functionally equivalent code
  fragments via random testing,'' in \emph{Proceedings of the Eighteenth
  International Symposium on Software Testing and Analysis}, ser. ISSTA
  '09.\hskip 1em plus 0.5em minus 0.4em\relax New York, NY, USA: ACM, 2009, pp.
  81--92. [Online]. Available: \url{http://doi.acm.org/10.1145/1572272.1572283}
\BIBentrySTDinterwordspacing

\bibitem{6676875}
W.~Qian, X.~Peng, Z.~Xing, S.~Jarzabek, and W.~Zhao, ``Mining logical clones in
  software: Revealing high-level business and programming rules,'' in
  \emph{2013 IEEE International Conference on Software Maintenance}, Sept 2013,
  pp. 40--49.

\bibitem{Basit:2007:ETB:1295014.1295029}
\BIBentryALTinterwordspacing
H.~A. Basit, S.~J. Puglisi, W.~F. Smyth, A.~Turpin, and S.~Jarzabek,
  ``Efficient token based clone detection with flexible tokenization,'' in
  \emph{The 6th Joint Meeting on European Software Engineering Conference and
  the ACM SIGSOFT Symposium on the Foundations of Software Engineering:
  Companion Papers}, ser. ESEC-FSE companion '07.\hskip 1em plus 0.5em minus
  0.4em\relax New York, NY, USA: ACM, 2007, pp. 513--516. [Online]. Available:
  \url{http://doi.acm.org/10.1145/1295014.1295029}
\BIBentrySTDinterwordspacing

\bibitem{Chen2016}
\BIBentryALTinterwordspacing
J.~Chen, T.~R. Dean, and M.~H. Alalfi, ``Clone detection in matlab stateflow
  models,'' \emph{Software Quality Journal}, vol.~24, no.~4, pp. 917--946,
  2016. [Online]. Available: \url{http://dx.doi.org/10.1007/s11219-015-9296-0}
\BIBentrySTDinterwordspacing

\bibitem{Kontogiannis1996}
\BIBentryALTinterwordspacing
K.~A. Kontogiannis, R.~Demori, E.~Merlo, M.~Galler, and M.~Bernstein, ``Pattern
  matching for clone and concept detection,'' \emph{Automated Software
  Engineering}, vol.~3, no.~1, pp. 77--108, 1996. [Online]. Available:
  \url{http://dx.doi.org/10.1007/BF00126960}
\BIBentrySTDinterwordspacing

\bibitem{7582748}
M.~White, M.~Tufano, C.~Vendome, and D.~Poshyvanyk, ``Deep learning code
  fragments for code clone detection,'' in \emph{2016 31st IEEE/ACM
  International Conference on Automated Software Engineering (ASE)}, Sept 2016,
  pp. 87--98.

\bibitem{6227881}
W.~Toomey, ``Ctcompare: Code clone detection using hashed token sequences,'' in
  \emph{2012 6th International Workshop on Software Clones (IWSC)}, June 2012,
  pp. 92--93.

\bibitem{7880510}
Y.~Yuki, Y.~Higo, and S.~Kusumoto, ``A technique to detect multi-grained code
  clones,'' in \emph{2017 IEEE 11th International Workshop on Software Clones
  (IWSC)}, Feb 2017, pp. 1--7.

\bibitem{Su:2016:CRD:2950290.2950321}
\BIBentryALTinterwordspacing
F.-H. Su, J.~Bell, K.~Harvey, S.~Sethumadhavan, G.~Kaiser, and T.~Jebara,
  ``Code relatives: Detecting similarly behaving software,'' in
  \emph{Proceedings of the 2016 24th ACM SIGSOFT International Symposium on
  Foundations of Software Engineering}, ser. FSE 2016.\hskip 1em plus 0.5em
  minus 0.4em\relax New York, NY, USA: ACM, 2016, pp. 702--714. [Online].
  Available: \url{http://doi.acm.org/10.1145/2950290.2950321}
\BIBentrySTDinterwordspacing

\bibitem{Brown:2010:CDE:1706356.1706378}
\BIBentryALTinterwordspacing
C.~Brown and S.~Thompson, ``Clone detection and elimination for haskell,'' in
  \emph{Proceedings of the 2010 ACM SIGPLAN Workshop on Partial Evaluation and
  Program Manipulation}, ser. PEPM '10.\hskip 1em plus 0.5em minus 0.4em\relax
  New York, NY, USA: ACM, 2010, pp. 111--120. [Online]. Available:
  \url{http://doi.acm.org/10.1145/1706356.1706378}
\BIBentrySTDinterwordspacing

\bibitem{6079771}
I.~Keivanloo, J.~Rilling, and P.~Charland, ``Internet-scale real-time code
  clone search via multi-level indexing,'' in \emph{2011 18th Working
  Conference on Reverse Engineering}, Oct 2011, pp. 23--27.

\bibitem{6225474}
I.~Keivanloo, C.~Forbes, and J.~Rilling, ``Similarity search plug-in: Clone
  detection meets internet-scale code search,'' in \emph{2012 4th International
  Workshop on Search-Driven Development: Users, Infrastructure, Tools, and
  Evaluation (SUITE)}, June 2012, pp. 21--22.

\bibitem{Lee:2005:SHP:1094855.1094903}
\BIBentryALTinterwordspacing
S.~Lee and I.~Jeong, ``Sdd: High performance code clone detection system for
  large scale source code,'' in \emph{Companion to the 20th Annual ACM SIGPLAN
  Conference on Object-oriented Programming, Systems, Languages, and
  Applications}, ser. OOPSLA '05.\hskip 1em plus 0.5em minus 0.4em\relax New
  York, NY, USA: ACM, 2005, pp. 140--141. [Online]. Available:
  \url{http://doi.acm.org/10.1145/1094855.1094903}
\BIBentrySTDinterwordspacing

\bibitem{5521760}
L.~Barbour, H.~Yuan, and Y.~Zou, ``A technique for just-in-time clone detection
  in large scale systems,'' in \emph{2010 IEEE 18th International Conference on
  Program Comprehension}, June 2010, pp. 76--79.

\bibitem{CHILOWICZ200947}
\BIBentryALTinterwordspacing
M.~Chilowicz, Étienne Duris, and G.~Roussel, ``Finding similarities in source
  code through factorization,'' \emph{Electronic Notes in Theoretical Computer
  Science}, vol. 238, no.~5, pp. 47 -- 62, 2009. [Online]. Available:
  \url{http://www.sciencedirect.com/science/article/pii/S1571066109003946}
\BIBentrySTDinterwordspacing

\bibitem{5609665}
B.~Hummel, E.~Juergens, L.~Heinemann, and M.~Conradt, ``Index-based code clone
  detection: incremental, distributed, scalable,'' in \emph{2010 IEEE
  International Conference on Software Maintenance}, Sept 2010, pp. 1--9.

\bibitem{Tekin2014406}
\BIBentryALTinterwordspacing
U.~Tekin and F.~Buzluca, ``A graph mining approach for detecting identical
  design structures in object-oriented design models,'' \emph{Science of
  Computer Programming}, vol. 95, Part 4, pp. 406 -- 425, 2014, special Issue
  on Software Clones (IWSC'12). [Online]. Available:
  \url{http://www.sciencedirect.com/science/article/pii/S0167642313002451}
\BIBentrySTDinterwordspacing

\bibitem{6631651}
J.~Feng, B.~Cui, and K.~Xia, ``A code comparison algorithm based on ast for
  plagiarism detection,'' in \emph{2013 Fourth International Conference on
  Emerging Intelligent Data and Web Technologies}, Sept 2013, pp. 393--397.

\bibitem{6631708}
G.~Tao, D.~Guowei, Q.~Hu, and C.~Baojiang, ``Improved plagiarism detection
  algorithm based on abstract syntax tree,'' in \emph{2013 Fourth International
  Conference on Emerging Intelligent Data and Web Technologies}, Sept 2013, pp.
  714--719.

\bibitem{5741248}
Y.~Higo and S.~Kusumoto, ``Code clone detection on specialized pdgs with
  heuristics,'' in \emph{2011 15th European Conference on Software Maintenance
  and Reengineering}, March 2011, pp. 75--84.

\bibitem{6463128}
D.~Kong, X.~Su, S.~Wu, T.~Wang, and P.~Ma, ``Detect functionally equivalent
  code fragments via k-nearest neighbor algorithm,'' in \emph{2012 IEEE Fifth
  International Conference on Advanced Computational Intelligence (ICACI)}, Oct
  2012, pp. 94--98.

\bibitem{deckard}
\BIBentryALTinterwordspacing
L.~Jiang, G.~Misherghi, Z.~Su, and S.~Glondu, ``Deckard: Scalable and accurate
  tree-based detection of code clones,'' in \emph{Proceedings of the 29th
  International Conference on Software Engineering}, ser. ICSE '07.\hskip 1em
  plus 0.5em minus 0.4em\relax Washington, DC, USA: IEEE Computer Society,
  2007, pp. 96--105. [Online]. Available:
  \url{http://dx.doi.org/10.1109/ICSE.2007.30}
\BIBentrySTDinterwordspacing

\bibitem{Nguyen2009}
\BIBentryALTinterwordspacing
H.~A. Nguyen, T.~T. Nguyen, N.~H. Pham, J.~M. Al-Kofahi, and T.~N. Nguyen,
  \emph{Accurate and Efficient Structural Characteristic Feature Extraction for
  Clone Detection}.\hskip 1em plus 0.5em minus 0.4em\relax Berlin, Heidelberg:
  Springer Berlin Heidelberg, 2009, pp. 440--455. [Online]. Available:
  \url{http://dx.doi.org/10.1007/978-3-642-00593-0_31}
\BIBentrySTDinterwordspacing

\bibitem{Leitao2004}
\BIBentryALTinterwordspacing
A.~M. Leit{\~a}o, ``Detection of redundant code using r2d2,'' \emph{Software
  Quality Journal}, vol.~12, no.~4, pp. 361--382, 2004. [Online]. Available:
  \url{http://dx.doi.org/10.1023/B:SQJO.0000039793.31052.72}
\BIBentrySTDinterwordspacing

\bibitem{Chen:2014:RRC:2667473.2667486}
\BIBentryALTinterwordspacing
X.~Chen, A.~Y. Wang, and E.~Tempero, ``A replication and reproduction of code
  clone detection studies,'' in \emph{Proceedings of the Thirty-Seventh
  Australasian Computer Science Conference - Volume 147}, ser. ACSC '14.\hskip
  1em plus 0.5em minus 0.4em\relax Darlinghurst, Australia, Australia:
  Australian Computer Society, Inc., 2014, pp. 105--114. [Online]. Available:
  \url{http://dl.acm.org/citation.cfm?id=2667473.2667486}
\BIBentrySTDinterwordspacing

\bibitem{Dou:2016:DTC:2950290.2950359}
\BIBentryALTinterwordspacing
W.~Dou, S.-C. Cheung, C.~Gao, C.~Xu, L.~Xu, and J.~Wei, ``Detecting table
  clones and smells in spreadsheets,'' in \emph{Proceedings of the 2016 24th
  ACM SIGSOFT International Symposium on Foundations of Software Engineering},
  ser. FSE 2016.\hskip 1em plus 0.5em minus 0.4em\relax New York, NY, USA: ACM,
  2016, pp. 787--798. [Online]. Available:
  \url{http://doi.acm.org/10.1145/2950290.2950359}
\BIBentrySTDinterwordspacing

\bibitem{Hermans:2013:DCD:2486788.2486827}
\BIBentryALTinterwordspacing
F.~Hermans, B.~Sedee, M.~Pinzger, and A.~v. Deursen, ``Data clone detection and
  visualization in spreadsheets,'' in \emph{Proceedings of the 2013
  International Conference on Software Engineering}, ser. ICSE '13.\hskip 1em
  plus 0.5em minus 0.4em\relax Piscataway, NJ, USA: IEEE Press, 2013, pp.
  292--301. [Online]. Available:
  \url{http://dl.acm.org/citation.cfm?id=2486788.2486827}
\BIBentrySTDinterwordspacing

\bibitem{7813733}
S.~Jadon, ``Code clones detection using machine learning technique: Support
  vector machine,'' in \emph{2016 International Conference on Computing,
  Communication and Automation (ICCCA)}, April 2016, pp. 399--303.

\bibitem{Joshi2015}
\BIBentryALTinterwordspacing
B.~Joshi, P.~Budhathoki, W.~L. Woon, and D.~Svetinovic, \emph{Software Clone
  Detection Using Clustering Approach}.\hskip 1em plus 0.5em minus 0.4em\relax
  Cham: Springer International Publishing, 2015, pp. 520--527. [Online].
  Available: \url{http://dx.doi.org/10.1007/978-3-319-26535-3_59}
\BIBentrySTDinterwordspacing

\bibitem{7087126}
R.~V. Patil, S.~D. Joshi, S.~V. Shinde, D.~A. Ajagekar, and S.~D. Bankar,
  ``Code clone detection using decentralized architecture and code reduction,''
  in \emph{2015 International Conference on Pervasive Computing (ICPC)}, Jan
  2015, pp. 1--6.

\bibitem{7062689}
B.~Priyambadha and S.~Rochimah, ``Case study on semantic clone detection based
  on code behavior,'' in \emph{2014 International Conference on Data and
  Software Engineering (ICODSE)}, Nov 2014, pp. 1--6.

\bibitem{7880502}
C.~Ragkhitwetsagul and J.~Krinke, ``Using compilation/decompilation to enhance
  clone detection,'' in \emph{2017 IEEE 11th International Workshop on Software
  Clones (IWSC)}, Feb 2017, pp. 1--7.

\bibitem{7880501}
K.~Uemura, A.~Mori, K.~Fujiwara, E.~Choi, and H.~Iida, ``Detecting and
  analyzing code clones in hdl,'' in \emph{2017 IEEE 11th International
  Workshop on Software Clones (IWSC)}, Feb 2017, pp. 1--7.

\bibitem{6613854}
T.~Kamiya, ``Agec: An execution-semantic clone detection tool,'' in \emph{2013
  21st International Conference on Program Comprehension (ICPC)}, May 2013, pp.
  227--229.

\bibitem{kraft2008cross}
N.~A. Kraft, B.~W. Bonds, and R.~K. Smith, ``Cross-language clone detection.''
  in \emph{Software Engineering and Knowledge Engineering}, 2008, pp. 54--59.

\bibitem{4222573}
S.~Livieri, Y.~Higo, M.~Matushita, and K.~Inoue, ``Very-large scale code clone
  analysis and visualization of open source programs using distributed
  ccfinder: D-ccfinder,'' in \emph{29th International Conference on Software
  Engineering (ICSE'07)}, May 2007, pp. 106--115.

\bibitem{Liu:2006:GDS:1150402.1150522}
\BIBentryALTinterwordspacing
C.~Liu, C.~Chen, J.~Han, and P.~S. Yu, ``Gplag: Detection of software
  plagiarism by program dependence graph analysis,'' in \emph{Proceedings of
  the 12th ACM SIGKDD International Conference on Knowledge Discovery and Data
  Mining}, ser. KDD '06.\hskip 1em plus 0.5em minus 0.4em\relax New York, NY,
  USA: ACM, 2006, pp. 872--881. [Online]. Available:
  \url{http://doi.acm.org/10.1145/1150402.1150522}
\BIBentrySTDinterwordspacing

\bibitem{Lillack:2014:DCC:2660190.2662116}
\BIBentryALTinterwordspacing
M.~Lillack, C.~Bucholdt, and D.~Schilling, ``Detection of code clones in
  software generators,'' in \emph{Proceedings of the 6th International Workshop
  on Feature-Oriented Software Development}, ser. FOSD '14.\hskip 1em plus
  0.5em minus 0.4em\relax New York, NY, USA: ACM, 2014, pp. 37--44. [Online].
  Available: \url{http://doi.acm.org/10.1145/2660190.2662116}
\BIBentrySTDinterwordspacing

\bibitem{iclonesthesis}
N.~G\''ode, ``Incremental clone detection,'' Ph.D. dissertation, Department of
  Mathematics and Computer Science, University of Bremen, 2008.

\bibitem{5645564}
B.~Biegel and S.~Diehl, ``Highly configurable and extensible code clone
  detection,'' in \emph{2010 17th Working Conference on Reverse Engineering},
  Oct 2010, pp. 237--241.

\bibitem{6377848}
S.~U. Rehman, K.~Khan, S.~Fong, and R.~Biuk-Aghai, ``An efficient new
  multi-language clone detection approach from large source code,'' in
  \emph{2012 IEEE International Conference on Systems, Man, and Cybernetics
  (SMC)}, Oct 2012, pp. 937--940.

\bibitem{Komondoor2001}
\BIBentryALTinterwordspacing
R.~Komondoor and S.~Horwitz, \emph{Tool Demonstration: Finding Duplicated Code
  Using Program Dependences}.\hskip 1em plus 0.5em minus 0.4em\relax Berlin,
  Heidelberg: Springer Berlin Heidelberg, 2001, pp. 383--386. [Online].
  Available: \url{http://dx.doi.org/10.1007/3-540-45309-1_25}
\BIBentrySTDinterwordspacing

\bibitem{Komondoor2001-2}
\BIBentryALTinterwordspacing
------, \emph{Using Slicing to Identify Duplication in Source Code}.\hskip 1em
  plus 0.5em minus 0.4em\relax Berlin, Heidelberg: Springer Berlin Heidelberg,
  2001, pp. 40--56. [Online]. Available:
  \url{http://dx.doi.org/10.1007/3-540-47764-0_3}
\BIBentrySTDinterwordspacing

\bibitem{Gitchell:1999:SUD:299649.299783}
\BIBentryALTinterwordspacing
D.~Gitchell and N.~Tran, ``Sim: A utility for detecting similarity in computer
  programs,'' in \emph{The Proceedings of the Thirtieth SIGCSE Technical
  Symposium on Computer Science Education}, ser. SIGCSE '99.\hskip 1em plus
  0.5em minus 0.4em\relax New York, NY, USA: ACM, 1999, pp. 266--270. [Online].
  Available: \url{http://doi.acm.org/10.1145/299649.299783}
\BIBentrySTDinterwordspacing

\bibitem{simian}
S.~Harris, ``Simian,'' \url{http://www.harukizaemon.com/simian/}.

\bibitem{Li2010}
\BIBentryALTinterwordspacing
H.~Li and S.~Thompson, \emph{Similar Code Detection and Elimination for Erlang
  Programs}.\hskip 1em plus 0.5em minus 0.4em\relax Berlin, Heidelberg:
  Springer Berlin Heidelberg, 2010, pp. 104--118. [Online]. Available:
  \url{http://dx.doi.org/10.1007/978-3-642-11503-5_10}
\BIBentrySTDinterwordspacing

\bibitem{Akhin2013}
\BIBentryALTinterwordspacing
M.~Akhin and V.~Itsykson, ``Tree slicing: Finding intertwined and gapped clones
  in one simple step,'' \emph{Automatic Control and Computer Sciences},
  vol.~47, no.~7, pp. 427--432, 2013. [Online]. Available:
  \url{http://dx.doi.org/10.3103/S0146411613070171}
\BIBentrySTDinterwordspacing

\bibitem{6140712}
A.~F.~M. Ali, S.~Sulaiman, and S.~M. Syed-Mohamad, ``An enhanced generic
  pipeline model for code clone detection,'' in \emph{2011 Malaysian Conference
  in Software Engineering}, Dec 2011, pp. 434--438.

\bibitem{Cordy:2004:PLD:1034914.1034915}
\BIBentryALTinterwordspacing
J.~R. Cordy, T.~R. Dean, and N.~Synytskyy, ``Practical language-independent
  detection of near-miss clones,'' in \emph{Proceedings of the 2004 Conference
  of the Centre for Advanced Studies on Collaborative Research}, ser. CASCON
  '04.\hskip 1em plus 0.5em minus 0.4em\relax IBM Press, 2004, pp. 1--12.
  [Online]. Available: \url{http://dl.acm.org/citation.cfm?id=1034914.1034915}
\BIBentrySTDinterwordspacing

\bibitem{Dandois2012}
\BIBentryALTinterwordspacing
C.~Dandois and W.~Vanhoof, \emph{Clones in Logic Programs and How to Detect
  Them}.\hskip 1em plus 0.5em minus 0.4em\relax Berlin, Heidelberg: Springer
  Berlin Heidelberg, 2012, pp. 90--105. [Online]. Available:
  \url{http://dx.doi.org/10.1007/978-3-642-32211-2_7}
\BIBentrySTDinterwordspacing

\bibitem{Lucca01cloneanalysis}
G.~D. Lucca, M.~D. Penta, A.~R. Fasolino, and P.~Granato, ``Clone analysis in
  the web era: an approach to identify cloned web pages,'' in \emph{IN
  PROCEEDINGS OF THE 7TH IEEE WORKSHOP ON EMPIRICAL STUDIES OF SOFTWARE
  MAINTENANCE (WESS’99}, 2001, pp. 107--113.

\bibitem{1134103}
E.~Burd and J.~Bailey, ``Evaluating clone detection tools for use during
  preventative maintenance,'' in \emph{Proceedings. Second IEEE International
  Workshop on Source Code Analysis and Manipulation}, 2002, pp. 36--43.

\bibitem{Dumas2013619}
\BIBentryALTinterwordspacing
M.~Dumas, L.~García-Bañuelos, M.~L. Rosa, and R.~Uba, ``Fast detection of
  exact clones in business process model repositories,'' \emph{Information
  Systems}, vol.~38, no.~4, pp. 619 -- 633, 2013, special section on \{BPM\}
  2011 conference. [Online]. Available:
  \url{http://www.sciencedirect.com/science/article/pii/S0306437912000993}
\BIBentrySTDinterwordspacing

\bibitem{Uba2011}
\BIBentryALTinterwordspacing
R.~Uba, M.~Dumas, L.~Garc{\'i}a-Ba{\~{n}}uelos, and M.~La~Rosa, \emph{Clone
  Detection in Repositories of Business Process Models}.\hskip 1em plus 0.5em
  minus 0.4em\relax Berlin, Heidelberg: Springer Berlin Heidelberg, 2011, pp.
  248--264. [Online]. Available:
  \url{http://dx.doi.org/10.1007/978-3-642-23059-2_20}
\BIBentrySTDinterwordspacing

\bibitem{5090048}
S.~Grant and J.~R. Cordy, ``Vector space analysis of software clones,'' in
  \emph{2009 IEEE 17th International Conference on Program Comprehension}, May
  2009, pp. 233--237.

\bibitem{6079769}
Y.~Higo, U.~Yasushi, M.~Nishino, and S.~Kusumoto, ``Incremental code clone
  detection: A pdg-based approach,'' in \emph{2011 18th Working Conference on
  Reverse Engineering}, Oct 2011, pp. 3--12.

\bibitem{5376091}
J.~l.~Huang and F.~p.~Li, ``Quick similarity measurement of source code based
  on suffix array,'' in \emph{2009 International Conference on Computational
  Intelligence and Security}, vol.~2, Dec 2009, pp. 308--311.

\bibitem{Hummel:2011:IMC:1985404.1985409}
\BIBentryALTinterwordspacing
B.~Hummel, E.~Juergens, and D.~Steidl, ``Index-based model clone detection,''
  in \emph{Proceedings of the 5th International Workshop on Software Clones},
  ser. IWSC '11.\hskip 1em plus 0.5em minus 0.4em\relax New York, NY, USA: ACM,
  2011, pp. 21--27. [Online]. Available:
  \url{http://doi.acm.org/10.1145/1985404.1985409}
\BIBentrySTDinterwordspacing

\bibitem{Johnson:1993:IRS:962289.962305}
\BIBentryALTinterwordspacing
J.~H. Johnson, ``Identifying redundancy in source code using fingerprints,'' in
  \emph{Proceedings of the 1993 Conference of the Centre for Advanced Studies
  on Collaborative Research: Software Engineering - Volume 1}, ser. CASCON
  '93.\hskip 1em plus 0.5em minus 0.4em\relax IBM Press, 1993, pp. 171--183.
  [Online]. Available: \url{http://dl.acm.org/citation.cfm?id=962289.962305}
\BIBentrySTDinterwordspacing

\bibitem{Johnson:1994:VTR:782185.782217}
\BIBentryALTinterwordspacing
------, ``Visualizing textual redundancy in legacy source,'' in
  \emph{Proceedings of the 1994 Conference of the Centre for Advanced Studies
  on Collaborative Research}, ser. CASCON '94.\hskip 1em plus 0.5em minus
  0.4em\relax IBM Press, 1994, pp. 32--. [Online]. Available:
  \url{http://dl.acm.org/citation.cfm?id=782185.782217}
\BIBentrySTDinterwordspacing

\bibitem{336783}
------, ``Substring matching for clone detection and change tracking,'' in
  \emph{Proceedings 1994 International Conference on Software Maintenance}, Sep
  1994, pp. 120--126.

\bibitem{7069882}
T.~Kamiya, ``An execution-semantic and content-and-context-based code-clone
  detection and analysis,'' in \emph{2015 IEEE 9th International Workshop on
  Software Clones (IWSC)}, March 2015, pp. 1--7.

\bibitem{Lavoie:2010:CCR:1808901.1808905}
\BIBentryALTinterwordspacing
T.~Lavoie, M.~Eilers-Smith, and E.~Merlo, ``Challenging cloning related
  problems with gpu-based algorithms,'' in \emph{Proceedings of the 4th
  International Workshop on Software Clones}, ser. IWSC '10.\hskip 1em plus
  0.5em minus 0.4em\relax New York, NY, USA: ACM, 2010, pp. 25--32. [Online].
  Available: \url{http://doi.acm.org/10.1145/1808901.1808905}
\BIBentrySTDinterwordspacing

\bibitem{6840038}
F.~M. Lazar and O.~Banias, ``Clone detection algorithm based on the abstract
  syntax tree approach,'' in \emph{2014 IEEE 9th IEEE International Symposium
  on Applied Computational Intelligence and Informatics (SACI)}, May 2014, pp.
  73--78.

\bibitem{6821712}
C.~Li, J.~Sun, and H.~Chen, ``An improved method for tree-based clone detection
  in web applications,'' in \emph{2014 Fourth International Conference on
  Digital Information and Communication Technology and its Applications
  (DICTAP)}, May 2014, pp. 363--367.

\bibitem{Pulkkinen:2015:RBP:2812428.2812471}
\BIBentryALTinterwordspacing
P.~Pulkkinen, J.~Holvitie, O.~S. Nevalainen, and V.~Lepp\"{a}nen, ``Reusability
  based program clone detection: Case study on large scale healthcare software
  system,'' in \emph{Proceedings of the 16th International Conference on
  Computer Systems and Technologies}, ser. CompSysTech '15.\hskip 1em plus
  0.5em minus 0.4em\relax New York, NY, USA: ACM, 2015, pp. 90--97. [Online].
  Available: \url{http://doi.acm.org/10.1145/2812428.2812471}
\BIBentrySTDinterwordspacing

\bibitem{Struber2016}
\BIBentryALTinterwordspacing
D.~Str{\"u}ber, J.~Pl{\"o}ger, and V.~Acre{\c{T}}oaie, \emph{Clone Detection
  for Graph-Based Model Transformation Languages}.\hskip 1em plus 0.5em minus
  0.4em\relax Cham: Springer International Publishing, 2016, pp. 191--206.
  [Online]. Available: \url{http://dx.doi.org/10.1007/978-3-319-42064-6_13}
\BIBentrySTDinterwordspacing

\bibitem{Udagawa:2014:ESR:2684200.2684290}
\BIBentryALTinterwordspacing
Y.~Udagawa, ``An empirical study on retrieving structural clones using sequence
  pattern mining algorithms,'' in \emph{Proceedings of the 16th International
  Conference on Information Integration and Web-based Applications and
  Services}, ser. iiWAS '14.\hskip 1em plus 0.5em minus 0.4em\relax New York,
  NY, USA: ACM, 2014, pp. 270--276. [Online]. Available:
  \url{http://doi.acm.org/10.1145/2684200.2684290}
\BIBentrySTDinterwordspacing

\bibitem{Udagawa:2016:MFS:3011141.3011160}
\BIBentryALTinterwordspacing
------, ``Maximal frequent sequence mining for finding software clones,'' in
  \emph{Proceedings of the 18th International Conference on Information
  Integration and Web-based Applications and Services}, ser. iiWAS '16.\hskip
  1em plus 0.5em minus 0.4em\relax New York, NY, USA: ACM, 2016, pp. 26--33.
  [Online]. Available: \url{http://doi.acm.org/10.1145/3011141.3011160}
\BIBentrySTDinterwordspacing

\bibitem{1386166}
V.~Wahler, D.~Seipel, J.~Wolff, and G.~Fischer, ``Clone detection in source
  code by frequent itemset techniques,'' in \emph{Source Code Analysis and
  Manipulation, Fourth IEEE International Workshop on}, Sept 2004, pp.
  128--135.

\bibitem{6982615}
F.~Zhang, D.~Wu, P.~Liu, and S.~Zhu, ``Program logic based software plagiarism
  detection,'' in \emph{2014 IEEE 25th International Symposium on Software
  Reliability Engineering}, Nov 2014, pp. 66--77.

\bibitem{kclone}
Y.~Jia, D.~Binkley, M.~Harman, J.~Krinke, and M.~Matsushita, ``Kclone: a
  proposed approach to fast precise code clone detection,'' in
  \emph{Proceedings of the 3rd intentional conference on software clones},
  2009.

\bibitem{6405285}
M.~H. Alalfi, J.~R. Cordy, T.~R. Dean, M.~Stephan, and A.~Stevenson, ``Models
  are code too: Near-miss clone detection for simulink models,'' in \emph{2012
  28th IEEE International Conference on Software Maintenance (ICSM)}, Sept
  2012, pp. 295--304.

\bibitem{7447218}
J.~R. Cordy, ``Simone: architecture-sensitive near-miss clone detection for
  simulink models,'' in \emph{2015 First International Workshop on Automotive
  Software Architecture (WASA)}, May 2015, pp. 1--2.

\bibitem{6227873}
M.~H. Alalfi, J.~R. Cordy, T.~R. Dean, M.~Stephan, and A.~Stevenson,
  ``Near-miss model clone detection for simulink models,'' in \emph{2012 6th
  International Workshop on Software Clones (IWSC)}, June 2012, pp. 78--79.

\bibitem{6726700}
A.~Agrawal and S.~K. Yadav, ``A hybrid-token and textual based approach to find
  similar code segments,'' in \emph{2013 Fourth International Conference on
  Computing, Communications and Networking Technologies (ICCCNT)}, July 2013,
  pp. 1--4.

\bibitem{7880511}
C.~M. Kamalpriya and P.~Singh, ``Enhancing program dependency graph based clone
  detection using approximate subgraph matching,'' in \emph{2017 IEEE 11th
  International Workshop on Software Clones (IWSC)}, Feb 2017, pp. 1--7.

\bibitem{Kaur:2014:CDS:2597716.2597723}
\BIBentryALTinterwordspacing
R.~Kaur and S.~Singh, ``Clone detection in software source code using
  operational similarity of statements,'' \emph{SIGSOFT Softw. Eng. Notes},
  vol.~39, no.~3, pp. 1--5, Jun. 2014. [Online]. Available:
  \url{http://doi.acm.org/10.1145/2597716.2597723}
\BIBentrySTDinterwordspacing

\bibitem{6227963}
I.~Keivanloo and J.~Rilling, ``Clone detection meets semantic web-based
  transitive closure computation,'' in \emph{2012 First International Workshop
  on Realizing AI Synergies in Software Engineering (RAISE)}, June 2012, pp.
  12--16.

\bibitem{5460547}
E.~Kodhai, S.~Kanmani, A.~Kamatchi, R.~Radhika, and B.~V. Saranya, ``Detection
  of type-1 and type-2 code clones using textual analysis and metrics,'' in
  \emph{2010 International Conference on Recent Trends in Information,
  Telecommunication and Computing}, March 2010, pp. 241--243.

\bibitem{Matsushita:2017:DCC:3018882.3018892}
\BIBentryALTinterwordspacing
T.~Matsushita and I.~Sasano, ``Detecting code clones with gaps by function
  applications,'' in \emph{Proceedings of the 2017 ACM SIGPLAN Workshop on
  Partial Evaluation and Program Manipulation}, ser. PEPM 2017.\hskip 1em plus
  0.5em minus 0.4em\relax New York, NY, USA: ACM, 2017, pp. 12--22. [Online].
  Available: \url{http://doi.acm.org/10.1145/3018882.3018892}
\BIBentrySTDinterwordspacing

\bibitem{7880355}
J.~Petrík, D.~Chudá, and B.~Steinmüller, ``Source code plagiarism detection:
  The unix way,'' in \emph{2017 IEEE 15th International Symposium on Applied
  Machine Intelligence and Informatics (SAMI)}, Jan 2017, pp.
  000\,467--000\,472.

\bibitem{7880503}
Y.~Sabi, Y.~Higo, and S.~Kusumoto, ``Rearranging the order of program
  statements for code clone detection,'' in \emph{2017 IEEE 11th International
  Workshop on Software Clones (IWSC)}, Feb 2017, pp. 1--7.

\bibitem{Tekchandani2016}
\BIBentryALTinterwordspacing
R.~Tekchandani, R.~Bhatia, and M.~Singh, ``Semantic code clone detection for
  internet of things applications using reaching definition and liveness
  analysis,'' \emph{The Journal of Supercomputing}, pp. 1--28, 2016. [Online].
  Available: \url{http://dx.doi.org/10.1007/s11227-016-1832-6}
\BIBentrySTDinterwordspacing

\bibitem{Bauhaus}
Bauhaus-Stuttgart, ``Projekt bauhaus,'' \url{http://www.bauhaus-stuttgart.de/}.

\bibitem{Raza2006}
\BIBentryALTinterwordspacing
A.~Raza, G.~Vogel, and E.~Pl{\"o}dereder, \emph{Bauhaus -- A Tool Suite for
  Program Analysis and Reverse Engineering}.\hskip 1em plus 0.5em minus
  0.4em\relax Berlin, Heidelberg: Springer Berlin Heidelberg, 2006, pp. 71--82.
  [Online]. Available: \url{http://dx.doi.org/10.1007/11767077_6}
\BIBentrySTDinterwordspacing

\bibitem{bulychev2008duplicate}
P.~Bulychev and M.~Minea, ``Duplicate code detection using anti-unification,''
  in \emph{Spring Young Researchers Colloquium on Software Engineering}, 2008,
  pp. 51--54.

\bibitem{Basit:2005:DHS:1081706.1081733}
\BIBentryALTinterwordspacing
H.~A. Basit and S.~Jarzabek, ``Detecting higher-level similarity patterns in
  programs,'' in \emph{Proceedings of the 10th European Software Engineering
  Conference Held Jointly with 13th ACM SIGSOFT International Symposium on
  Foundations of Software Engineering}, ser. ESEC/FSE-13.\hskip 1em plus 0.5em
  minus 0.4em\relax New York, NY, USA: ACM, 2005, pp. 156--165. [Online].
  Available: \url{http://doi.acm.org/10.1145/1081706.1081733}
\BIBentrySTDinterwordspacing

\bibitem{4658086}
Y.~Zhang, H.~A. Basit, S.~Jarzabek, D.~Anh, and M.~Low, ``Query-based filtering
  and graphical view generation for clone analysis,'' in \emph{2008 IEEE
  International Conference on Software Maintenance}, Sept 2008, pp. 376--385.

\bibitem{4796208}
H.~A. Basit and S.~Jarzabek, ``A data mining approach for detecting
  higher-level clones in software,'' \emph{IEEE Transactions on Software
  Engineering}, vol.~35, no.~4, pp. 497--514, July 2009.

\bibitem{Basit:2011:VSC:1985404.1985406}
\BIBentryALTinterwordspacing
H.~A. Basit, U.~Ali, and S.~Jarzabek, ``Viewing simple clones from structural
  clones' perspective,'' in \emph{Proceedings of the 5th International Workshop
  on Software Clones}, ser. IWSC '11.\hskip 1em plus 0.5em minus 0.4em\relax
  New York, NY, USA: ACM, 2011, pp. 1--6. [Online]. Available:
  \url{http://doi.acm.org/10.1145/1985404.1985406}
\BIBentrySTDinterwordspacing

\bibitem{Sager:2006:DSJ:1137983.1138000}
\BIBentryALTinterwordspacing
T.~Sager, A.~Bernstein, M.~Pinzger, and C.~Kiefer, ``Detecting similar java
  classes using tree algorithms,'' in \emph{Proceedings of the 2006
  International Workshop on Mining Software Repositories}, ser. MSR '06.\hskip
  1em plus 0.5em minus 0.4em\relax New York, NY, USA: ACM, 2006, pp. 65--71.
  [Online]. Available: \url{http://doi.acm.org/10.1145/1137983.1138000}
\BIBentrySTDinterwordspacing

\bibitem{Jacob:2010:ACC:1808901.1808903}
\BIBentryALTinterwordspacing
F.~Jacob, D.~Hou, and P.~Jablonski, ``Actively comparing clones inside the code
  editor,'' in \emph{Proceedings of the 4th International Workshop on Software
  Clones}, ser. IWSC '10.\hskip 1em plus 0.5em minus 0.4em\relax New York, NY,
  USA: ACM, 2010, pp. 9--16. [Online]. Available:
  \url{http://doi.acm.org/10.1145/1808901.1808903}
\BIBentrySTDinterwordspacing

\bibitem{6779537}
G.~Mahajan and M.~Bharti, ``Implementing a 3-way approach of clone detection
  and removal using pc detector tool,'' in \emph{2014 IEEE International
  Advance Computing Conference (IACC)}, Feb 2014, pp. 1435--1441.

\bibitem{Yamamoto2005}
\BIBentryALTinterwordspacing
T.~Yamamoto, M.~Matsushita, T.~Kamiya, and K.~Inoue, \emph{Measuring Similarity
  of Large Software Systems Based on Source Code Correspondence}.\hskip 1em
  plus 0.5em minus 0.4em\relax Berlin, Heidelberg: Springer Berlin Heidelberg,
  2005, pp. 530--544. [Online]. Available:
  \url{http://dx.doi.org/10.1007/11497455_41}
\BIBentrySTDinterwordspacing

\bibitem{Li2012}
\BIBentryALTinterwordspacing
N.~Li, M.~Shen, S.~Li, L.~Zhang, and Z.~Li, \emph{STVsm: Similar Structural
  Code Detection Based on AST and VSM}.\hskip 1em plus 0.5em minus 0.4em\relax
  Berlin, Heidelberg: Springer Berlin Heidelberg, 2012, pp. 15--21. [Online].
  Available: \url{http://dx.doi.org/10.1007/978-3-642-35267-6_3}
\BIBentrySTDinterwordspacing

\bibitem{Al-Batran:2011:SCD:2050655.2050681}
\BIBentryALTinterwordspacing
B.~Al-Batran, B.~Sch\"{a}tz, and B.~Hummel, ``Semantic clone detection for
  model-based development of embedded systems,'' in \emph{Proceedings of the
  14th International Conference on Model Driven Engineering Languages and
  Systems}, ser. MODELS'11.\hskip 1em plus 0.5em minus 0.4em\relax Berlin,
  Heidelberg: Springer-Verlag, 2011, pp. 258--272. [Online]. Available:
  \url{http://dl.acm.org/citation.cfm?id=2050655.2050681}
\BIBentrySTDinterwordspacing

\bibitem{6671325}
E.~P. Antony, M.~H. Alalfi, and J.~R. Cordy, ``An approach to clone detection
  in behavioural models,'' in \emph{2013 20th Working Conference on Reverse
  Engineering (WCRE)}, Oct 2013, pp. 472--476.

\bibitem{Ballarin2016}
\BIBentryALTinterwordspacing
M.~Ballarin, R.~Lape{\~{n}}a, and C.~Cetina, \emph{Leveraging Feature Location
  to Extract the Clone-and-Own Relationships of a Family of Software
  Products}.\hskip 1em plus 0.5em minus 0.4em\relax Cham: Springer
  International Publishing, 2016, pp. 215--230. [Online]. Available:
  \url{http://dx.doi.org/10.1007/978-3-319-35122-3_15}
\BIBentrySTDinterwordspacing

\bibitem{7158423}
S.~Chodarev, E.~Pietriková, and J.~Kollár, ``Haskell clone detection using
  pattern comparing algorithm,'' in \emph{2015 13th International Conference on
  Engineering of Modern Electric Systems (EMES)}, June 2015, pp. 1--4.

\bibitem{Choi2009862}
\BIBentryALTinterwordspacing
S.~Choi, H.~Park, H.~il~Lim, and T.~Han, ``A static \{API\} birthmark for
  windows binary executables,'' \emph{Journal of Systems and Software},
  vol.~82, no.~5, pp. 862 -- 873, 2009. [Online]. Available:
  \url{http://www.sciencedirect.com/science/article/pii/S0164121208002689}
\BIBentrySTDinterwordspacing

\bibitem{daveycd}
N.~Davey, P.~Barson, S.~Field, and R.~Frank, ``The development of a software
  clone detector,'' \emph{International Journal of Applied Software
  Technology}, vol.~1, pp. 219--236, 1995.

\bibitem{6178896}
F.~Deissenboeck, L.~Heinemann, B.~Hummel, and S.~Wagner, ``Challenges of the
  dynamic detection of functionally similar code fragments,'' in \emph{2012
  16th European Conference on Software Maintenance and Reengineering}, March
  2012, pp. 299--308.

\bibitem{5941574}
D.~G. Devi and M.~Punithavalli, ``A hierarchical method for detecting
  codeclone,'' in \emph{2011 3rd International Conference on Electronics
  Computer Technology}, vol.~1, April 2011, pp. 126--128.

\bibitem{6227874}
R.~Elva and G.~T. Leavens, ``Semantic clone detection using method
  ioe-behavior,'' in \emph{2012 6th International Workshop on Software Clones
  (IWSC)}, June 2012, pp. 80--81.

\bibitem{6308805}
J.~He, ``Detecting c source code clones in college students' homework,'' in
  \emph{2012 International Conference on Computer Science and Information
  Processing (CSIP)}, Aug 2012, pp. 104--107.

\bibitem{Higo:2014:WMF:2635868.2635886}
\BIBentryALTinterwordspacing
Y.~Higo and S.~Kusumoto, ``How should we measure functional sameness from
  program source code? an exploratory study on java methods,'' in
  \emph{Proceedings of the 22Nd ACM SIGSOFT International Symposium on
  Foundations of Software Engineering}, ser. FSE 2014.\hskip 1em plus 0.5em
  minus 0.4em\relax New York, NY, USA: ACM, 2014, pp. 294--305. [Online].
  Available: \url{http://doi.acm.org/10.1145/2635868.2635886}
\BIBentrySTDinterwordspacing

\bibitem{7880504}
K.~Ito, T.~Ishio, and K.~Inoue, ``Web-service for finding cloned files using
  b-bit minwise hashing,'' in \emph{2017 IEEE 11th International Workshop on
  Software Clones (IWSC)}, Feb 2017, pp. 1--2.

\bibitem{juillerat2007}
N.~Juillerat and B.~Hirsbrunner, ``Detecting and removing clones in an
  automated way,'' in \emph{3rd Workshop on Software Evolution through
  Transformations: Embracing the Change}.\hskip 1em plus 0.5em minus
  0.4em\relax SeTra, 2006.

\bibitem{6649853}
I.~Keivanloo and J.~Rilling, ``Semantic-enabled clone detection,'' in
  \emph{2013 IEEE 37th Annual Computer Software and Applications Conference},
  July 2013, pp. 393--398.

\bibitem{6976163}
K.~Kumar, ``Detecting collaborative patterns in programs,'' in \emph{2014 IEEE
  International Conference on Software Maintenance and Evolution}, Sept 2014,
  pp. 664--664.

\bibitem{Kumar:2014:DDS:2660252.2660397}
\BIBentryALTinterwordspacing
K.~Kumar and S.~Jarzabek, ``Detecting design similarity patterns using program
  execution traces,'' in \emph{Proceedings of the Companion Publication of the
  2014 ACM SIGPLAN Conference on Systems, Programming, and Applications:
  Software for Humanity}, ser. SPLASH '14.\hskip 1em plus 0.5em minus
  0.4em\relax New York, NY, USA: ACM, 2014, pp. 55--56. [Online]. Available:
  \url{http://doi.acm.org/10.1145/2660252.2660397}
\BIBentrySTDinterwordspacing

\bibitem{Lee:2009:TDC:1651309.1651312}
\BIBentryALTinterwordspacing
H.-S. Lee and K.-G. Doh, ``Tree-pattern-based duplicate code detection,'' in
  \emph{Proceedings of the ACM First International Workshop on Data-intensive
  Software Management and Mining}, ser. DSMM '09.\hskip 1em plus 0.5em minus
  0.4em\relax New York, NY, USA: ACM, 2009, pp. 7--12. [Online]. Available:
  \url{http://doi.acm.org/10.1145/1651309.1651312}
\BIBentrySTDinterwordspacing

\bibitem{maeda}
\BIBentryALTinterwordspacing
K.~Maeda, ``Syntax sensitive and language independent detection of code
  clones,'' \emph{International Journal of Computer, Electrical, Automation,
  Control and Information Engineering}, vol.~3, no.~12, pp. 2845 -- 2849, 2009.
  [Online]. Available: \url{http://waset.org/Publications?p=36}
\BIBentrySTDinterwordspacing

\bibitem{989796}
A.~Marcus and J.~I. Maletic, ``Identification of high-level concept clones in
  source code,'' in \emph{Proceedings 16th Annual International Conference on
  Automated Software Engineering (ASE 2001)}, Nov 2001, pp. 107--114.

\bibitem{6478318}
M.~Dong, H.~Zhuang, R.~Zhang, S.~Bi, X.~Zeng, S.~Guo, W.~Cai, and Z.~Tang, ``A
  new method of software clone detection based on binary instruction structure
  analysis,'' in \emph{2012 8th International Conference on Wireless
  Communications, Networking and Mobile Computing}, Sept 2012, pp. 1--4.

\bibitem{6832302}
K.~Raheja and R.~K. Tekchandani, ``An efficient code clone detection model on
  java byte code using hybrid approach,'' in \emph{Confluence 2013: The Next
  Generation Information Technology Summit (4th International Conference)},
  Sept 2013, pp. 16--21.

\bibitem{6420770}
D.~Rattan, R.~Bhatia, and M.~Singh, ``Model clone detection based on tree
  comparison,'' in \emph{2012 Annual IEEE India Conference (INDICON)}, Dec
  2012, pp. 1041--1046.

\bibitem{Rodrigues2010}
\BIBentryALTinterwordspacing
N.~Rodrigues and J.~L. Vila{\c{c}}a, \emph{Identifying Clones in Functional
  Programs for Refactoring}.\hskip 1em plus 0.5em minus 0.4em\relax Berlin,
  Heidelberg: Springer Berlin Heidelberg, 2010, pp. 309--317. [Online].
  Available: \url{http://dx.doi.org/10.1007/978-3-642-16402-6_33}
\BIBentrySTDinterwordspacing

\bibitem{Santone:2011:CDT:1985404.1985422}
\BIBentryALTinterwordspacing
A.~Santone, ``Clone detection through process algebras and java bytecode,'' in
  \emph{Proceedings of the 5th International Workshop on Software Clones}, ser.
  IWSC '11.\hskip 1em plus 0.5em minus 0.4em\relax New York, NY, USA: ACM,
  2011, pp. 73--74. [Online]. Available:
  \url{http://doi.acm.org/10.1145/1985404.1985422}
\BIBentrySTDinterwordspacing

\bibitem{Singh:2014:CDU:2597716.2597726}
\BIBentryALTinterwordspacing
S.~Singh and R.~Kaur, ``Clone detection in uml class models using class
  metrics,'' \emph{SIGSOFT Softw. Eng. Notes}, vol.~39, no.~3, pp. 1--3, Jun.
  2014. [Online]. Available: \url{http://doi.acm.org/10.1145/2597716.2597726}
\BIBentrySTDinterwordspacing

\bibitem{SINGH2015915}
\BIBentryALTinterwordspacing
M.~Singh and V.~Sharma, ``Detection of file level clone for high level
  cloning,'' \emph{Procedia Computer Science}, vol.~57, pp. 915 -- 922, 2015.
  [Online]. Available:
  \url{http://www.sciencedirect.com/science/article/pii/S1877050915020384}
\BIBentrySTDinterwordspacing

\bibitem{SUDHAMANI2015892}
\BIBentryALTinterwordspacing
M.~Sudhamani and L.~Rangarajan, ``Structural similarity detection using
  structure of control statements,'' \emph{Procedia Computer Science}, vol.~46,
  pp. 892 -- 899, 2015. [Online]. Available:
  \url{http://www.sciencedirect.com/science/article/pii/S1877050915002239}
\BIBentrySTDinterwordspacing

\bibitem{6141393}
A.~Surendran, P.~Samuel, and K.~P. Jacob, ``Code clones in program test
  sequence identification,'' in \emph{2011 World Congress on Information and
  Communication Technologies}, Dec 2011, pp. 1050--1055.

\bibitem{Sutton:2005:HEA:1068009.1068191}
\BIBentryALTinterwordspacing
A.~Sutton, H.~Kagdi, J.~I. Maletic, and L.~G. Volkert, ``Hybridizing
  evolutionary algorithms and clustering algorithms to find source-code
  clones,'' in \emph{Proceedings of the 7th Annual Conference on Genetic and
  Evolutionary Computation}, ser. GECCO '05.\hskip 1em plus 0.5em minus
  0.4em\relax New York, NY, USA: ACM, 2005, pp. 1079--1080. [Online].
  Available: \url{http://doi.acm.org/10.1145/1068009.1068191}
\BIBentrySTDinterwordspacing

\bibitem{Tairas:2006:PCD:1185448.1185597}
\BIBentryALTinterwordspacing
R.~Tairas and J.~Gray, ``Phoenix-based clone detection using suffix trees,'' in
  \emph{Proceedings of the 44th Annual Southeast Regional Conference}, ser.
  ACM-SE 44.\hskip 1em plus 0.5em minus 0.4em\relax New York, NY, USA: ACM,
  2006, pp. 679--684. [Online]. Available:
  \url{http://doi.acm.org/10.1145/1185448.1185597}
\BIBentrySTDinterwordspacing

\bibitem{Tairas2012}
\BIBentryALTinterwordspacing
R.~Tairas and J.~Cabot, \emph{Cloning in DSLs: Experiments with OCL}.\hskip 1em
  plus 0.5em minus 0.4em\relax Berlin, Heidelberg: Springer Berlin Heidelberg,
  2012, pp. 60--76. [Online]. Available:
  \url{http://dx.doi.org/10.1007/978-3-642-28830-2_4}
\BIBentrySTDinterwordspacing

\bibitem{6832306}
R.~Tekchandani, R.~K. Bhatia, and M.~Singh, ``Semantic code clone detection
  using parse trees and grammar recovery,'' in \emph{Confluence 2013: The Next
  Generation Information Technology Summit (4th International Conference)},
  Sept 2013, pp. 41--46.

\bibitem{Tekin:2012:MOD:2664398.2664405}
\BIBentryALTinterwordspacing
U.~Tekin, U.~Erdemir, and F.~Buzluca, ``Mining object-oriented design models
  for detecting identical design structures,'' in \emph{Proceedings of the 6th
  International Workshop on Software Clones}, ser. IWSC '12.\hskip 1em plus
  0.5em minus 0.4em\relax Piscataway, NJ, USA: IEEE Press, 2012, pp. 43--49.
  [Online]. Available: \url{http://dl.acm.org/citation.cfm?id=2664398.2664405}
\BIBentrySTDinterwordspacing

\bibitem{4556127}
C.~K. Roy and J.~R. Cordy, ``Scenario-based comparison of clone detection
  techniques,'' in \emph{2008 16th IEEE International Conference on Program
  Comprehension}, June 2008, pp. 153--162.

\bibitem{1342759}
F.~V. Rysselberghe and S.~Demeyer, ``Evaluating clone detection techniques from
  a refactoring perspective,'' in \emph{Proceedings. 19th International
  Conference on Automated Software Engineering, 2004.}, Sept 2004, pp.
  336--339.

\bibitem{Tiarks2011}
\BIBentryALTinterwordspacing
R.~Tiarks, R.~Koschke, and R.~Falke, ``An extended assessment of type-3 clones
  as detected by state-of-the-art tools,'' \emph{Software Quality Journal},
  vol.~19, no.~2, pp. 295--331, 2011. [Online]. Available:
  \url{http://dx.doi.org/10.1007/s11219-010-9115-6}
\BIBentrySTDinterwordspacing

\bibitem{5279980}
------, ``An assessment of type-3 clones as detected by state-of-the-art
  tools,'' in \emph{2009 Ninth IEEE International Working Conference on Source
  Code Analysis and Manipulation}, Sept 2009, pp. 67--76.

\bibitem{Ragkhitwetsagul2016}
\BIBentryALTinterwordspacing
C.~Ragkhitwetsagul, M.~Paixao, M.~Adham, S.~Busari, J.~Krinke, and J.~H. Drake,
  \emph{Searching for Configurations in Clone Evaluation -- A Replication
  Study}.\hskip 1em plus 0.5em minus 0.4em\relax Cham: Springer International
  Publishing, 2016, pp. 250--256. [Online]. Available:
  \url{http://dx.doi.org/10.1007/978-3-319-47106-8_20}
\BIBentrySTDinterwordspacing

\bibitem{1542064}
M.~Bruntink, A.~van Deursen, R.~van Engelen, and T.~Tourwe, ``On the use of
  clone detection for identifying crosscutting concern code,'' \emph{IEEE
  Transactions on Software Engineering}, vol.~31, no.~10, pp. 804--818, Oct
  2005.

\bibitem{Wang:2013:SBC:2491411.2491420}
\BIBentryALTinterwordspacing
T.~Wang, M.~Harman, Y.~Jia, and J.~Krinke, ``Searching for better
  configurations: A rigorous approach to clone evaluation,'' in
  \emph{Proceedings of the 2013 9th Joint Meeting on Foundations of Software
  Engineering}, ser. ESEC/FSE 2013.\hskip 1em plus 0.5em minus 0.4em\relax New
  York, NY, USA: ACM, 2013, pp. 455--465. [Online]. Available:
  \url{http://doi.acm.org/10.1145/2491411.2491420}
\BIBentrySTDinterwordspacing

\bibitem{6613045}
S.~Schulze and D.~Meyer, ``On the robustness of clone detection to code
  obfuscation,'' in \emph{2013 7th International Workshop on Software Clones
  (IWSC)}, May 2013, pp. 62--68.

\bibitem{7332459}
J.~Svajlenko and C.~K. Roy, ``Evaluating clone detection tools with
  {B}ig{C}lone{B}ench,'' in \emph{2015 IEEE International Conference on
  Software Maintenance and Evolution (ICSME)}, Sept 2015, pp. 131--140.

\bibitem{7816530}
C.~Ragkhitwetsagul, ``Measuring code similarity in large-scaled code corpora,''
  in \emph{2016 IEEE International Conference on Software Maintenance and
  Evolution (ICSME)}, Oct 2016, pp. 626--630.

\bibitem{7781805}
C.~Ragkhitwetsagul, J.~Krinke, and D.~Clark, ``Similarity of source code in the
  presence of pervasive modifications,'' in \emph{2016 IEEE 16th International
  Working Conference on Source Code Analysis and Manipulation (SCAM)}, Oct
  2016, pp. 117--126.

\end{thebibliography}



\end{document}